\documentclass[twocolumn,twocolappendix]{aastex63}

\newcommand{\Angstrom}{\mathring{\rm A}}
\newcommand{\degsq}{\mbox{$\rm deg^{2}$}}

\newcommand{\psec}{\mbox{$\rm s^{-1}$}}

\newcommand{\erg}{\mbox{$\rm erg$}}
\newcommand{\pcmsq}{\mbox{$\rm cm^{-2}$}}

\usepackage{amsmath}

\shorttitle{Optical to X-ray Signatures of CSM Interaction}
\shortauthors{Margalit, Quataert, \& Ho}
\graphicspath{{./}{figures/}}

\begin{document}

\title{Optical to X-ray Signatures of Dense Circumstellar Interaction in Core-Collapse Supernovae}

\correspondingauthor{Ben Margalit}
\email{benmargalit@berkeley.edu}

\author[0000-0001-8405-2649]{Ben Margalit}
\altaffiliation{NASA Einstein Fellow}
\affiliation{Astronomy Department and Theoretical Astrophysics Center, University of California, Berkeley, Berkeley, CA 94720, USA}

\author[0000-0001-9185-5044]{Eliot Quataert}
\affiliation{Department of Astrophysical Sciences, Princeton University, Princeton, NJ 08544, USA}

\author[0000-0002-9017-3567]{Anna Y. Q.~Ho}
\affiliation{Department of Astronomy, University of California, Berkeley, 501 Campbell Hall, Berkeley, CA, 94720, USA}
\affiliation{Miller Institute for Basic Research in Science, 468 Donner Lab, Berkeley, CA 94720, USA}



\begin{abstract}
Progenitors of core-collapse supernovae (SNe) can shed significant mass to circumstellar material (CSM) in the months--years preceding core-collapse. The ensuing SN explosion launches ejecta that may subsequently collide with this CSM, producing shocks that can power emission across the electromagnetic spectrum. In this work we explore the thermal signatures of dense CSM interaction, when the CSM density profile is truncated at some outer radius. CSM with optical depth $>c/v$ (where $v$ is the shock velocity) will produce primarily $\sim$blackbody optical/UV emission whereas lower optical-depth CSM will power bremsstrahlung X-ray emission. Focusing on the latter, we derive light-curves and spectra of the resulting X-ray transients, that include a detailed treatment of Comptonization. Due to strong photoelectric absorption, the X-ray light-curve is dominated by the `post-interaction' phase that occurs after the shock reaches the CSM truncation radius. We treat this regime here for the first time. Using these results, we present the phase-space of optical, UV, and X-ray transients as a function of CSM properties, and discuss detectability prospects. We find that ROSAT would not have been sensitive to CSM X-ray transients but that eROSITA is expected to detect many such events. Future wide-field UV missions such as ULTRASAT will dramatically enhance sensitivity to large optical-depth CSM configurations. Finally, we present a framework within which CSM properties may be directly inferred from observable features of X-ray transients. This can serve as an important tool for studying stellar mass loss using SN X-ray detections.
\end{abstract}



\section{Introduction} 
\label{sec:intro}

Towards the end of their lives, massive stars can shed significant mass through winds or eruptions, thus polluting their immediate environment with dense circumstellar material (CSM). The death of these stars results in a violent supernova (SN) explosion, whose observational features may be markedly shaped by such CSM. First light produced by the SN (at shock breakout from the stellar surface) can flash-ionize nearby CSM and produce narrow emission lines \citep[e.g.][]{Yaron+17,Bruch+21}. As material ejected from the SN expands, it will ultimately shock the surrounding CSM. This  CSM interaction is observable in a myriad of ways.

Type IIn SNe exhibit narrow emission lines indicative of ionization of the surrounding CSM, and their optical light-curves have long been modelled as being powered by CSM interaction \citep[e.g.][]{Smith17}.
Radio SNe are another well-studied class of events, whose bright radio luminosity are powered by non-thermal emission in SN-ejecta--CSM shocks \citep{Chevalier82,Weiler+86,Chevalier98,Weiler+02,Chevalier&Fransson17}.

More recently, wide-field optical surveys are discovering rare optical transients whose light-curves cannot be powered by radioactivity \citep{Quimby+07,Drout+14}. CSM-interaction is an appealing alternative energy source.
One class of such optical transients are superluminous SNe (SLSNe; \citealt{GalYam12}). These SNe are extremely bright, are found preferentially in low-metalicity dwarf galaxies similar to the hosts of long gamma-ray bursts (GRBs; \citealt{Lunnan+14,Perley+16}), and are typically divided into Type-I/II subclasses depending on whether hydrogen is observed in their spectra.
Two leading theories for the mechanism powering SLSNe have emerged: CSM interaction \citep[e.g.][]{Chevalier&Irwin11,Ginzburg&Balberg12}, and spin-down of a rapidly-rotating highly-magnetized NS (a millisecond magnetar) that may have been born in the stellar explosion \citep{Kasen&Bildsten10,Woosley10,Metzger+15}.\footnote{Note that other (non-magnetar) engine-powered models have also been proposed \citep[e.g.][]{Dexter&Kasen13}.}
Interaction models face challenges with SLSN-I, as it is difficult to explain the lack of hydrogen features given the large CSM masses inferred. Instead, these events have been extensively modelled within the magnetar model \citep[e.g.][]{Inserra+13,Nicholl+17}. CSM-interaction remains an appealing model for SLSNe-II, and has been used to model the light-curves of these events \citep{Chatzopoulos+13,Inserra+18}.

In recent years, another class of optical transients called Fast Blue Optical Transients (FBOTs)\footnote{
These have also been termed Fast Evolving Luminous Transients (FELTs), or just `rapidly evolving transients'. In lieu of a standardized naming convention, we adopt FBOT in this work.
} 
are being discovered (\citealt{Ofek+10,Drout+14,Arcavi+16,Pursiainen+18,Rest+18,Prentice+18,Ho+21}).
These events have short ($\sim$several day) duration and cannot be explained by standard $^{56}$Ni radioactive decay. Instead, they have been successfully modelled by interaction of fast SN ejecta with dense CSM \citep[e.g.][]{Rest+18}.
Interestingly, this modeling suggests that a `shell'-like CSM structure with an abrupt outer truncation radius is necessary to explain the observations.   Such an outer edge to the CSM is in fact expected theoretically if mass-loss is enhanced in late stages of massive stellar evolution (e.g., \citealt{Quataert2012}). 
A truncated CSM profile may therefore also be relevant to other classes of interacting SNe.

Motivated by these recent discoveries at optical bands, we here address the timely question: {\it what are the signatures of dense CSM-interaction at other wavelengths?}
In this work, we systematically explore the parameter-space of CSM interaction and investigate signatures that can arise in different regions of parameter space, with a particular focus on X-rays. 
Readers that are interested only in our primary results and their relation to observations may wish to skip ahead to \S\ref{sec:results} and \S\ref{sec:inferring_CSM}.

We begin by considering the relevant processes and their associated timescales (\S\ref{sec:GeneralConsiderations}). In \S\ref{sec:ParameterSpace} we present the parameter-space of shocks in dense CSM, which can be fully specified in terms of the shock velocity and CSM Thompson optical depth. We then discuss X-ray emission produced during the interaction phase (\S\ref{sec:InteractionPhase}), including Comptonization, bound-free absorption, and non-thermal synchrotron emission. We continue by discussing the post-interaction phase, and derive X-ray light-curves in the various regimes of interest (\S\ref{sec:ExpansionPhase}). In \S\ref{sec:UVOptical} we briefly review properties of optical/UV emission produced by CSM with optical depth $>c/v$. In \S\ref{sec:results} we use our results to discuss the phase-space of CSM-interaction powered thermal transients, and estimate detectability prospects by X-ray and UV surveys. Finally, we use our results to present a framework within which CSM properties can be directly inferred from X-ray transient observations (\S\ref{sec:inferring_CSM}). We summarize our findings and conclude in \S\ref{sec:conclusions}.

\section{General Considerations}
\label{sec:GeneralConsiderations}

We consider the following physical setup: a constant-density CSM shell of mass $M$ and width $\Delta R \sim R$ is located at radius $R$. A shock propagates through this CSM with velocity $v$, reaching the outer CSM edge at $t=0$. This shock converts bulk kinetic energy into thermal energy, efficiently heating the post-shock CSM material which can subsequently radiate some of this energy. Our assumption of a top-hat (constant) density profile for the CSM shell is somewhat arbitrary, motivated primarily by convenience. However our results do not depend sensitively on this assumption so long as the density profile is less steep than $\propto r^{-3}$ and most of the mass is located at $r \sim R$. In particular, our results can be easily applied to the case of a truncated wind density profile $\rho_{\rm w} = \dot{M}/4\pi v_{\rm w}r^2$ ($r<R$) under the simple transformation $M \to \Delta R \dot{M}/v_{\rm w}$. The top-hat approach reasonably describes the intrinsic emission so long as the density external to the CSM shell (at $r>R$) is $\ll$ than that at $r \lesssim R$, and the total mass of any such ``circum-shell'' material is $\ll M$. For most purposes we therefore assume $\rho \sim 0$ at $r>R$, however in Appendix~\ref{sec:Appendix_Wind_IonizationBreakout} we also consider the effects of bound-free absorption by an ambient low-density circum-shell wind, finding that such effects are typically negligible in regions of interest.

We consider separately emission during the `interaction phase', while the shock is still propagating within the dense CSM shell (at $t<0$; \S\ref{sec:InteractionPhase}), and the subsequent `post-interaction phase' at $t>0$ that accounts for expansion and cooling of the (now fully-shocked) CSM shell (\S\ref{sec:ExpansionPhase}).

We begin with general considerations of the shocked CSM properties, focusing first on the low optical-depth regime.
In this regime,
a gas-pressure dominated collisionless shock forms with a post-shock temperature
\begin{equation}
\label{eq:Te}
k_B T_e 
= \epsilon_T \frac{3}{16} \mu m_p v^2
\approx 120 \, {\rm keV} \,  \epsilon_T v_9^2 
.
\end{equation}
Here $v_9 \equiv v / 10^9 \, {\rm cm \, s}^{-1}$ is the shock velocity normalized to $10,000 \, {\rm km \, s}^{-1}$, $\epsilon_T$ is a fudge factor $\lesssim 1$ that allows for non-equilibrium electron temperatures\footnote{
Coulomb collisions will equilibrate electrons with ions over a timescale $t_{ei} \approx 5.7 \times 10^4 \, {\rm s} \, M_{-1}^{-1} R_{15}^{3} v_9^3 \left(\ln\Lambda/10\right)^{-1}$, where $\ln\Lambda$ is the Coulomb logarithm
and $t_{ei} \propto n_e^{-1} T_e^{3/2}$ (\citealt{Spitzer56}; eqs.~\ref{eq:Te},\ref{eq:ne}). If 
$t_{ei}$ is long ($> \min \left[t_{\rm dyn},t_{\rm cool}\right]$)
then Coulomb processes cannot fully equilibrate  electrons and ions and $T_e$ may be lower than implied by eq.~(\ref{eq:Te}), i.e., $\epsilon_T \lesssim 1$. However, this depends on the relative heating of electrons and ions by collisionless shocks and the possible role of plasma instabilities in enhancing post-shock electron-ion coupling.   Throughout this paper we account for these uncertainties via the parameter $\epsilon_T$ in eq. \ref{eq:Te}.
}
(electron-ion equilibrium implies $\epsilon_T=1$) and we have adopted a mean molecular weight of $\mu \simeq 0.62$ appropriate for fully-ionized material with Solar composition.
In general, the shock velocity may change as it propagates through the CSM. Here we use $v$ to denote the velocity of the shock near the outer CSM edge ($r \sim R$) and neglect order-unity corrections that may arise due to shock deceleration.

The immediate post-shock electron density is
\begin{equation}
\label{eq:ne}
n_e = \frac{\tilde{r} M / \mu_e m_p}{4\pi R^2 \Delta R }
\approx 3.2 \times 10^{10} \, {\rm cm}^{-3} \, \left(\frac{\tilde{r}}{4}\right) M_{-1} R_{15}^{-3}  
\end{equation}
where $\tilde{r}$ is the shock compression ratio,
$M_{-1} \equiv M / 0.1 M_\odot$, $R_{15} \equiv R/10^{15} \, {\rm cm}$ are the CSM mass and radius normalized to fiducial values, and $\mu_e \simeq 1.18$. Above and in the following we take the CSM shell width to be $\Delta R = R$ and omit explicit dependence on $\Delta R$. Smaller values of $\Delta R < R$ can be easily accounted for by replacing $M \to \left(\Delta R/R\right)^{-1} M$ in most expressions below.

The characteristics of thermal emission from the shocked shell are fully specified by the shock velocity $v$ along with two CSM parameters (e.g. any combination of $R$, $M$, $n_e$). 
Although quantitative values will depend on this full three-parameter family, we show below that the shock phase-space is effectively separated into qualitatively distinct regions with only two variables: the shock velocity, and CSM column density ($\propto n_e R$). A useful parameterization of the phase-space is therefore given in terms of the CSM Thompson optical depth,
\begin{equation}
\label{eq:tau_T}
\tau_T = n_e \sigma_T R / \tilde{r}
\approx 5.3 \, M_{-1} R_{15}^{-2} .
\end{equation}
The resulting phase-space is discussed in \S\ref{sec:ParameterSpace} and illustrated in Fig.~\ref{fig:PhaseSpace}, where different curves separate regions based on the hierarchy of timescales in the problem.
These timescales are described below.

\subsection{Timescales}

The first relevant timescale is the dynamical time $t_{\rm dyn}$ in which the shock crosses the CSM shell,
\begin{equation}
\label{eq:t_dyn}
t_{\rm dyn} = R/v 
\approx 10^6 \, {\rm s}
\, R_{15} v_9^{-1} .
\end{equation}
A second important timescale is the time it takes photons to escape the medium,
\begin{equation}
\label{eq:t_esc}
t_{\rm esc} = \max\left(1, \tau_T\right) \frac{fR}{c} 
\approx 
3.3 \times 10^4 \, {\rm s} \, 
\max\left(1,\tau_T\right) fR_{15}
,
\end{equation}
where $f \leq 1$ is a parameter that is useful in cases where the effective width of the medium is less than the CSM shell width (e.g. for radiative shocks). Otherwise, $f=1$.

Turning to radiative processes, we show in \S\ref{sec:Comptonization} that inverse Compton scattering can affect both the emergent X-ray spectrum and the electron cooling rate.
The characteristic timescale for soft photons to Compton-upscatter on hot thermal electrons is
\begin{equation}
\label{eq:t_IC}
t_{\rm IC} = \frac{1}{n_e \sigma_T c} \left(\frac{4 k_B T_e}{m_e c^2}\right)^{-1}
\approx 1.6 \times 10^3 \, {\rm s} \, M_{-1}^{-1} R_{15}^3 \epsilon_T^{-1} v_9^{-2}
.
\end{equation}
Finally, bremsstrahlung is the primary photon-production mechanism and free-free emission governs cooling of shock-heated electrons in much of the parameter space. The free-free electron cooling time is therefore another important timescale. It is
\begin{equation}
\label{eq:t_ff}
t_{\rm ff} = \frac{\frac{3}{2} n_e k_B T_e}{\Lambda_{\rm ff}(n_e,T_e)} 
\approx 5.7 \times 10^4 \,{\rm s} \, M_{-1}^{-1} R_{15}^3 \epsilon_T^{1/2} v_9
,
\end{equation}
where $\Lambda_{\rm ff} \propto n_e^2 T_e^{1/2} \bar{g}(T_e)$ is the free-free cooling rate \citep[e.g.][]{Draine} and in the final expression we have neglected the weak (logarithmic) dependence of the frequency-averaged gaunt factor $\bar{g}$ on temperature.\footnote{$\bar{g}(T_e) \simeq 2.97$ for our reference parameters (eq.~\ref{eq:Te}), and spans between $\simeq$2--4 within the full parameter-space of interest.}
We note that our treatment of Comptonization and free-free emission assumes non-relativistic electron temperatures. In Appendix~\ref{sec:Appendix_Pairs} we extend this with more detailed calculations that show that, at shock velocities of interest here ($v_9 \sim 1$), our non-relativistic treatment will suffice.

\subsection{Parameter Space}
\label{sec:ParameterSpace}

\begin{figure}
    \centering
    \includegraphics[width=0.5\textwidth]{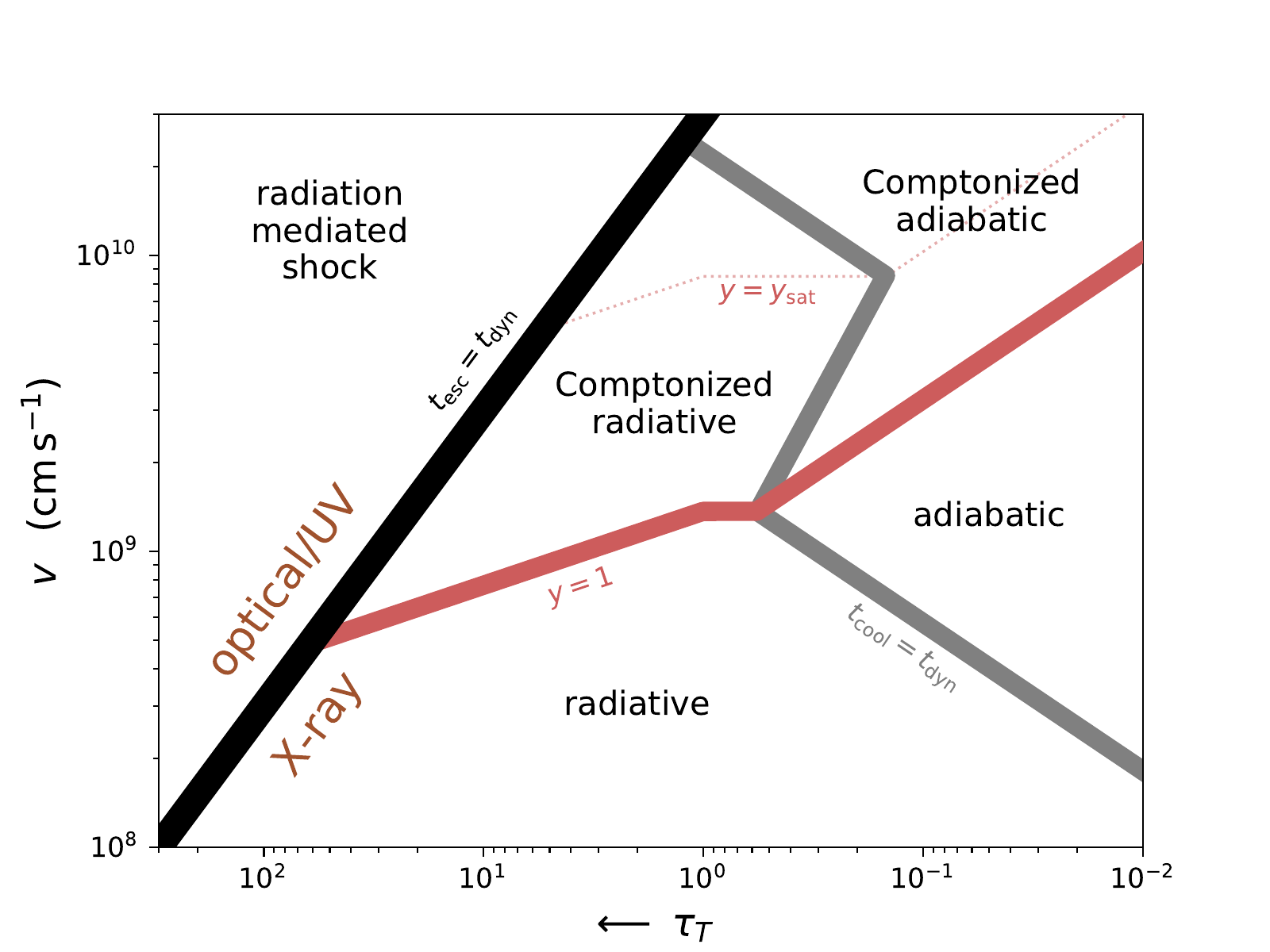}
    \caption{
    Parameter space for non-relativistic shocks in confined CSM shells. The phase space is fully characterized by the shock velocity $v$ and the CSM column density, here parameterized in terms of the Thompson optical depth $\tau_T$ (eq.~\ref{eq:tau_T}; note that $\tau_T$ is plotted on an inverse scale to facilitate comparison with Fig.~\ref{fig:Results}).
    The black curve separates collisionless shocks (below) whose emission peaks in the X-rays from radiation-mediated shocks (above) that produce predominantly optical/UV emission. Collisionless shocks can be characterized as either adiabatic or radiative, and may be further separated into regions where inverse-Compton scattering is important or negligible. See \S\ref{sec:ParameterSpace} for further details.
    Appendix~\ref{sec:Appendix_Pairs} shows that this schematic is in line with more detailed calculations that include relativistic corrections (important at high shock velocities).
    }
    \label{fig:PhaseSpace}
\end{figure}

Figure~\ref{fig:PhaseSpace} shows the CSM shock parameter space, which is separated into several distinct regions.
Starting from low optical depth (the right end in Fig.~\ref{fig:PhaseSpace}), the shock is optically thin and slow-cooling (/adiabatic), i.e. the cooling time is long compared to the dynamical timescale. As one moves towards higher optical depths (e.g. by increasing the CSM density $n_e$, or decreasing the shell radius $R$) the free-free cooling time decreases with respect to the dynamical timescale, and eventually they come to equal one another once (see eqs.~\ref{eq:tau_T},\ref{eq:t_dyn},\ref{eq:t_ff})
\begin{equation}
\label{eq:tff_equal_tdyn}
t_{\rm ff} = t_{\rm dyn}
:~~~
\tau_T \approx 0.30 \epsilon_T^{1/2} v_9^2 
.
\end{equation}
This is depicted by the lower solid-grey curve in Fig.~\ref{fig:PhaseSpace} (below the red curve).
To the left/below this curve, $t_{\rm ff} < t_{\rm dyn}$ and the shock is radiative.

For sufficiently high shock velocities, one encounters the solid-red curve. To the left/above this curve $t_{\rm IC} < t_{\rm esc}$ and photons Comptonize before escaping the scattering medium. This is equivalent to the condition $y>1$ on the Compton-y parameter 
(eqs.~\ref{eq:t_esc},\ref{eq:t_IC}), 
\begin{equation}
\label{eq:y}
y 
= \frac{t_{\rm esc}}{t_{\rm IC}}
= 
f \tau_T \max \left(1,\tau_T\right) 
\left(\frac{4 k_B T_e}{m_e c^2}\right) .
\end{equation}
The limiting case is schematically illustrated by the red curve in Fig.~\ref{fig:PhaseSpace} along which $y=1$. In the adiabatic regime $f=1$ and this reduces to
\begin{equation}
\label{eq:tIC_equal_tesc}
{\rm adiabatic}
:~
y=1 ~\leftrightarrow~
\max \left(\tau_T,\tau_T^2\right) \approx 1.06 \epsilon_T^{-1} v_9^{-2}
.
\end{equation}
In the radiative regime, post-shock electrons remain hot only within a small fraction $\sim t_{\rm cool}/t_{\rm dyn} < 1$ of the CSM width. 
This can be accounted for by taking $f = t_{\rm ff}/t_{\rm dyn}$ in the Compton-y parameter expression (eq.~\ref{eq:y}).
The condition for Comptonization in the radiative regime is therefore
\begin{equation}
{\rm radiative}
:~~
y=1 ~\leftrightarrow~
v_9 \approx 1.37 \epsilon_T^{-3/8} \max \left(1,\tau_T\right)^{-1/4}
,
\end{equation}
which takes precedence over eq.~(\ref{eq:tIC_equal_tesc}) for $\tau_T \lesssim 0.57$ (at which $t_{\rm ff} < t_{\rm dyn}$ when $y=1$; eq.~\ref{eq:tff_equal_tdyn}). This is shown by the horizontal branch of the red curve in Fig.~\ref{fig:PhaseSpace}.

Compton scattering is important anywhere above the red curves. A significant by-product of efficient Compton scattering is an enhanced cooling rate. The subsequent discussion in \S\ref{sec:Comptonization} shows that, beyond some transition region of intermediate Compton-y, the $y > y_{\rm sat}$ cooling rate saturates at an approximately constant value and can be described as an enhancement by factor $\mathcal{C} \sim$several hundred to the bremsstrahlung cooling rate. 
In this saturated regime, the electron cooling rate is simply $t_{\rm ff} / \mathcal{C}$. Therefore, the transition between fast- and slow-cooling shocks in the saturated Comptonized region of parameter space is given by an analog of eq.~(\ref{eq:tff_equal_tdyn}) with $t_{\rm ff}$ replaced by $t_{\rm ff} / \mathcal{C}$. This condition is shown as the top solid-grey curve in Fig.~\ref{fig:PhaseSpace} which separates the radiative and adiabatic Comptonized regions.
For purposes of this schematic figure we set $y_{\rm sat} = 10$ (see Appendix~\ref{sec:Appendix_Comptonization} for more detailed estimates).

Finally, as one moves to even higher optical depths the condition $t_{\rm esc} = t_{\rm dyn}$ is reached and photons become trapped within the CSM until shock breakout. This condition is depicted by the solid-black curve in Fig.~\ref{fig:PhaseSpace} along which (eqs.~\ref{eq:t_dyn},\ref{eq:t_esc})
\begin{equation}
\label{eq:tesc_equal_tdyn}
t_{\rm esc} = t_{\rm dyn}
:~~~
\tau_T \approx \frac{c}{v} \approx 30 v_9^{-1}
.
\end{equation}

Shocks propagating in CSM shells whose optical depth exceeds eq.~(\ref{eq:tesc_equal_tdyn}; to the left of the black curve) form a radiation mediated shock
in which $U_{\rm rad} > U_{\rm gas}$ and radiation pressure rather than plasma instabilities mediate the shock over large length-scales $\Delta \tau \sim c/v$. In this regime the post-shock temperature (assuming equilibrium) is $T \sim \left(\rho v^2 / a_{\rm BB}\right)^{1/4} \sim 10^5$--$10^6 \, {\rm K}$,
where $\rho$ is the downstream density and $a_{\rm BB}$ is the radiation constant. This temperature is much smaller 
than implied by eq.~(\ref{eq:Te}) and non-thermal particle acceleration is likely ineffective (the shock width---over which hydrodynamical variables change---is $\gg$ than particle gyro-radii, thus limiting standard Fermi acceleration).
Such conditions are unfavorable for producing X-ray emission
and will instead produce bright thermal optical-UV sources.
Condition~(\ref{eq:tesc_equal_tdyn}; black curve) therefore separates the CSM parameter space into two regions that are qualitatively distinct in terms of their observable emission: $\sim$blackbody thermal optical/UV emission for high column densities and/or shock velocities versus (potentially Comptonized) bremsstrahlung X-ray emission in the opposite regime.
We note  that even an initially optically-thick radiation-mediated shock may transition into a collisionless shock when it reaches the outer edge of the CSM shell where $\tau_T(r) < c/v$ \citep{Katz+11}; however this remains uncertain (e.g. it is dependent on whether radiation forces are able to ``pre-accelerate'' the outer CSM material effectively).

\section{The Interaction Phase}
\label{sec:InteractionPhase}

We now turn to calculating the emission produced by collisionless shocks ($\tau_T < c/v$), beginning with the `interaction phase' that occurs while the shock is still propagating within the CSM shell.
This scenario has been considered by previous authors in the context of extended CSM with wind-like profiles \citep{Chevalier&Irwin12,Svirski+12}.
A major issue pointed out in these studies is the severe inhibition of X-rays produced at the shock by photoelectric absorption and Compton downscattering in the unshocked upstream CSM.
Here we recap several key points of these studies, and expand on issues related to ionization-breakout, Comptonization, and the emergent X-ray spectrum. 
Later, in \S\ref{sec:ExpansionPhase} we consider the additional scenario of X-ray emission produced after the CSM has been fully shocked and there is no more continued interaction.

When inverse Compton scattering is negligible, the total bremsstrahlung luminosity produced by the shock during the interaction phase is
\begin{equation}
\label{eq:L_ff}
L_{\rm ff} = \min\left( \eta \frac{t_{\rm dyn}}{t_{\rm ff}}, 1 \right) L_{\rm sh} ,
\end{equation}
which equals the kinetic shock power $L_{\rm sh}$,
\begin{equation}
\label{eq:Lsh}
L_{\rm sh} = 2\pi R^2 \rho v^3 
= \frac{1}{2} M v^2 / t_{\rm dyn}
\approx 10^{44} \, {\rm erg \, s}^{-1} \, M_{-1} R_{15}^{-1} v_9^3
,
\end{equation}
in the radiative regime ($t_{\rm ff} \ll t_{\rm dyn}$),
and where
\begin{equation}
\eta \equiv \frac{\frac{3}{2} n_e k_B T_e}{\frac{1}{2} \rho v^2}
= \frac{9}{16} \frac{\mu}{\mu_e} \epsilon_T
\approx 0.3 \epsilon_T
\end{equation}
is a factor
that relates the post-shock thermal energy density to the kinetic energy density of the shock.
This factor is only relevant in the slow-cooling regime ($t_{\rm ff} > t_{\rm dyn}$) where $L_{\rm ff}$ is governed by the 
instantaneous bremsstrahlung luminosity $\sim \Lambda_{\rm ff} R^3$.
Although the total radiated power can be substantial (eq.~\ref{eq:Lsh}), the luminosity emitted in the observing band $h\nu \sim 1 \, {\rm keV}$ is\footnote{
There is an additional logarithmic dependence on frequency due to the free-free Gaunt factor, which we neglect here.
}
\begin{align}
\label{eq:LX_unComptonized}
L_X^{\rm ff} 
&= \left. \nu L_\nu \right.
\approx x e^{-x} L_{\rm ff}
\\ \nonumber
&\approx 
\min 
\begin{cases}
4.3 \times 10^{42} \, {\rm erg \, s}^{-1} \, M_{-1}^2 R_{15}^{-3} v_9^{-1} \epsilon_T^{-1/2} \nu_{\rm keV} 
\\
8.2 \times 10^{41} \, {\rm erg \, s}^{-1} \, M_{-1} R_{15}^{-1} v_9 \nu_{\rm keV} 
\end{cases}
,
\end{align}
only a small fraction 
\begin{equation}
\label{eq:x}
x \equiv h\nu / k_B T_e 
\approx 8.2 \times 10^{-3} \epsilon_T^{-1} v_9^{-2} \nu_{\rm keV}
\end{equation}
of the bolometric luminosity, because the bremsstrahlung emissivity peaks at high frequencies $\approx k_B T_e \sim 100 \, {\rm keV}$ (eq.~\ref{eq:Te}).
The top case in eq.~(\ref{eq:LX_unComptonized}) corresponds to the adiabatic regime, whereas the lower case is applicable in the radiative regime.

In the radiative regime, electrons behind the shock cool within a layer of width $\sim \left(t_{\rm ff} / t_{\rm dyn}\right) R$. Although most of the thermal energy is radiated by electrons of temperature $T_e$ (eq.~\ref{eq:Te}) within this layer, if the emission spectrum is strongly peaked at $\sim k_B T_e$ then colder electrons further downstream could potentially contribute more to the luminosity at low frequencies.
The condition for this is that the emission spectrum $\nu L_\nu$ rise faster than $h\nu/k_B T_e$. Thus, for bremsstrahlung emission the contribution to $L_X$ (measured at frequency $\nu$) from electrons at different temperatures is roughly the same
so long as $k_B T_e > h\nu$.
This can be seen by considering that the contribution to $L_X$ of electrons with different temperatures $T_e$ is $L_X \sim \frac{3}{2} x N_e k_B T_e / t_{\rm ff} \sim \frac{3}{2} \dot{N}_e h\nu$, where $N_e(T_e) \sim \dot{N}_e t_{\rm ff}(T_e)$ is the number of electrons at a given temperature and $\dot{N}_e \sim 4\pi R^2 v n_e$ is the rate at which electrons cross the shock.

\subsection{Comptonization}
\label{sec:Comptonization}

The estimates above do not account for inverse Compton scattering and apply only in the un-Comptonized adiabatic and radiative regimes (see Fig.~\ref{fig:PhaseSpace}). In the following, we consider higher-optical depth CSM configurations where Comptonization plays an important role. 
Importantly, we show that previous treatments of Comptonization in this context are inapplicable in the $y \gtrsim 1$ regime.

Soft photons are effectively inverse Compton scattered by the hot post-shock electrons if the Compton-y parameter is $\gtrsim 1$ (eq.~\ref{eq:tIC_equal_tesc}).
This process modifies the emergent spectrum, increasing the power at higher frequencies.
In the limit $y \to \infty$, escaping photons form a Wien spectrum with $\nu L_\nu \propto x^4 e^{-x}$ (where $x$ is the photon frequency normalized by $k_B T_e$; eq.~\ref{eq:x}), and emission at low frequencies ($x \ll 1$) is significantly inhibited.
However, for sufficiently small $y < y_{\rm crit}(x)$, the emergent spectrum is $\sim$flat in $L_\nu$ and is not appreciably changed with respect to un-Comptonized free-free emission.
In Appendix~\ref{sec:Appendix_Comptonization} we derive approximate expressions for the emergent spectrum of Comptonized bremsstrahlung with arbitrary Compton-y parameter and show that, at low frequencies this can be expressed in terms of a multiplicative ``correction factor'' $\Psi(x,y)$ to the free-free source function (eq.~\ref{eq:Appendix_Psi}),
\begin{equation}
\Psi \left( x , y \right)
\equiv \left. \frac{L_X^{\rm C}}{L_X^{\rm ff}} \right\vert_{\rm adiabatic}
\sim 
\begin{cases}
1 
&, y < y_{\rm crit}
\\
\left(\frac{y}{y_{\rm crit}}\right)^{-1/2}
&, y > y_{\rm crit}
\end{cases}
\end{equation}
where
$y_{\rm crit}(x) \sim \mathcal{O}(10^3)$ (eq.~\ref{eq:Appendix_y_crit}) so that typically $y \ll y_{\rm crit}$ and $\Psi \sim 1$.
A more precise estimate of $\Psi$ is given by eq.~(\ref{eq:Appendix_Psi_lowy}) and is consistent with this picture.
This implies that $L_X^{\rm C} \sim L_X^{\rm ff}$ in the adiabatic regime and that Comptonization does not significantly alter the X-ray luminosity 
at frequencies $x \ll 1$ of interest (see Fig.~\ref{fig:Appendix_spectrum}).

Compton scattering can also significantly enhance the electron cooling rate. 
Compton cooling is proportional to the {\it number} of photons that can be Compton upscattered, and their mean energy gain $\sim e^y h\nu$. Previous studies \citep[e.g.][]{Chevalier&Irwin12} adopted the familiar expression $\Lambda_{\rm C} = n_e \sigma_T c \left( 4 k_B T_e / m_e c^2 \right) U_{\rm rad}$ for the Compton cooling rate which is proportional to the photon energy density $U_{\rm rad}$, but this expression is only valid in regimes where $y \ll 1$. In a scattering medium with large Compton-y (and large Thompson optical-depth), the radiation energy density increases with the number of scatterings and the incident $U_{\rm rad}$ is no longer an appropriate parameter. 

Because in our scenario, the seed photon density is produced by bremsstrahlung emission from the very same electrons that Compton upscatter these photons, inverse-Compton and free-free cooling are inherently coupled to one another (in Appendix~\ref{sec:Appendix_Synch_Comptonization} we show that Comptonization of soft synchrotron photons is less effective).
It is therefore convenient to express the Compton cooling rate in terms of a correction factor $\mathcal{C}(y)$ to the free-free cooling rate
such that the total cooling rate is $\Lambda_{\rm C} + \Lambda_{\rm ff} \equiv \mathcal{C}  \Lambda_{\rm ff}$ (eq.~\ref{eq:Appendix_Cy_final}).

For low $y \lesssim 1$, $\mathcal{C} \approx 1$ and the total cooling rate is set by free-free emission. For larger Compton-y, electron losses are instead dominated by Compton cooling. This cooling rate is a factor $\mathcal{C} \approx \frac{3}{2} y_{\rm sat}^2 \sim \mathcal{O}(10^2)$ larger than $\Lambda_{\rm ff}$ in the saturated regime, $y \gtrsim y_{\rm sat}$, in which $\mathcal{C}$ is roughly constant and only weakly (logarithmically) dependant on system parameters. Between the two regimes there is a transition region, $1 \lesssim y \lesssim y_{\rm sat}$, where $\mathcal{C}$ rises from unity to its asymptotic value (approximately as $\propto y^2$; eq.~\ref{eq:Appendix_Cy_final}).
The enhanced cooling rate in the Comptonized regime affects the conditions at which the shock transitions from adiabatic to radiative, as illustrated in Fig.~\ref{fig:PhaseSpace}.
Accounting for these effects, the total electron cooling timescale is modified from $t_{\rm ff}$ (eq.~\ref{eq:t_ff}) and is
\begin{equation}
\label{eq:t_cool}
t_{\rm cool} = t_{\rm ff} / \mathcal{C}(y)
.
\end{equation}
Likewise, the Comptonized free-free X-ray luminosity is 
\begin{align}
\label{eq:LX_Compton}
L_X^{\rm C} 
&= L_{\rm sh} \min\left( \eta \frac{t_{\rm dyn}}{t_{\rm ff} /\mathcal{C}(y)} , 1 \right) x e^{-x} \frac{\Psi\left(x,y\right)}{\mathcal{C}(y)}
\\ \nonumber
&=
L_X^{\rm ff} 
\Psi\left(x,y\right)
\begin{cases}
1 &, ~{\rm adiabatic}
\\
1/\mathcal{C}(y) &, ~{\rm radiative}
\end{cases}
,
\end{align}
a modification to the un-Comptonized free-free luminosity $L_X^{\rm ff}$ (eq.~\ref{eq:LX_unComptonized}).
The spectrum within the Comptonized radiative regime may differ from that assumed above because cold post-shock electrons may contribute non-negligibly to the luminosity at $x \ll 1$. These details require solving the spatially dependent coupled electron cooling and Comptonization equations, which is outside the scope of our present work. We note that the Comptonized radiative expression above can be considered a lower-bound on the true luminosity in this regime.

\subsection{Non-thermal Emission}

In addition to bremsstrahlung photons produced by the thermal pool of shock-heated electrons, a collisionless shock may produce non-thermal emission from a relativistic population of electrons accelerated through diffusive shock (first-order Fermi) acceleration \citep[e.g.][]{Blandford&Eichler87}.
We assume that electrons in this non-thermal population carry a fraction $\epsilon_e$ of the shock energy and are distributed as a power-law in momentum, $\partial N / \partial (\beta \gamma) \propto (\beta \gamma)^{-p}$, above some Lorentz factor $\gamma_m$,\footnote{
For the non-relativistic shocks we consider, $\gamma_m \approx 1$, and this formalism follows the `deep-Newtonian regime' of \cite{Sironi&Giannios13}. 
}
and that the shock amplifies magnetic fields with efficiency $\epsilon_B$.

Synchrotron emission by these relativistic electrons can contribute to the X-ray luminosity of the shock. 
In the X-ray band, these electrons are fast cooling. Their luminosity is therefore
\begin{equation}
\label{eq:LX_syn}
L_X^{\rm syn} \sim
\gamma \left(\frac{\partial N }{ \partial \gamma}\right) \frac{\gamma m_e c^2 }{t_{\rm dyn}}
\equiv \epsilon_X^{\rm syn} L_{\rm sh}
,
\end{equation}
where $L_{\rm sh}$ is given by eq.~(\ref{eq:Lsh}), and we have defined
\begin{align}
\label{eq:epsilon_syn}
\epsilon_X^{\rm syn} 
&= 
\frac{p-2}{p-1} \frac{\epsilon_e}{\mu_e} \gamma^{2-p}
= \frac{p-2}{p-1} \frac{\epsilon_e}{\mu_e} \left(\frac{\pi m_e c \nu R^{3/2}}{e \epsilon_B^{1/2} M^{1/2} v} \right)^{-\frac{p-2}{2}}
\\ \nonumber
&\underset{p=3}{\approx}
2.4 \times 10^{-6} \epsilon_{e,-1} \epsilon_{B,-1}^{1/4} M_{-1}^{1/4} R_{15}^{-3/4} v_9^{1/2} \nu_{\rm keV}^{-1/2} 
\end{align}
as the fractional shock power emitted as synchrotron radiation at frequency $\nu$. In the second line of eq.~(\ref{eq:epsilon_syn}) have taken $p=3$, often inferred for non-relativistic shocks in radio SNe \citep{Weiler+02}.

Comparison of eqs.~(\ref{eq:LX_unComptonized},\ref{eq:epsilon_syn}; see also~\ref{eq:epsilon_ff}) illustrates that the synchrotron X-ray luminosity is subdominant to the thermal bremsstrahlung component, unless the shock velocity is very large, or $p<3$ (note that in any case it is required that $p>2$).

\subsection{Propagation Effects}
\label{sec:Propagation}

As first pointed out by \cite{Chevalier&Irwin12} and \cite{Svirski+12}, X-rays produced during the interaction phase are significantly inhibited due to propagation effects in the upstream (unshocked) CSM. X-ray photons passing through the cold upstream medium are subject to Compton downscattering and photoelectric absorption.
Compton downscattering will inhibit high-energy photons above a frequency $\sim m_e c^2 / \tau_T^2$ and is therefore only important at $\sim$keV frequencies for dense CSM with $\tau_T \gtrsim 23 \nu_{\rm keV}^{-1/2}$.
Typical CSM configurations of interest will be below this threshold (eq.~\ref{eq:tau_T}).\footnote{We note that by scattering high energy photons down to $h\nu_{\rm Cds} \sim m_e c^2 / \tau_T^2$, Compton downscattering can enhance the luminosity at this frequency by a modest factor $\sim 1 + \ln\left( k_B T_e / h\nu_{\rm Cds} \right)$ if $h\nu_{\rm Cds} < k_B T_e$. This is estimated by assuming that $\sim$all bremsstrahlung photons above $\nu_{\rm Cds}$ are scattered down to this critical frequency (neglecting photoelectric absorption). This effect is only relevant for particularly high $\tau_T$ CSM, and we neglect it throughout the rest of this paper.}

Photoelectric absorption has a more severe effect on keV photons given the large bound-free opacity, $\kappa_{\rm bf} \approx 106 \, {\rm cm}^2 \, {\rm g}^{-1} \, \nu_{\rm keV}^{-8/3}$ \citep[e.g.][]{Chevalier&Fransson17}. The optical depth to bound-free absorption throughout the bulk of the CSM is $\gg 1$,
\begin{equation}
\label{eq:tau_bf}
\tau_{\rm bf} \approx
1700 M_{-1} R_{15}^{-2} \nu_{\rm keV}^{-8/3} ,
\end{equation}
and this implies that keV photons would not be able to escape to an observer.
There are two caveats to this conclusion. First, eq.~(\ref{eq:tau_bf}) applies for X-rays propagating a distance $\Delta R \sim R$ through the upstream medium. If the CSM density drops sharply at some outer edge (as we consider here) then the column density of cold gas ahead of the shock will drop such that $\tau_{\rm bf} \lesssim 1$ once the shock is a distance $d \ll R$ from the outer edge. Second, the estimate above neglects the important back reaction that photoelectrically absorbed X-ray photons have on the upstream medium---such photons photoionize the upstream medium and reduce the bound-free optical depth (which counts only unionized species).
A shock that is sufficiently luminous in ionizing X-ray photons, can thus lead to an ionization breakout of keV photons.

The full ionization state of photoionized gas depends on detailed atomic processes, CSM composition, and incident radiation spectrum, and can be modeled numerically (e.g. \citealt{Margalit+18}). Here we instead adopt a simplified analytic approach,
that is detailed in Appendix~\ref{sec:Appendix_IonizationBreakout} (see also \citealt{Metzger+14}). Our primary result is a condition on the minimum shock velocity required for ionization breakout to occur (eq.~\ref{eq:Appendix_v_breakout}),
\begin{equation}
\label{eq:v_ionization_bo}
v_9 >
5.1
\, 
M_{-1}^{1/3} R_{15}^{-2/3} X_{A,-2}^{1/3} \alpha_{{\rm rr},-12}^{1/3} \nu_{\rm keV}^{1/3} \epsilon_{X,-3}^{-1/3}
.
\end{equation}
Here $\alpha_{\rm rr} = 10^{-12} \alpha_{{\rm rr},-12} \, {\rm cm}^3 \, {\rm s}^{-1}$ is the radiative recombination rate of the species whose bound-free transition is closest below the observing frequency, $X_A = 10^{-2} X_{A,-2}$ the fractional number density of this species, and $\epsilon_X \equiv L_X/L_{\rm sh}$ is the fractional shock power emitted at this frequency.
Equation~(\ref{eq:v_ionization_bo}) is an implicit relation because $\epsilon_X$ can itself depend on velocity.
For the case where synchrotron emission dominates, $\epsilon_X$ is given by eq.~(\ref{eq:epsilon_syn}).
For (potentially Comptonized) bremsstrahlung emission this is determined by eqs.~(\ref{eq:Lsh},\ref{eq:LX_unComptonized},\ref{eq:LX_Compton},\ref{eq:Appendix_Psi}),
\begin{equation}
\label{eq:epsilon_ff}
\epsilon_X^{\rm C}
\equiv \frac{L_X^{\rm C}}{L_{\rm sh}}
\approx \min 
\begin{cases}
4.3 \times 10^{-2} \, M_{-1} R_{15}^{-2} \epsilon_T^{-1/2} v_9^{-4} \nu_{\rm keV}
\\
8.2 \times 10^{-3} \, v_9^{-2} \nu_{\rm keV} \mathcal{C}(y)^{-1}
\end{cases}
.
\end{equation}

Equation~(\ref{eq:v_ionization_bo}) illustrates that ionization breakout of keV X-rays from the bulk of the CSM (i.e. for a significant fraction of the interaction phase duration, $\sim t_{\rm dyn}$) is unlikely for the non-relativistic shocks we consider here.
For thermal emission alone (eq.~\ref{eq:epsilon_ff}), the most optimistic (radiative) case implies that $v_9  \gtrsim 3 \tau_T X_{A,-2} \alpha_{{\rm rr},-12}$ is required for ionization breakout, which may not be satisfied by typical CCSNe
(see Fig.~\ref{fig:PhaseSpace}).
Even if ionizing radiation cannot burrow its way out of the entirety of the CSM shell, X-ray photons will still propagate a distance $d_{\rm p.i.} = R/\tau_{\rm p.i.}$ ahead of the shock. The duration of the interaction-phase X-ray light curve is therefore
\begin{equation}
\label{eq:t_int}
t_{\rm int} = t_{\rm dyn} \min \left( 1, \tau_{\rm p.i}^{-1} \right) ,
\end{equation}
where $\tau_{\rm p.i.}$ is a pseudo optical depth to photoionizing photons, given by eq.~(\ref{eq:Appendix_taupi}). It is set by (the larger of) the mean-free-path to bound-free absorption or the photoionizing (Stromgren) depth of photons. This is described in greater detail in Appendix~\ref{sec:Appendix_IonizationBreakout}.

\subsection{The Reverse Shock}

In our discussion to this point we have considered only the forward shock that propagates into the CSM shell. The ejecta-CSM collision will also drive a reverse shock that propagates into the ejecta material. This reverse shock is often also considered as a source of X-ray emission in interacting SNe \citep[e.g.][]{Chevalier&Fransson94,Fransson+96,Nymark+06}. We briefly consider this possibility below, and find that the reverse shock will typically not contribute appreciably to the X-ray signature.

Pressure balance at the contact discontinuity that separates shocked CSM from shocked ejecta implies that the temperature $T_{\rm rev}$ behind the reverse shock is a factor $\sim \left(\rho_{\rm csm}/\rho_{\rm ej}\right) \ll 1$ lower than the forward shock (eq.~\ref{eq:Te}). The free-free timescale $t_{\rm ff} \propto \rho^{-1} T^{1/2}$ of shocked ejecta material is accordingly a factor $\sim \left(\rho_{\rm csm}/\rho_{\rm ej}\right)^{3/2}$ shorter than the shocked CSM free-free timescale (eq.~\ref{eq:t_ff}).
We may crudely estimate the ejecta-CSM density ratio by assuming a uniform-density ejecta of mass $M_{\rm ej}$, such that $\rho_{\rm ej} \sim 3 M_{\rm ej}/4\pi R^3$ and $\rho_{\rm csm}/\rho_{\rm ej} \sim M_{\rm csm} / 3 M_{\rm ej}$. The reverse shock is therefore radiative if (eqs.~\ref{eq:t_dyn},\ref{eq:t_ff})
\begin{equation}
\label{eq:rev_shock_radiative_req}
    3.1 \times 10^{-5}\, M_{-1}^{1/2} R_{15}^2 v_9^2 \epsilon_T^{1/2} \left(\frac{M_{\rm ej}}{5\,M_\odot}\right)^{-3/2} < 1 .
\end{equation}
This condition is satisfied throughout almost the entire relevant parameter-space, so that the reverse shock is nearly always radiative.

The kinetic power of the reverse shock is a factor $\sim \left(\rho_{\rm csm}/\rho_{\rm ej}\right)^{1/2}$ smaller than that of the forward shock (eq.~\ref{eq:Lsh}), 
\begin{equation}
\label{eq:L_rev}
    L_{\rm rev} 
    \approx 8.1 \times 10^{42} \, {\rm erg\,s}^{-1}\, M_{-1}^{3/2} R_{15}^{-1} v_9^3 \left(\frac{M_{\rm ej}}{5\,M_\odot}\right)^{-1/2}
    .
\end{equation}
However, the lower temperature of the reverse shock implies that its band-limited luminosity 
can exceed the forward shock's contribution (eq.~\ref{eq:LX_unComptonized}), so long as $T_{\rm rev}$ is still above the observing frequency. The latter condition is only satisfied for sufficiently massive CSM, $M > 0.12\,M_\odot\,\epsilon_T^{-1} v_9^{-2} \nu_{\rm keV} \left(M_{\rm ej}/5\,M_\odot\right)$ (eq.~\ref{eq:x}).
More importantly---because the reverse shock is radiative, a cold dense layer of material will accumulate downstream. This layer will photoelectrically absorb X-ray photons produced at the reverse shock and severely inhibit any continuum X-ray emission by this component (see \S\ref{sec:Propagation}). 
Instead, the reverse-shock dissipated power (eq.~\ref{eq:L_rev}) may emerge as line-cooling emission from the cold dense shell (see e.g. \citealt{Nymark+06} for detailed discussion and calculations).

We therefore conclude that the reverse shock does not contribute appreciably to the emergent X-ray luminosity for typical parameters. As a caveat, we note that in the estimates above we have assumed that the ejecta density is characterized by its bulk average value, but that a more accurate ejecta density structure would place a small amount of ejecta material at lower densities \citep[e.g.][]{Matzner&Mckee99}. This would lower the ejecta-CSM density ratio at early times and may change some of our estimates above (in a time-dependent manner). Given our conclusion that the post-interaction phase is in any case more important to the resulting X-ray light-curve (\S\ref{sec:Propagation} and \S\ref{sec:ExpansionPhase}), and that the reverse shock does not persist long after interaction ceases, we find it reasonable to neglect complications related to the ejecta density structure at present.

\section{Post-interaction Phase}
\label{sec:ExpansionPhase}

In the previous section we discussed X-ray emission produced by the shock as it crosses the bulk of the CSM shell. In the keV band, this emission is severely inhibited by bound-free absorption of X-rays in the cold upstream, limiting the observable phase-space for such X-rays.
This motivates us to consider emission produced when the shock crosses the outer edge of the CSM, and during subsequent expansion of the hot CSM gas.
The benefit of this scenario is that photoelectric absorption by the (now fully shocked and highly ionized) CSM shell would be negligible, allowing X-ray photons to escape unattenuated.
Before proceeding we first note that the above argument is only correct if the environment surrounding the shell is sufficiently ``clean''.
In particular, because the bound-free opacity is so large, even a small amount of material external to the shell could potentially inhibit the observed X-ray signal.
In Appendix~\ref{sec:Appendix_Wind_IonizationBreakout} we study this scenario and show that, for typical stellar mass-loss rates and for the X-ray luminosities produced by SNe shocking a dense CSM shell, keV photons manage to photoionize their way out of an initially bound-free optically-thick wind that might surround the dense CSM shell. Therefore, the presence of such a wind should not affect our estimates below.

We organize this section by discussing each of the shock regimes shown in Fig.~\ref{fig:PhaseSpace}, moving from low to high optical depth. The main result of this section is the characterization of the expected X-ray light-curves in the different regimes, and is summarized in Fig.~\ref{fig:lightcurves}.

\subsection{Adiabatic Shock}

In the adiabatic regime, $t_{\rm ff} > t_{\rm dyn}$ and the shocked CSM doubles in radius (over an expansion timescale $\sim t_{\rm dyn}$) before any radiative losses occur.
This implies that $\tau_T$, $T_e$ and other relevant properties remain $\sim$constant during this first radius-doubling timescale.
All shocked electrons therefore remain hot and contribute to the free-free luminosity in this regime, $L_{\rm ff} \sim \frac{3}{2} N_e k_B T_e / t_{\rm ff}$. 
The fraction of this luminosity that is emitted in the observing band is $x \ll 1$ (eq.~\ref{eq:x}), and the resulting X-ray luminosity during the first expansion timescale is equivalent to that during the interaction phase (eq.~\ref{eq:LX_unComptonized}).

Following the initial (first) expansion timescale $\sim t_{\rm dyn}$, the hot gas will undergo adiabatic cooling.
Assuming homologous expansion, $R \propto 1+t/t_{\rm dyn}$, the electron temperature at $t \gg t_{\rm dyn}$ drops as $T_e \propto t^{-2}$ 
(this assumes an adiabatic index $5/3$ appropriate for non-relativistic electrons).
In the meantime, the free-free timescale increases due to this expansion, $t_{\rm ff} \propto n_e^{-1} T_e^{1/2} \propto t^2$.
This implies that gas that starts off slow-cooling will remain slow-cooling throughout the expansion phase.
At temperatures $\lesssim 10^6\,{\rm K}$ line-cooling will take over free-free and the situation changes, however this does not affect emission in the X-ray band, which is contributed by electrons whose temperature exceeds $\gtrsim {\rm keV} \sim 10^7 \, {\rm K}$.

The X-ray light-curve is therefore $\sim$constant over a duration $\sim t_{\rm dyn}$, $L_X \approx \frac{3}{2} N_e h \nu / t_{\rm ff}$ (and this is equivalent to eq.~\ref{eq:LX_unComptonized} in the adiabatic regime), and subsequently declines as $L_X \propto t^{-2}$.
This persists until the electron temperature drops below the observing band, at time 
\begin{equation}
\label{eq:t_T}
{\rm adiabatic}:~~
t_{T} 
\equiv t \left( x=1 \right)
= t_{\rm dyn} \left( x_0^{-1/2} - 1 \right)
,
\end{equation}
where $x_0 \equiv x(t=0)$ is the initial temperature-normalized observing frequency (eq.~\ref{eq:x}).
Subsequent to this time, the light-curve declines exponentially. 
This light-curve evolution is illustrated schematically by the black curve in  Fig.~\ref{fig:lightcurves}.
The red curves show Comptonized light-curves, as discussed in the following subsection. 

Before proceeding onward, we note an interesting point in the un-Comptonized adiabatic regime: that cancellations of $T_e$ due to the flat bremsstrahlung spectrum lead to the fact that the fluence at frequency $\nu$ is simply proportional to the number of radiating electrons, $4\pi D^2 F_X = \frac{3}{2} N_e h \nu \propto M$ (see eq.~\ref{eq:LX_unComptonized}). Observations of such events are therefore able to uniquely probe the CSM shell mass (see  \S\ref{sec:inferring_CSM}).

\subsection{Comptonized Adiabatic Shock}

\begin{figure*}
    \includegraphics[width=0.5\textwidth]{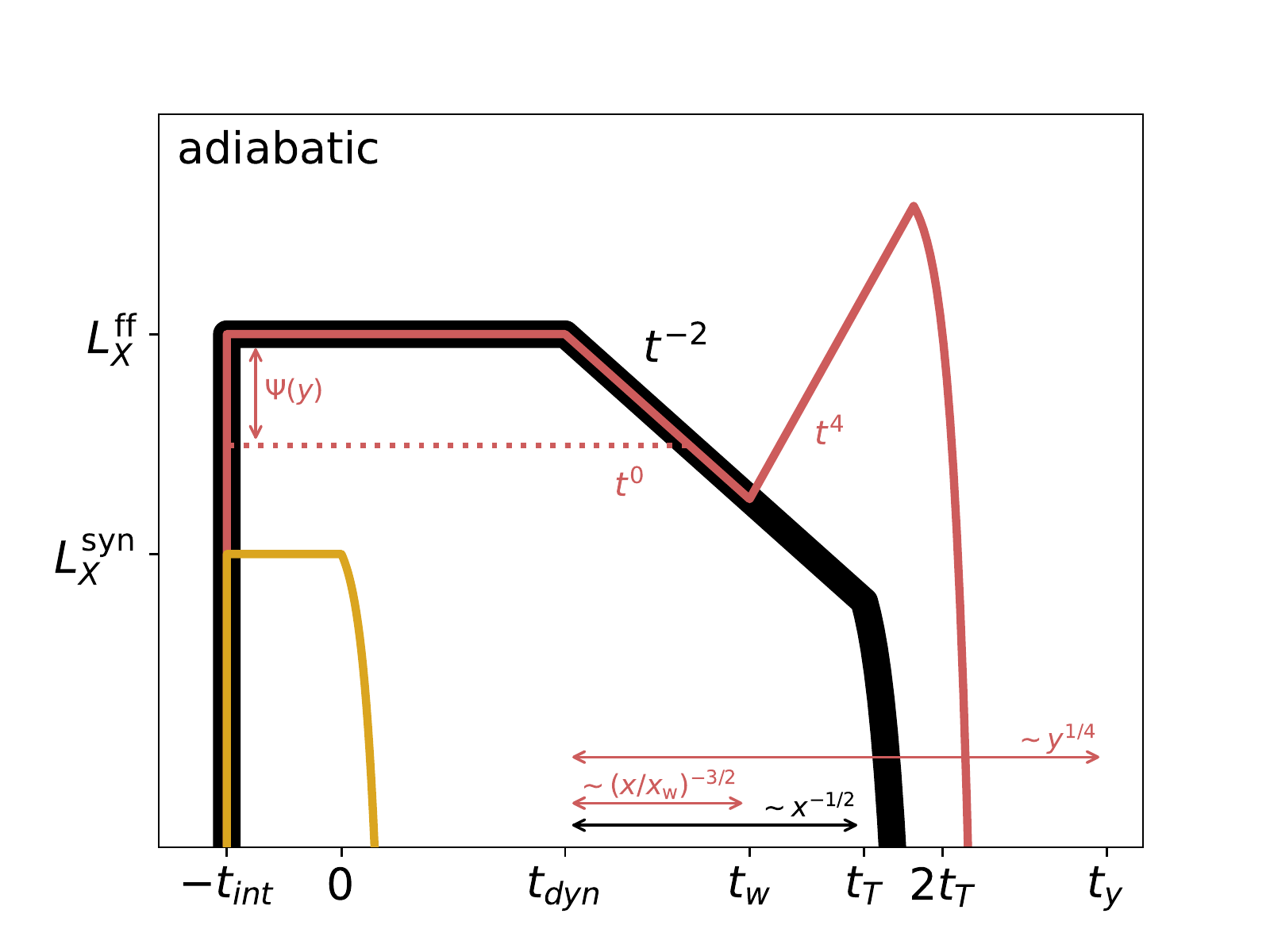}
    \includegraphics[width=0.5\textwidth]{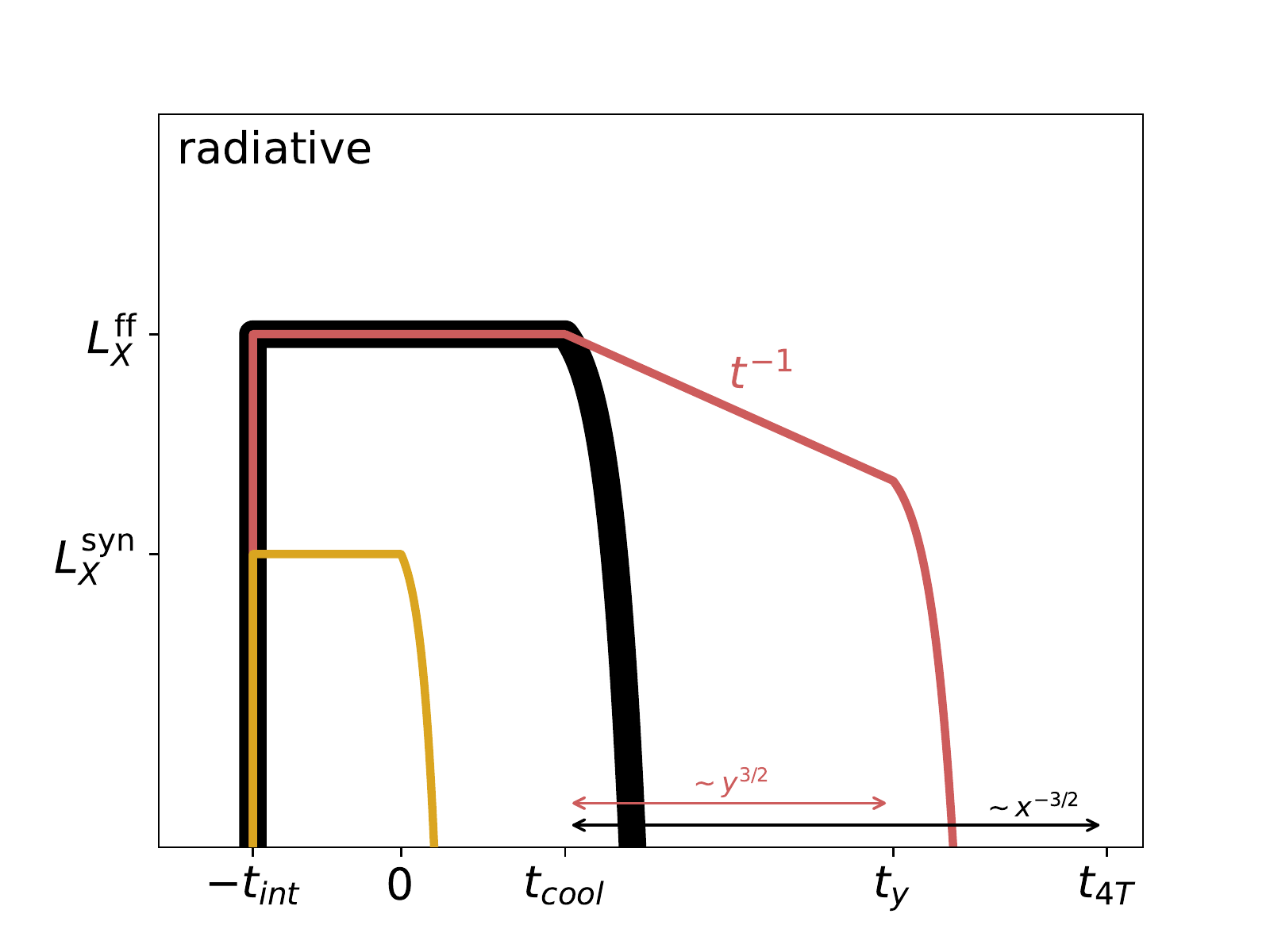}
    \caption{{\it Left}: Schematic X-ray light curves in the adiabatic regime. The black curve shows the un-Comptonized light-curve in the expansion phase. At $t>t_{\rm dyn}$ this decays as $\propto t^{-2}$ from the initial luminosity (eq.~\ref{eq:LX_unComptonized}), and exponentially at $t>t_T$ (eq.~\ref{eq:t_T}). Red curves show light-curves in the Comptonized adiabatic regime for $x \ll 1$ and $y_{\rm sat} < y < y_{\rm crit}$ (solid red), $y > y_{\rm crit}$ (dotted red).
    These light-curves show the case where Comptonization is important throughout the entire evolution. In practice, the Compton-y parameter drops dramatically at $t>t_{\rm dyn}$ and Comptonization ceases to play a role at time $\sim t_y$ (eq.~\ref{eq:t_y}; in this figure it is assumed that $t_y > t_T$). At times $t \gtrsim t_y$ the light-curve in the Comptonized regime of parameter space will therefore track the un-Comptonized light-curve (solid black). The yellow curve illustrates a possible synchrotron component to the light-curve, that begins a short duration $t_{\rm int} \lesssim t_{\rm dyn}$ (eq.~\ref{eq:t_int}) before the shock exits the CSM shell. Throughout much of the parameter space $L_X^{\rm syn} < L_X^{\rm ff}$ so that this component is subdominant (eqs.~\ref{eq:LX_Compton},\ref{eq:LX_syn}).
    {\it Right}: Same as left panel, but in the radiative regime $t_{\rm cool} < t_{\rm dyn}$ (eq.~\ref{eq:t_cool}). The un-Comptonized light-curve (black) decays exponentially over a cooling time, whereas the light-curve in the Comptonized regime ($y>1$; red) will decay more slowly, over a timescale $\sim \min \left(t_y,t_{4T}\right)$, here given by eqs.~(\ref{eq:t_y_radiative},\ref{eq:t_T_radiative}). This neglects light-crossing effects. If the light-crossing time $t_{\rm lc}$ is longer than any of the above timescales, the observed light-curve will be ``smeared'' over $\sim t_{\rm lc}$ (eq.~\ref{eq:t_lc}).
    }
    \label{fig:lightcurves}
\end{figure*}

In the Comptonized regime, the expansion-phase light curve can take on more interesting morphologies. As the expanding gas dilutes and cools adiabatically, the Compton-y parameter (eq.~\ref{eq:y}) drops dramatically, $y \propto \left(1 + t/t_{\rm dyn}\right)^{-4}$. 
For this scaling we have assumed that $\tau_T < 1$ as relevant throughout most of the Comptonized adiabatic regime (see Figures~\ref{fig:PhaseSpace},\ref{fig:Appendix_pairs}).
An important timescale is therefore the time $t_y$ at which Comptonization ceases to become important,
\begin{equation}
\label{eq:t_y}
{\rm adiabatic}:~~
t_y 
\equiv t\left(y=1\right)
= t_{\rm dyn} \left( y_0^{1/4} - 1 \right)
,
\end{equation}
where $y_0 \equiv y(t=0)$ is the initial Compton-y parameter (eq.~\ref{eq:y}).
At times $t>t_y$ the light-curve reverts to that in the un-Comptonized adiabatic regime discussed in the previous subsection, and will track the black curve in Fig.~\ref{fig:lightcurves}.

At sufficiently low frequencies, the Comptonized light-curve will be nearly indistinguishable from that of the un-Comptonized regime even at times $t<t_y$ (when $y>1$ and Comptonization is still important). This is because the low-frequency spectrum is not dramatically affected by Comptonization ($\Psi \sim 1$; see eq.~\ref{eq:Appendix_Psi}) so long as $y < y_{\rm crit}$. For parameters of interest, this is usually the case.
Comptonization will start affecting the light-curve once 
the temperature drops sufficiently such that the Wien peak dominates emission in the observed band, $x \gtrsim x_{\rm w}$. This occurs at time
\begin{equation}
\label{eq:t_w}
t_{\rm w} 
\equiv t\left(x=x_{\rm w}\right)
\sim t_{\rm dyn} \left( \frac{x_{{\rm w},0}}{x_0} - 1 \right)^{3/2}
,
\end{equation}
where we used 
$x_{\rm w} \sim \left(y_{\rm sat} y/16\right)^{-1/6} \propto t^{2/3}$ 
based on eq.~(\ref{eq:Appendix_xw}), an approximate expression for $x_{\rm w}$ in the saturated regime ($y > y_{\rm sat}$; eq.~\ref{eq:Appendix_ysat}).
At times $t \gtrsim t_{\rm w}$, the X-ray light curve at fixed frequency {\it rises} as $L_X \propto \mathcal{C}(y) x^4 L_{\rm ff} \propto t^4$, where $\mathcal{C}(y)$ is the Compton cooling correction to the bremsstrahlung luminosity (eq.~\ref{eq:Appendix_Cy_final}), and we assumed $y > y_{\rm sat}$.
This rise terminates once the Compton-y parameter drops below $y_{\rm sat}$ or when the temperature drops below the observing band. In the Comptonized regime, the latter occurs when $h\nu \approx 4k_B T_e$, at time $\sim 2t_T$ (eq.~\ref{eq:t_T}).
Time $t_{\rm w}$ is smaller than $2t_T$ (eq.~\ref{eq:t_T}) only if the observing frequency is not too low, 
$x_0 \gtrsim \left(y_{\rm sat}y\right)^{-1/4}$.
Fig.~\ref{fig:lightcurves} shows light-curves in this scenario.
If however $t_{\rm w} > 2t_T$, then the Wien portion of the evolution is irrelevant.
Similarly, the Wien light curve is not realized if $t_{\rm w} \gtrsim t_y y_{\rm sat}^{-1/4}$ such that the Compton-y drops below its saturation value before time $t_{\rm w}$ (or if initially $y<y_{\rm sat}$).

If initially $y>y_{\rm sat}$ and $t_{\rm w} < 0$ then the light-curve is already sampling the Wien peak, and the $\propto t^4$ rise in $L_X$ commences immediately at $t_{\rm dyn}$. The duration of this rise ($\sim 2t_T$) will be short given that the thermal peak is not much above the observing band during the interaction phase (these conditions are only realized for $x_0 \sim 1$).

Another scenario, though of less practical interest, occurs if the Compton-y parameter can reach extremely large values $> y_{\rm crit}$. In this case, the initial X-ray luminosity is inhibited by a factor $\Psi \sim \left(y/y_{\rm crit}\right)^{-1/2}$ with respect to the un-Comptonized regime (eq.~\ref{eq:LX_Compton}).
The critical Compton-y parameter (eq.~\ref{eq:Appendix_y_crit}) is only logarithmically dependent on time,
and can be approximated as roughly constant.
In the expansion phase, the light curve therefore evolves as $L_X \sim \left(y/y_{\rm crit}\right)^{-1/2} x L_{\rm ff} \propto t^0$ while $y>y_{\rm crit}$. This evolution terminates 
at time $2t_T$ when the temperature drops below the observing band, or earlier---at time $\sim t_y y_{\rm crit}^{-1/4}$, when the Compton-y parameter drops below $y_{\rm crit}$ and the X-ray light-curve reverts to the previously discussed regimes above.
The dotted red curve in Fig.~\ref{fig:lightcurves} shows the light-curve in this regime, $y>y_{\rm crit}$, and assuming that $t_y y_{\rm crit}^{-1/4} < 2t_T$. 

\subsection{Radiative Shock}

As discussed in \S\ref{sec:InteractionPhase}, the bremsstrahlung X-ray luminosity in the radiative regime has $\sim$equal contribution from hot electrons (at temperature $T_e$) immediately post-shock and cooler electrons further downstream, a result of the flat free-free spectrum.\footnote{If radiating electrons were all heated ``instantaneously'' (at the same time) then they would cool in tandem, and the resulting X-ray luminosity, $L_X \sim N_e h \nu / t_{\rm ff}$, would in fact peak after several cooling $e$-fold times when $t_{\rm ff}$ is small. The fact that radiating electrons are not heated instantaneously by the shock changes things dramatically: electrons that were heated at earlier times have had time to cool, and the number of electrons at different temperatures is $N_e(T_e) \sim \dot{N}_e t_{\rm ff}(T_e)$, so that $t_{\rm ff}$ cancels out. Most importantly, this subtlety is only important for runaway cooling, where this fine ($<t_{\rm ff}$) timing becomes relevant.}
The downstream gas is thermally unstable in this regime because the free-free cooling time, $t_{\rm ff} \propto n_e^{-1} T_e^{1/2}$, becomes unceasingly shorter
 as $T_e$ drops (if cooling is isobaric then $n_e \propto T_e^{-1}$ making matters even worse).

Radiating electrons therefore produce X-ray emission for a duration $\sim t_{\rm ff}(T_e)$ set by the first cooling $e$-fold time (at initial temperature $T_e$; eq.~\ref{eq:Te}). 
An additional timescale necessary to consider in this regime is the light-crossing time of the CSM shell,
\begin{equation}
\label{eq:t_lc}
t_{\rm lc} \sim R/c \approx 3.3 \times 10^4 \, {\rm s} \, R_{15} .
\end{equation}
Even if electrons radiate their energy over very short timescales, $t_{\rm ff} \ll t_{\rm lc}$, the resulting light curve---which is contributed by electrons distributed around the shell---would be ``smeared'' over the light-crossing time.
The observed X-ray light-curve will therefore last a duration $\sim \max\left(t_{\rm ff},t_{\rm lc}\right)$ after the shock exits the CSM outer edge (when the interaction phase stops).
The light curve drops dramatically following this time.
Finally, note that $t_{\rm ff} < t_{\rm dyn}$ in the radiative regime (by definition), so the light-curve falls off before the shocked gas manages to double in radius.

\subsection{Comptonized Radiative Shock}

The Comptonized radiative regime behaves similarly to the un-Comptonized case so long as the (Compton modified) electron cooling time, $t_{\rm cool} \approx t_{\rm ff}/\mathcal{C}(y)$, decreases with temperature. 
The additional dependence of $t_{\rm cool}$ on the Compton-y parameter will however change things throughout much of the parameter space.
The full time dependent spatial problem of cooling electrons coupled to a (self-consistently generated) bath of Comptonizing photons is outside the scope of this work. Instead, we adopt an approximate approach and assume that the separation of timescales $t_{\rm IC} \ll t_{\rm cool}$ is maintained, namely that Comptonization acts rapidly with respect to electron cooling. In this regime, Comptonization achieves a steady-state at each epoch of electron cooling, and the results derived in Appendix~\ref{sec:Appendix_Comptonization} under the assumption of fixed $T_e$ can be used.

As gas cools (radiatively) the Compton-y parameter decreases proportionally to the electron temperature, $y \propto T_e$ (eq.~\ref{eq:y}; here and in the following we assume isochoric radiative cooling). 
The Compton cooling correction scales as $\mathcal{C} \propto y^2$ in the regime $1 < y < y_{\rm sat}$ (eq.~\ref{eq:Appendix_Cy_final}), which implies $t_{\rm cool} \propto t_{\rm ff} / y^2 \propto T_e^{-3/2}$ for this range of Compton-y. In this regime, Comptonization has a significant effect---it changes the cooling rate such that $t_{\rm cool}$ {\it increases} as gas cools down. This is markedly different from the runaway cooling that applies in the un-Comptonized regime, and implies a slow (power-law) decline in temperature, $T_e \propto \left( 1 + 3t/2t_{{\rm cool},0} \right)^{-2/3}$.

The cooling time in this regime increases as $t_{\rm cool} \propto t$ and
the light-curve for $1<y<y_{\rm sat},y_{\rm crit}$ (in which $\Psi \sim 1$; eq.~\ref{eq:Appendix_Psi}), $L_X \sim N_e h\nu / t_{\rm cool}$, therefore decays as $L_X \propto t^{-1}$.
This continues until the Compton-y parameter drops below unity at which point $\mathcal{C} \sim 1$ and the cooling reverts back to the un-Comptonized runaway process described in the previous subsection. This occurs at time
\begin{equation}
\label{eq:t_y_radiative}
{\rm radiative},~ y \leq y_{\rm sat}: ~~
t_y = t_{{\rm cool},0} \frac{2}{3} \left( y_0^{3/2} - 1 \right) ,
\end{equation}
where we used the temperature temporal evolution and the fact that $y \propto T_e$.
Another relevant timescale that can terminate the power-law X-ray light-curve occurs when the temperature drops below the observing band ($h\nu = 4 k_B T_e$),
\begin{equation}
\label{eq:t_T_radiative}
{\rm radiative},~ y \leq y_{\rm sat}: ~~
t_{4T} = t_{{\rm cool},0} \frac{2}{3} \left( 8 x_0^{-3/2} - 1 \right) .
\end{equation}
This is analogous to eq.~(\ref{eq:t_T}) except in the radiative (rather than adiabatic) cooling regime of interest.
Typically $y_0<4x_0^{-1}$ so that $t_y < t_{4T}$ for parameters of interest ($1 < y < y_{\rm sat}$). In this hierarchy, $t_{4T}$ does not influence the resulting light-curve.

Finally, note that our assumption that $t_{\rm IC} \ll t_{\rm cool}$, if valid at $t=0$, remains valid subsequently as well. This is because the cooling time increases as $t_{\rm cool} \propto t$ while the Comptonization timescale (eq.~\ref{eq:t_IC}) increases only as $t_{\rm IC} \propto T_e^{-1} \propto t^{2/3}$.
Another implicit assumption of our steady-state treatment of the Comptonizing radiation field is that $t_{\rm esc} \ll t_{\rm cool}$. Because $t_{\rm esc}$ does not change throughout the cooling process (again, assuming isochoric cooling) while $t_{\rm cool} \propto t$: if $t_{\rm esc} < t_{{\rm cool},0}$ initially, then this condition will also be satisfied at any later point. Using eq.~(\ref{eq:t_esc}) with $f = t_{{\rm cool},0}/t_{\rm dyn}$ it is easy to show that $t_{\rm esc} < t_{{\rm cool},0}$ throughout the entire parameter-space of non-relativistic radiative collisionless shocks ($v<c$, $t_{{\rm cool},0}<t_{\rm dyn}$, $\tau_T < c/v$).

For $y > y_{\rm sat}$ Comptonization saturates and $\mathcal{C} \sim const$ (eq.~\ref{eq:Appendix_Cy_final}). In this highly-Comptonized case $t_{\rm cool} \propto T_e^{1/2}$ and again there is runaway cooling. This runaway process halts once $y = y_{\rm sat}$ and one enters the regime discussed in the preceding paragraphs. 

Finally, we note again that the observed light-curve will be ``smeared'' over light-crossing timescales (eq.~\ref{eq:t_lc}) so that the temporal evolution described above may not be directly observable if $t_y < t_{\rm lc}$.

\section{Thermal Optical/UV Signature of Dense Shells}
\label{sec:UVOptical}

In the previous sections we described the X-ray signatures of CSM interaction, and showed that these (both thermal and non-thermal) are expected to occur only for shells whose initial optical depth is $\tau_T < c/v$. In the case of dense shells where $\tau_T > c/v$, a radiation-mediated shock replaces the collisionless shock so that the characteristic (post-shock) temperature (e.g. \citealt{Svirski+12})
\begin{equation}
\label{eq:T_RMS}
T_{\rm RMS} = \left( \frac{18 \rho v^2 }{ 7 a_{\rm BB} } \right)^{1/4} \approx 
23 \, {\rm eV} \, 
R_{14}^{-3/4} M_{-1}^{1/4} v_9^{1/2} 
\end{equation}
is much lower than that of a collisionless shock (eq.~\ref{eq:Te}; note that above we normalize to $R_{14} \equiv R/10^{14}\,{\rm cm}$). This brings the resulting thermal transient into the UV/optical range, and indeed---such dense-shell circum-stellar interaction has been commonly invoked as a model for bright optical transients \citep[e.g.][]{Ofek+10,Chevalier&Irwin11,Ginzburg&Balberg12}.

The optical/UV light-curve that results from such CSM interaction can be broadly separated into two phases. The first is the initial shock-breakout phase during which photons first manage to escape ahead of the shock. This occurs when the shock nears the outer CSM edge, and the radiation temperature (assuming LTE) is $\sim T_{\rm RMS}$ (eq.~\ref{eq:T_RMS}). For sufficiently dense shells (small $R$ and/or large $M$) the breakout flash will occur in the far-UV or soft X-ray bands, and may not be observable with optical facilities.
Such configurations however may still be detectable in the optical during the cooling-envelope phase that follows the initial shock-breakout phase. Cooling-envelope emission is produced once the fully-shocked CSM begins to expand and cool (through a combination of adiabatic and radiative losses), which causes the characteristic radiation temperature to cascade down as a function of time. 
Note that the shock-breakout and cooling-envelope phases are analogous to the `interaction' and `post-interaction' phases we have discussed in \S\ref{sec:InteractionPhase},\ref{sec:ExpansionPhase}. The former are relevant to optically-thick $\tau_T > c/v$ shocks (radiation-mediated shocks) whereas the latter to $\tau_T < c/v$ shocks (collisionless shocks).

In the following, we adopt the formalism of \cite{Margalit21} to calculate the optical/UV signature of CSM shells with $\tau > c/v$. This work derived a full analytic solution to the light-curve of such CSM shells starting from shock-breakout and through the subsequent cooling-envelope phase, and we refer readers to \cite{Margalit21} for further details. In calculating band-limited properties (such as peak luminosity and duration) we assume a thermal (blackbody) spectrum.

\section{Transient Phase Space and Detectability}
\label{sec:results}

In the preceding sections \S\ref{sec:InteractionPhase}-\ref{sec:ExpansionPhase} we have described the detailed X-ray light curve in various regimes. Using these results, we are now in a position to show the main observable features of X-ray counterparts to CSM interaction as a function of the shock and CSM parameters.

\begin{figure}
    \includegraphics[width=0.5\textwidth]{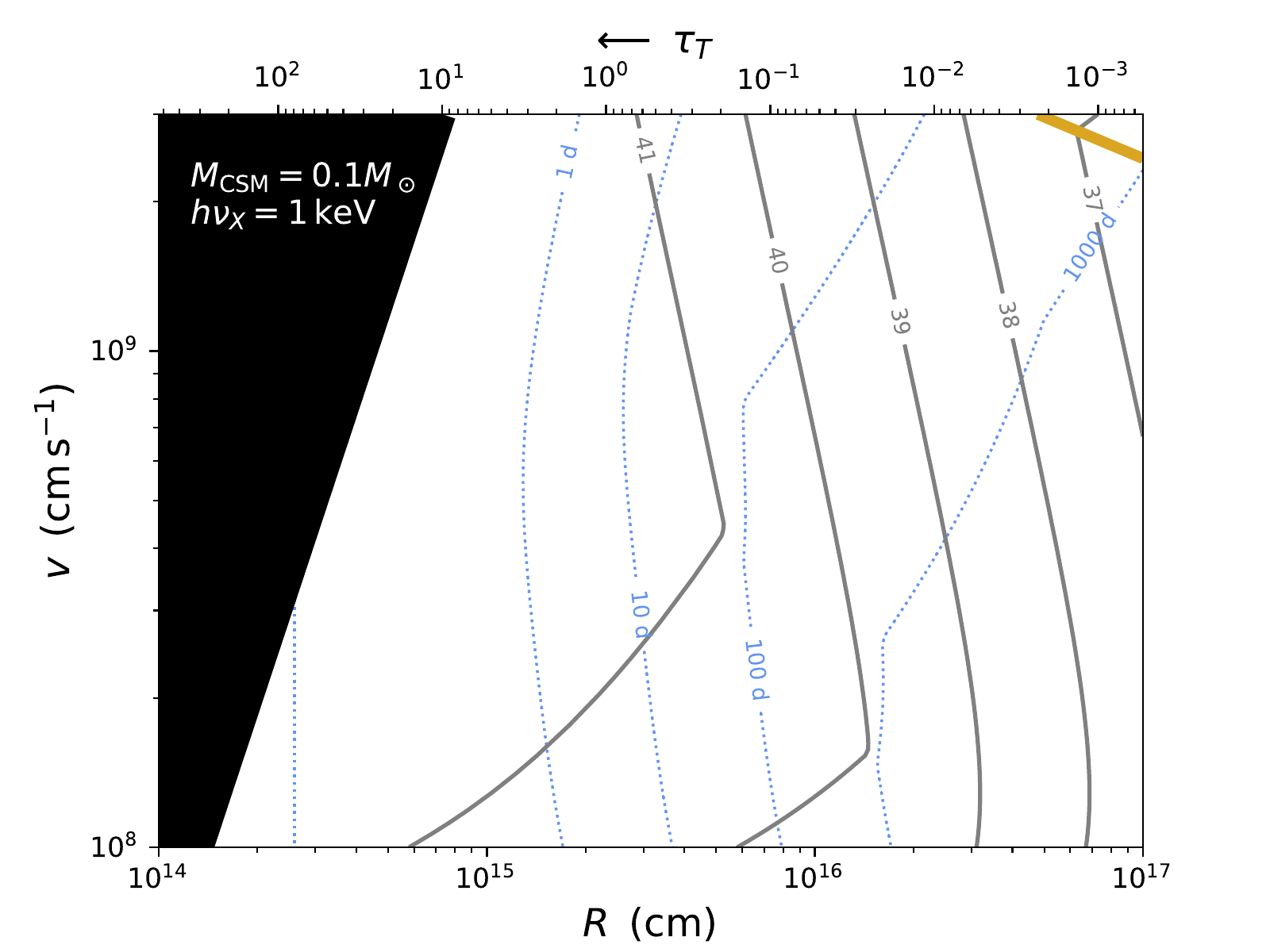}
    \caption{
    Contours of 1~keV X-ray peak luminosity (grey; labeled in logarithmic units of ${\rm erg \, s}^{-1}$) and timescale (eq.~\ref{eq:t_duration}; dotted blue) resulting from a shock of velocity $v$ driven into a compact CSM shell at radius $R$ (and width $\Delta R = R$). 
    The CSM mass is fixed at  $0.1 M_\odot$, so that the shell position ($R$) determines the Thompson optical depth $\tau_T$ (top axis). This allows direct comparison with Fig.~\ref{fig:PhaseSpace}.
    The peak luminosity is dominated by (potentially Comptonized) bremsstrahlung emission below the solid yellow curve,
    and synchrotron emission above this curve.
    Within the black shaded region a radiation-mediated shock is formed (see \S\ref{sec:UVOptical}).
    }
    \label{fig:Results}
\end{figure}

Figure~\ref{fig:Results} shows contours of peak luminosity (solid grey) and duration (dotted blue) in the phase space of shock velocity $v$ and CSM radius $R$. The CSM mass is fixed at $0.1 M_\odot$ and we assume a shell width $\Delta R = R$.
The yellow curve shows the contour along which $L_X^{\rm syn} = L_X^{\rm C}$ (eqs.~\ref{eq:LX_Compton},\ref{eq:LX_syn}; assuming $p=3$, $\epsilon_e = \epsilon_B = 0.1$). Above this curve, synchrotron emission dominates the (interaction phase) X-ray light curve,
whereas below it the peak of the light-curve is dominated by free-free emission.
The duration of the interaction phase depends on photoelectric absorption of X-ray photons (eq.~\ref{eq:t_int}; Appendix~\ref{sec:Appendix_IonizationBreakout}), and is subject to some uncertainty given our simplified analytic treatment. 
More precisely, the duration of the X-ray transient is taken to be
\begin{equation}
\label{eq:t_duration}
t_X = 
\begin{cases}
\max\left[ t_{\rm int} + \max\left(t_{\rm dyn},t_{\rm cool}\right) , t_{\rm lc} \right] &, {\rm brem}
\\
\max\left[ t_{\rm int} , t_{\rm lc} \right] &, {\rm syn}
\end{cases}
\end{equation}
where $t_{\rm dyn}$, $t_{\rm cool}$, $t_{\rm int}$, and $t_{\rm lc}$ are the dynamical, cooling, interaction, and light-crossing times (eqs.~\ref{eq:t_dyn},\ref{eq:t_cool},\ref{eq:t_int},\ref{eq:t_lc}; respectively) and the two cases correspond to the bremsstrahlung and synchrotron components (see Fig.~\ref{fig:lightcurves}).

The top horizontal axis shows the Thompson optical depth $\tau_T$ of the CSM shell (increasing from right to left; eq.~\ref{eq:tau_T}), facilitating comparison with the schematic phase-space illustrated in Fig.~\ref{fig:PhaseSpace}.
For visual clarity, we do not plot in Fig.~\ref{fig:Results} curves delineating the different regimes shown in Fig.~\ref{fig:PhaseSpace}; however, the transition between radiative and adiabatic shocks is clearly noticeable as a kink in the luminosity/duration contours.
Finally, we note that within the black shaded region $\tau_T > c/v$ and a radiation-mediated shock that produces primarily optical/UV emission replaces the X-ray-producing collisionless shock. This regime is discussed in \S\ref{sec:UVOptical} and further below.

Fig.~\ref{fig:Results} shows that X-ray emission is brightest for fast shock velocities and CSM configurations that lie closest to the $\tau_T = c/v$ boundary. The peak duration is also shortest for these parameters. We note that there exists a parameter-space around $R \sim 10^{15} \, {\rm cm}$, $v \sim 10^9 \, {\rm cm \, s}^{-1}$ where the peak X-ray luminosity is not very sensitive to either $R$ or $v$, but in which the duration can vary substantially (primarily as a function of $R$). 
This is due to the weak (linear) scaling of $L_X$ on parameters within the radiative regime compared to the stronger $t_{\rm ff} \propto R^3$ scaling of duration within this regime (eqs.~\ref{eq:t_ff},\ref{eq:LX_syn}).
For our fiducial $M=0.1M_\odot$ this motivates searches for $\sim 10^{41} \, {\rm erg \, s}^{-1}$ ($10^{42} \, {\rm erg \, s}^{-1}$) transients at $1\,{\rm keV}$ ($10\,{\rm keV}$), with durations spanning $\lesssim$day to $\sim$month timescales. These quoted luminosities roughly scale as $\propto M$ for different CSM masses.

\begin{figure*}
    \includegraphics[width=\textwidth]{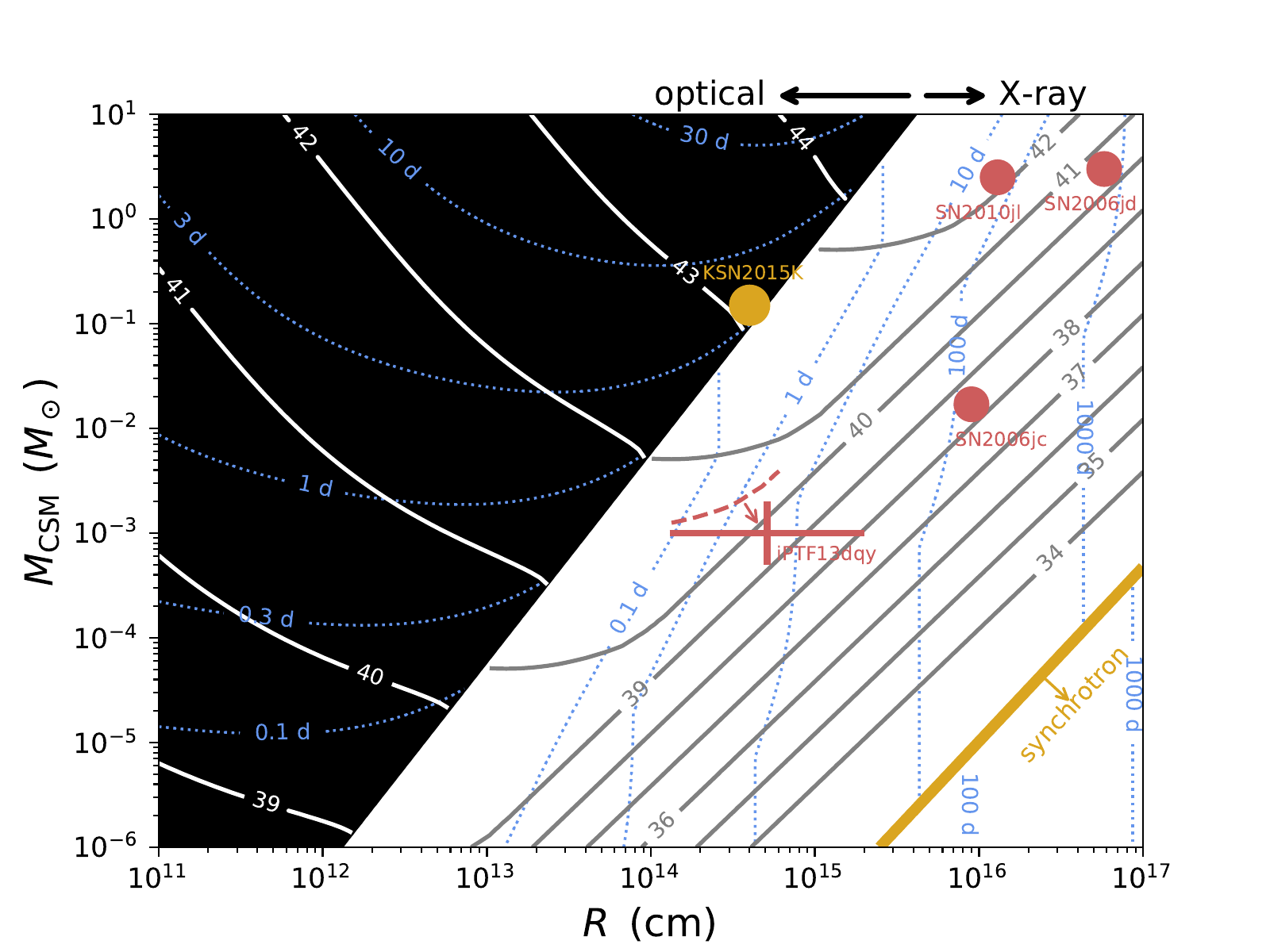}
    \caption{
    Similar to Fig.~\ref{fig:Results} but for fixed velocity $v=10,000 \, {\rm km \, s}^{-1}$, and varying CSM mass, $M_{\rm CSM}$. Within the black shaded region, $\tau_T > c/v$ and a fast optical transient is expected, consistent e.g. with the FBOT KSN2015K (yellow circle; \citealt{Rest+18}). Contours of peak optical luminosity (white; at $5000\Angstrom$) and duration (blue) of such transients is shown within this region following the model described by (\citealt{Margalit21}; see \S\ref{sec:UVOptical}). To the right of the black region, CSM interaction is expected to power primarily hard X-ray emission (grey/blue contours show luminosity/timescale of this emission at $1 \, {\rm keV}$). CSM properties inferred from flash-ionization spectroscopy of the IIp SN iPTF13dqy are shown with the red errorbar \citep{Yaron+17},
    while a few SNe with detected X-ray emission and inferred CSM properties are shown with red circles (IIn SN2006jd; \citealt{Chandra+12}; IIn SN2010jl; \citealt{Ofek+14,Chandra+15}; Ibn SN2006jc; \citealt{Immler+08}). 
    }
    \label{fig:MR}
\end{figure*}

We can similarly plot 
the phase-space of interaction-powered transients as a function of CSM shell mass and position while keeping the shock velocity fixed. This is shown in Fig.~\ref{fig:MR} and helps illustrate the dependence on CSM properties alone. Plotting results in this phase-space is also motivated by the expected narrower dynamical range in shock velocity compared to CSM mass.
In Fig.~\ref{fig:MR} we fix the velocity to a fiducial $v = 10^9 \, {\rm cm \, s}^{-1}$, and take $\Delta R = R$.
As in Fig.~\ref{fig:Results}, grey (blue) curves on the right-hand-side of the figure show contours of constant X-ray luminosity (duration) at $1 \, {\rm keV}$. As previously discussed, a long-lasting X-ray signature is primarily expected outside of the black shaded region. Within this shaded region $\tau_T > c/v$ and a radiation-mediated shock forms. The equilibrium temperature of such shocks is several orders of magnitude lower than eq.~(\ref{eq:Te}; it falls within the UV, rather than the hard X-ray band, cf. eq.~\ref{eq:T_RMS})
resulting in bright optical/UV emission, without appreciable X-rays.

This dichotomous behaviour is important in interpreting observational constraints on dense shell stellar mass-loss. As we discuss in \S\ref{sec:detectability} below, X-ray surveys are not currently sensitive to the short duration keV emission of shells just right of this boundary ($\tau_T \lesssim c/v$). Optical facilities are, however, becoming increasingly sensitive to short-duration transients such as FBOTs, and are thus beginning to probe the parameter-space of CSM shells with $\tau_T \gtrsim c/v$.
As one example, we show in Fig.~\ref{fig:MR} CSM parameters inferred from optical observations of the FBOT KSN2015K marked with a yellow circle \citep{Rest+18}. 

The fact that KSN2015K (as well as other FBOTs) lies so close to the interface $\tau_T \sim c/v$ may, at first glance, seem surprising. However, as we now discuss, this can be understood by considering the properties of optical transients within the parameter-space. To illustrate this, we plot contours of constant optical luminosity $\nu L_\nu$ (solid white; at a fiducial wavelength of $5000 \Angstrom$) and duration (dotted blue) within the allowed parameter-space (black shaded region), as discussed in \S\ref{sec:UVOptical}. 
At fixed shell mass $M$, the luminosity in the optical band drops for CSM shells located at smaller radii. The transient duration is typically also shorter for CSM located at small radii (though not always; see the curvature in the dotted-blue curves).
Optical surveys are less sensitive to shorter-duration less-luminous transients, and are therefore generally biased towards finding events with large CSM mass that are closer to the interface $\tau_T \sim c/v$.

Similarly, Fig.~\ref{fig:MR} illustrates why CSM shells have not typically been detected at much larger radii $\gg 10^{14} \, {\rm cm}$. In this regime $\tau_T < c/v$ and most of the energy is radiated in the hard X-ray band. 
As discussed in \S\ref{sec:detectability}, the sensitivity of current X-ray instruments to such events in blind searches is limited. Optical imaging surveys would also be blind to such CSM configurations simply because no appreciable optical continuum emission is expected in this regime (as opposed to optically bright $\tau \gtrsim c/v$ CSM shells).
Such CSM is however increasingly being probed by rapid SN follow-up efforts. In particular, narrow emission lines revealed by early-time flash-ionization spectroscopy indicate that dense CSM may be ubiquitous. Recent work by \cite{Bruch+21} analyzed a systematic sample of Type II SNe detected by ZTF where early spectra were obtained, and concluded that $\gtrsim 30\%$ of such events harbour dense CSM at $\lesssim 10^{15} \, {\rm cm}$ scales.

One well observed example of early flash-ionized emission lines is the Type IIp SN iPTF13dqy \citep{Yaron+17}. Detailed observations of this event allowed \cite{Yaron+17} to constrain properties of the surrounding CSM, finding that a CSM of mass $\sim 10^{-3} M_\odot$ truncated around $R \sim$several$\times 10^{14} \, {\rm cm}$ is required to explain the multi-wavelength data.
The inferred CSM properties for iPTF13dqy are show with red markings in Fig.~\ref{fig:MR}.
Our present work predicts a $\sim 10^{40} \, {\rm erg \, s}^{-1}$ X-ray counterpart to such CSM. It is interesting to note that \cite{Yaron+17} present X-ray upper-limits based on non-detections from {\it Swift} follow-up (shown as the red-dashed luminosity contour in Fig.~\ref{fig:MR}), and that the predicted emission falls below these limits. We therefore conclude that despite the deep X-ray follow-up in this event, these observations would not have been sensitive enough to detect the X-ray counterpart to such CSM interaction.

Figure~\ref{fig:MR} also shows a handful of SNe with X-ray follow-up detections from which CSM masses and radii have been estimated (red circles). 
These give a sense for the range of CSM properties inferred from core-collapse SNe.
Only a small fraction of Type IIn SNe have detected X-ray emission \citep{Chandra17}, and even a smaller fraction of other classes of SNe \citep{Ofek+13}. Here we show the Type IIn SN2006jd \citep{Chandra+12}, the well observed IIn SN2010jl \citep{Ofek+14,Chandra+15}, and the Ibn SN2006jc \citep{Immler+08}. These occupy regions with CSM radii $\gtrsim 10^{16}\, {\rm cm}$ and CSM masses $\gtrsim M_\odot$ for the IIn SNe ($\sim 10^{-2} M_\odot$ for the Ibn).
While there is good evidence for an outer CSM truncation radius for iPTF2013dqy and SN2006jc, the situation is less clear for other core-collapse SNe (though deviations from an $r^{-2}$ wind have been claimed by some authors for both SN2006jd and SN2010jl). Our current modeling assumes a constant density shell, but would similarly apply to other truncated CSM profiles.

\begin{figure*}
    \includegraphics[width=0.5\textwidth]{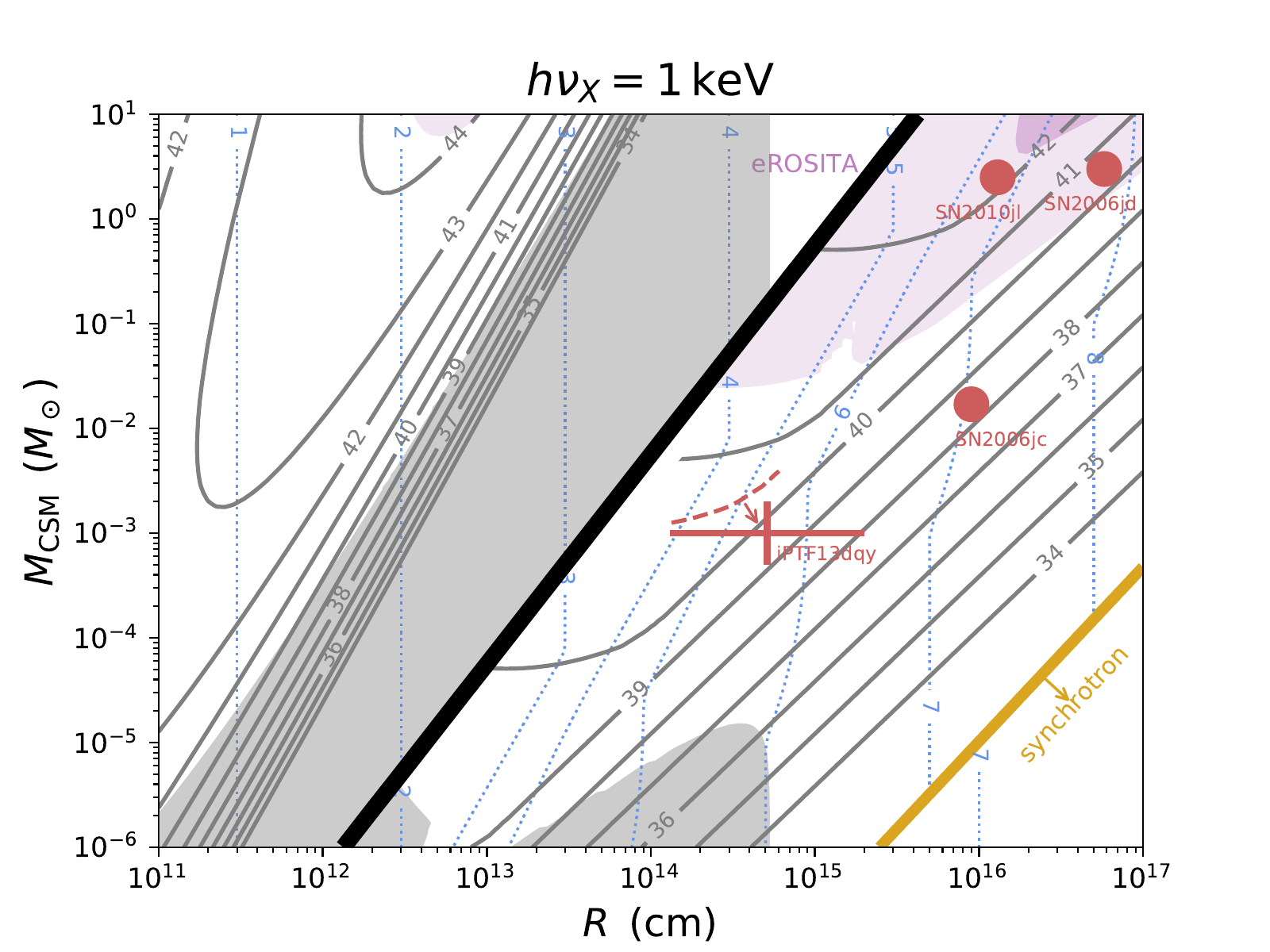}
    \includegraphics[width=0.5\textwidth]{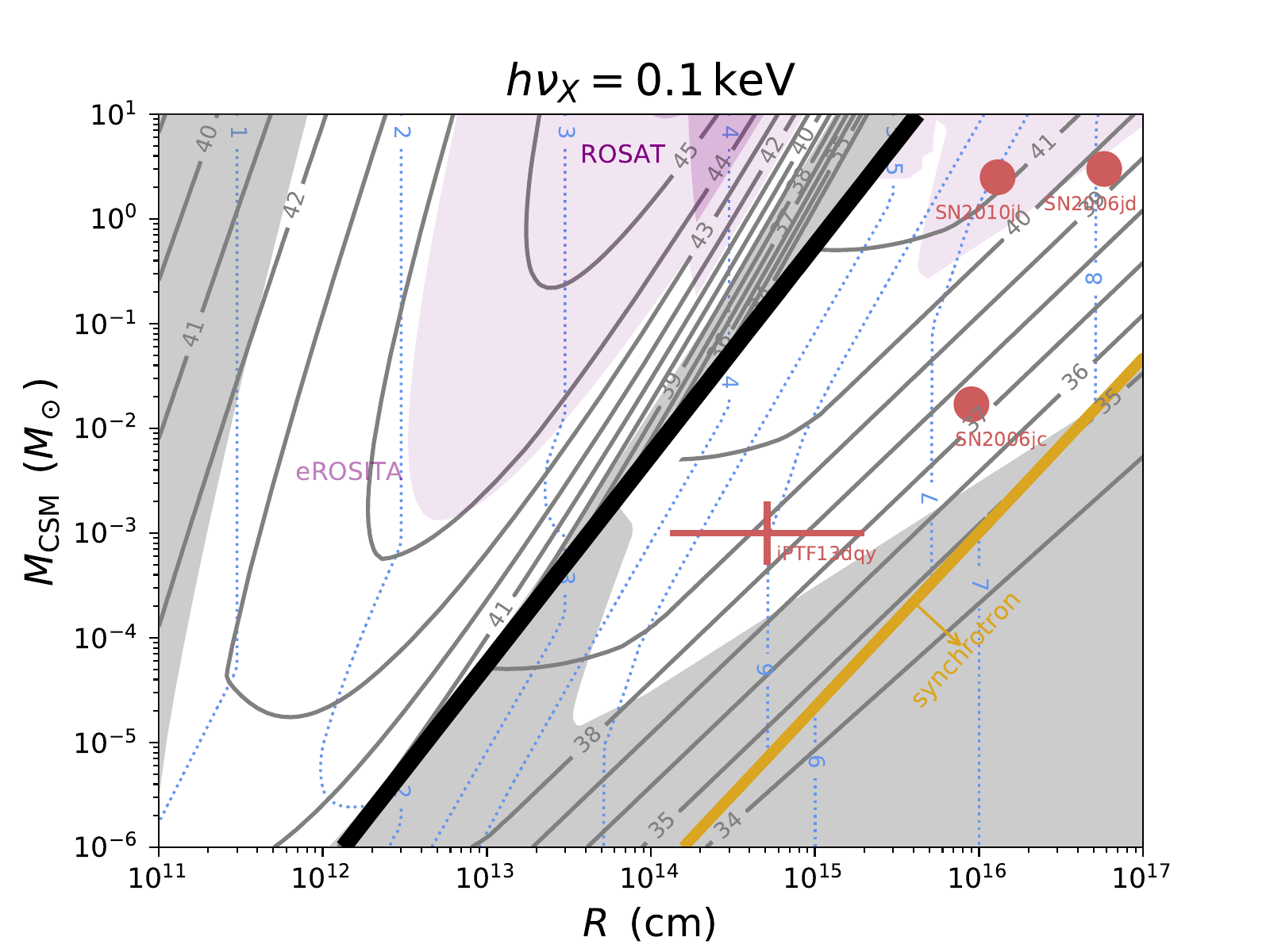}
    \caption{
    Same as Fig.~\ref{fig:MR}, but with the $\tau > c/v$ region of parameter space (left of the solid black curve) now showing the expected X-ray luminosity (solid grey; in logarithmic units of ${\rm erg \, s}^{-1}$) and duration (dotted blue; in log seconds) within this regime. The shock-breakout and cooling-envelope X-ray emission are calculated following \cite{Margalit21}. Light (dark) shaded purple regions show the CSM parameter-space where eROSITA (ROSAT) is expected to detect $N_{\rm det}>1$ events, as described in \S\ref{sec:detectability}.
    This shows that the ROSAT All-Sky Survey would not have been sensitive to such CSM-powered X-ray transients, however eROSITA may be expected to detect many such events (see also Table~\ref{tab:rates}).
    X-rays can be susceptible to photoelectric absorption in a low-density wind that may surround the CSM shell. Grey shaded areas show regions where such absorption is relevant, after accounting for ionization breakout (Appendix~\ref{sec:Appendix_Wind_IonizationBreakout}). This shows that an external wind would not inhibit X-rays in regions where detectability prospects are most promising. Left panel, for $h\nu_X = 1~{\rm keV}$, and right panel, for $h\nu_X = 0.1~{\rm keV}$.
    }
    \label{fig:Xray_both_sides}
\end{figure*}

Above we have focused on the dichotomy between CSM shells where $\tau_T$ is larger or smaller than $c/v$, pointing out that optical emission is dominant at large optical depth and X-ray emission at low optical depth. Although this is true for sufficiently long $\gtrsim$day duration transients that are produced by CSM shells at characteristic radii $\sim 10^{14} \, {\rm cm}$, bright X-ray emission can also be produced within the $\tau_T > c/v$ region if the CSM is sufficiently dense and confined.
Eq.~(\ref{eq:T_RMS}) shows that the post-shock temperature of radiation-mediated shocks can be pushed into the keV band if $R \lesssim 10^{12} \,{\rm cm}$. If this temperature is above the observing band then shock-breakout and cooling-envelope emission (that occurs within the $\tau_T >c/v$ region) will also produce X-ray emission that may be detectable.

Figure~\ref{fig:Xray_both_sides} illustrates this point. Similar to Fig.~\ref{fig:MR} we plot contours of X-ray luminosity and duration in the CSM mass-radius phase space, however here we extend these contours (in the X-ray band) to within the $\tau_T > c/v$ region of parameter space (to the left of the solid black curve). At keV frequencies (left panel) the X-ray signature within this region is negligible unless $R \lesssim 10^{12} \, {\rm cm}$ ($\tau_T \gg c/v$), consistent with our discussion above. The luminosity contours roughly peak when $T_{\rm RMS} \sim$keV, along the track 
$R \approx 7 \times 10^{11} \, {\rm cm} \, M_{-1}^{1/3} v_9^{2/3} \nu_{\rm keV}^{-4/3}$
(eq.~\ref{eq:T_RMS}).\footnote{We have verified that thermal equilibrium is established for these parameters, and therefore the observed radiation temperature equals $T_{\rm RMS}$. Specifically, using eq.~(10) of \cite{Nakar&Sari10} we find that $\eta_0 \approx 0.05 v_9^4 \nu_{\rm keV}^{-1/2}$ along the $T_{\rm RMS} =$keV track, where $\eta_0<1$ implies thermal equilibrium.} 
At larger radii the effective temperature is always below the observing band and the luminosity falls off exponentially.
The expression above also illustrates the sensitive dependence on frequency. At $0.1\,{\rm keV}$, even relatively extended CSM at $R \sim 10^{13} \, {\rm cm}$ can produce bright X-ray emission. This is shown in the right panel of Fig.~\ref{fig:Xray_both_sides}. 
Although the peak luminosity of such transients would be very large, their duration is expected to be very short, $\sim$minutes-hours, limited by the light-crossing time (eq.~\ref{eq:t_lc}).

A potential concern is whether X-rays from such compact shells could be bound-free absorbed by even a small amount of surrounding circum-shell material (e.g. a `standard' stellar wind), if present. This would seemingly occur if $R < R_{\rm bf} \approx 5\times10^{14}\,{\rm cm}\,\nu_{\rm keV}^{-8/3}$ (eq.~\ref{eq:Appendix_Rbf}).  However, we show in Appendix~\ref{sec:Appendix_Wind_IonizationBreakout} that photoionization of the surrounding medium allows X-rays to break out in much of the $R < R_{\rm bf}$ parameter-space. Grey shaded patches in Fig.~\ref{fig:Xray_both_sides} show regions where photoelectric absorption in a surrounding wind would be able to quench the X-ray signature, assuming a fiducial wind mass-loss rate $\dot{M} = 10^{-5}\,M_\odot\,{\rm yr}^{-1}$ and velocity $v_{\rm w} = 100\,{\rm km\,s}^{-1}$.
This shows that bound-free absorption in an external wind is only effective at mitigating X-rays in regions where the X-ray luminosity is extremely low, and where observational prospects are in any case less promising (see \S\ref{sec:detectability} below).

\subsection{Detectability in X-ray Surveys}
\label{sec:detectability}

Above we have discussed the various X-ray signatures of CSM interaction.
We can set a limit on the rate of such X-ray transients
using the ROSAT all-sky survey (RASS; \citealt{Truemper1982}), the most sensitive wide-field survey before the ongoing extended ROentgen Survey with an Imaging Telescope Array (eROSITA) on the Spektrum-Roentgen-Gamma (SRG) mission \citep{Predehl2010}.
ROSAT completed one all-sky survey over six months, in addition to (more sensitive) pointed observations.
Roughly 10\% of the RASS area had a previous or subsequent pointed observation, so this sets the fraction of the RASS fields in which a day- to month-timescale transient could have been identified. \citet{Donley2002} conducted a thorough search for X-ray transients using the RASS Bright Source Catalog \citep{Voges1999},
focusing on transients in Galactic nuclei.
However, the search criteria were very generic, and would have identified a transient regardless of whether it was actually located in a Galactic nucleus.
Despite this, only five extragalactic X-ray transients were discovered, and all were consistent with nuclear AGN activity.

We can use the non-detection of such transients in the RASS data to estimate their rate.
ROSAT scanned a 2-degree wide 360-degree circle with a 96-min orbit,
and over the course of one day this circle shifted by 1 degree \citep{Belloni1994}.
The effective exposure for a given source position was 10--30\,s.
In this way, the full sky was mapped after 6 months,
but the cadence at any given part of the sky was very sensitive to the latitude.
In what follows we refer to a single orbit as a `scan' and the accumulation of multiple scans of a given region a `visit'. 

The \citet{Donley2002} search is only relevant for transients with durations significantly longer than $\sim2\,$d (the time spent on a given visit; \citealt{Belloni1994}).
For such transients the
expected number detected in the six-month sky survey $N_\mathrm{det}$ is:
\begin{align}
    &N_\mathrm{det} \left(t_X>2\,{\rm d}\right) = f_\mathrm{rep} N_\mathrm{visit} \frac{\Omega_{\rm scan}}{4\pi} \mathcal{R} V_{\rm lim} t_X
    \\ \nonumber
    &\approx 
    0.02 \,
    \left(\frac{\mathcal{R}}{700 \, {\rm Gpc}^{-3} \, {\rm yr}^{-1}}\right) \left(\frac{L_X}{10^{41}\,{\rm erg \, s}^{-1}}\right)^{3/2} \left(\frac{t_X}{100\,{\rm d}}\right)
    .
\end{align}
Above, $f_\mathrm{rep} \approx 0.1$ is the fractional area with previous or subsequent repeat pointed observations, $N_\mathrm{visit}=180$ is the total number of visits of a given strip, $\Omega_{\rm scan}/4\pi$ 
is the fraction of sky covered in a single scan 
where $\Omega_{\rm scan} = 720\,\degsq$, 
$t_X$ is the transient duration, $\mathcal{R}$ the volumetric rate of such transients, and $V_{\rm lim}$ the volume out to which the transient would be detectable by the survey.
We assume a Euclidean geometry such that $V_{\rm lim} = (4\pi/3) \left({L_X}/{4\pi F_{\rm lim}}\right)^{3/2}$ and have adopted a fiducial volumetric rate of the same order of magnitude as the transients detected by optical surveys, 1\% of the core-collapse supernova rate \citep{Drout+14},
or 
$\mathcal{R} = 700 \, {\rm Gpc}^{-3} \, {\rm yr}^{-1}$ 
\citep{Li2011}.
For the survey sensitivity, we use the count rate in \citet{Donley2002}: the search was estimated to be complete to 0.031\,ct\,\psec.
We used WebPIMMS\footnote{\url{https://heasarc.gsfc.nasa.gov/cgi-bin/Tools/w3pimms/w3pimms.pl}}
with a synchrotron spectrum and and a photon index $\Gamma=1$ to convert this to an approximate limiting unabsorbed 0.2--2.4\,keV flux density of $F_{\rm lim} = 5\times10^{-13}\,\erg\,\pcmsq\,\psec$
(the result does not change assuming a thermal bremsstrahlung spectrum).

The above reasoning only applies to sources with duration longer than 2\,d and shorter than the time between RASS and the pointed observations, which ranged from months to several years \citep{Donley2002}.
So, we caution that the results do not apply to sources with timescales of several years or longer, as they would not be recognized as transients.

Next we consider the case
where the event duration is shorter than 2\,d (the visit time close to the ecliptic plane) but longer than the 96\,min of a single orbit.
In this case,
all transients that explode in the strip area over the course of those two days should be detected, and the expected number of detections is not sensitive to the transient duration.
The expected number is
\begin{align}
    N_\mathrm{det} &\left(96\,{\rm min} < t_X < 2\,{\rm d}\right) 
    = N_\mathrm{visit} \frac{\Omega_{\rm scan}}{4\pi} \mathcal{R} V_{\rm lim} t_{\rm visit} 
    \\ \nonumber 
    &\approx 0.1 
    \, \left(\frac{\mathcal{R}}{700 \, {\rm Gpc}^{-3} \, {\rm yr}^{-1}}\right) \left(\frac{L_X}{
    10^{43}\,{\rm erg \, s}^{-1}}\right)^{3/2}
    ,
\end{align}
where 
$t_{\rm visit} = 2\, {\rm d}$ is the visit time at the ecliptic plane. The number detected is set only by the luminosity of the transient.
In this case, we use the single-scan sensitivity, reported in \citet{Greiner1999} to be 0.3\,ct\,\psec\ (as per their Fig.~1).
Using WebPIMMS we find that this corresponds to an unabsorbed 0.2--2.4\,keV flux density of $F_{\rm lim} = 5\times10^{-12}\,\erg\,\pcmsq\,\psec$.

Finally, we consider the regime where the transient duration is less than the 96\,min orbital period during which a full 720 \degsq\, is surveyed. For each strip of sky, ROSAT performed $N_{\rm orb}=15$ such orbits over the course of 1\,d, effectively viewing each ROSAT $\Omega_{\rm FOV} = \pi \,\degsq$ field-of-view (FOV; assuming a 2\,deg FOV diameter) 15 times. 
Therefore, any detectable event with duration $<$96\,min would be seen in at most one ROSAT orbit but not in any of the other 14, and could in-principle be flagged as a transient. However, the number detected by RASS is sensitive to the event duration,
\begin{align}
    &N_\mathrm{det} \left(30\,{\rm s} < t_X < 96\,{\rm min} \right) 
    = N_{\rm orb} N_\mathrm{visit} \frac{\Omega_{\rm FOV}}{4\pi} \mathcal{R} V_{\rm lim} t_X 
    \\ \nonumber
    &\approx 
    10^{-3} \, \left(\frac{\mathcal{R}}{700 \, {\rm Gpc}^{-3} \, {\rm yr}^{-1}}\right) \left(\frac{L_X}{10^{44}\,{\rm erg \, s}^{-1}}\right)^{3/2} \left(\frac{t_X}{10^3\,{\rm s}}\right)
    .
\end{align}

Figure~\ref{fig:Xray_both_sides} shows the CSM parameter space that ROSAT could have been sensitive to: the dark purple shaded area in this figure shows regions where the number of events detectable by ROSAT is $N_{\rm det} > 1$. 
This area is almost non-existent, showing that ROSAT would not have been sensitive to X-ray transients of this type (see also Table~\ref{tab:rates}).
This is in line with the non-detection of such transients in the RASS data as discussed above \citep{Donley2002}.

We now repeat the calculation for eROSITA.
eROSITA uses a similar survey strategy to the RASS, but over a longer period of time:
the full sky every six months for four years.
Again, we caution that sources with durations significantly longer than four years might not be recognizable in the survey, so our calculations here do not apply.

In a given six-month survey, the sensitivity in the more sensitive soft (0.5--2\,keV) band is roughly $5\times10^{-14}\,\erg\,\pcmsq\,\psec$ \citep{Merloni2012},
or $\sim 10$ times more sensitive than the RASS.
For sources with durations between 2\,d and six months,
the expected number of detections exceeds that of ROSAT by $10^{3/2} \approx 32$ from the sensitivity, and a factor of 10 from the fact that the full sky will have repeat visits (that is $f_{\rm rep}=1$ instead of $0.1$), for a factor of $\sim 300$ in total.

The eROSITA single-scan sensitivity is $10^{-13}\,\erg\,\pcmsq\,\psec$ in the 0.5--10\,keV range \citep{Merloni2012}, which is a factor of 50 better than the single-scan sensitivity of ROSAT.
So, the improvement in the expected number of sources detected (for durations $<2\,$d) is a factor of $50^{3/2} \approx 350$ from the sensitivity,
with an additional factor of eight from the number of all-sky surveys, for a total of $\sim 2800$.

\begin{deluxetable}{lrrrrr}[]
\tablecaption{Estimated number of sources $N_{\rm det}$ detected by RASS and eROSITA for a few representative CSM shell configurations and a shock velocity $v=10,000\,{\rm km\,s}^{-1}$. Calculations assume an event rate of 1\% the CCSN rate, $\mathcal{R} \approx 700 \, {\rm Gpc}^{-3}\,{\rm yr}^{-1}$, and peak sensitivity at 1\,keV. See \S\ref{sec:detectability} for further details.
\label{tab:rates}} 
\tablewidth{0pt} 
\tablehead{ \colhead{$M_\mathrm{CSM}$} & \colhead{$R_\mathrm{CSM}$} & \colhead{$N_{\rm det}^\mathrm{RASS}$} & \colhead{$N_{\rm det}^\mathrm{eROSITA}$}} 
\tabletypesize{\small} 
\startdata 
$10\,M_\odot$ & $3\times 10^{16}\,$cm & $3$ & $10^3$\\ 
$1\,M_\odot$ & $3\times10^{16}\,$cm & $9\times10^{-3}$ & $3$ \\ 
$1\,M_\odot$ & $10^{16}\,$cm & $0.2$ & $50$\\ 
$1\,M_\odot$ & $3\times10^{15}\,$cm & $6\times10^{-3}$ & $16$\\ 
$0.1\,M_\odot$ & $3\times10^{15}\,$cm & $10^{-2}$ & $3$ \\ 
$0.1\,M_\odot$ & $10^{15}\,$cm & $10^{-3}$ & $3$ \\ 
\enddata 
\end{deluxetable}

These results are summarized in Fig.~\ref{fig:Xray_both_sides} and Table~\ref{tab:rates}. Shaded light purple regions in Fig.~\ref{fig:Xray_both_sides} show the CSM parameter space where $N_{\rm det}>1$ for eROSITA while dark purple shading shows the region where the number of events in the RASS is $N_{\rm det}>1$, as discussed above. These figures show that the improved sensitivity of eROSITA will open up the possibility of probing a new region of the CSM parameter space at $\tau_T \lesssim c/v$: eROSITA may detect many events produced by $\gtrsim 0.1 \, M_\odot$ CSM at $\sim 10^{15}\,{\rm cm}$ scales. In the softer X-ray band, eROSITA may detect many shock-breakout events (left of the black curve in Fig.~\ref{fig:Xray_both_sides}, where $\tau_T > c/v$) of $\sim$hr duration, although we have not accounted here for the rapid drop in instrument sensitivity at these low frequencies. 
Finally, we note that we have chosen to normalize $N_{\rm det}$ using a modest volumetric event rate of $\sim1\%$ the CCSN rate. This is motivated by the inferred rate of fast-optical transients that are thought to be powered by dense CSM interaction (\citealt{Drout+14}; Fig.~\ref{fig:MR}), however recent flash-ionization spectroscopy indicates that perhaps most Type II SNe are surrounded by such CSM \citep{Bruch+21}. In this case our estimates of $N_{\rm det}$ should be scaled up by a factor of $\sim 100$, significantly improving detectability prospects.
Note that our discussion above does not address complications related the possibility of foreground (imposter) events, which is beyond the scope of our present order-of-magnitude estimates.

\subsection{UV Phase-Space}
\label{sec:UV}

\begin{figure}
    \includegraphics[width=0.5\textwidth]{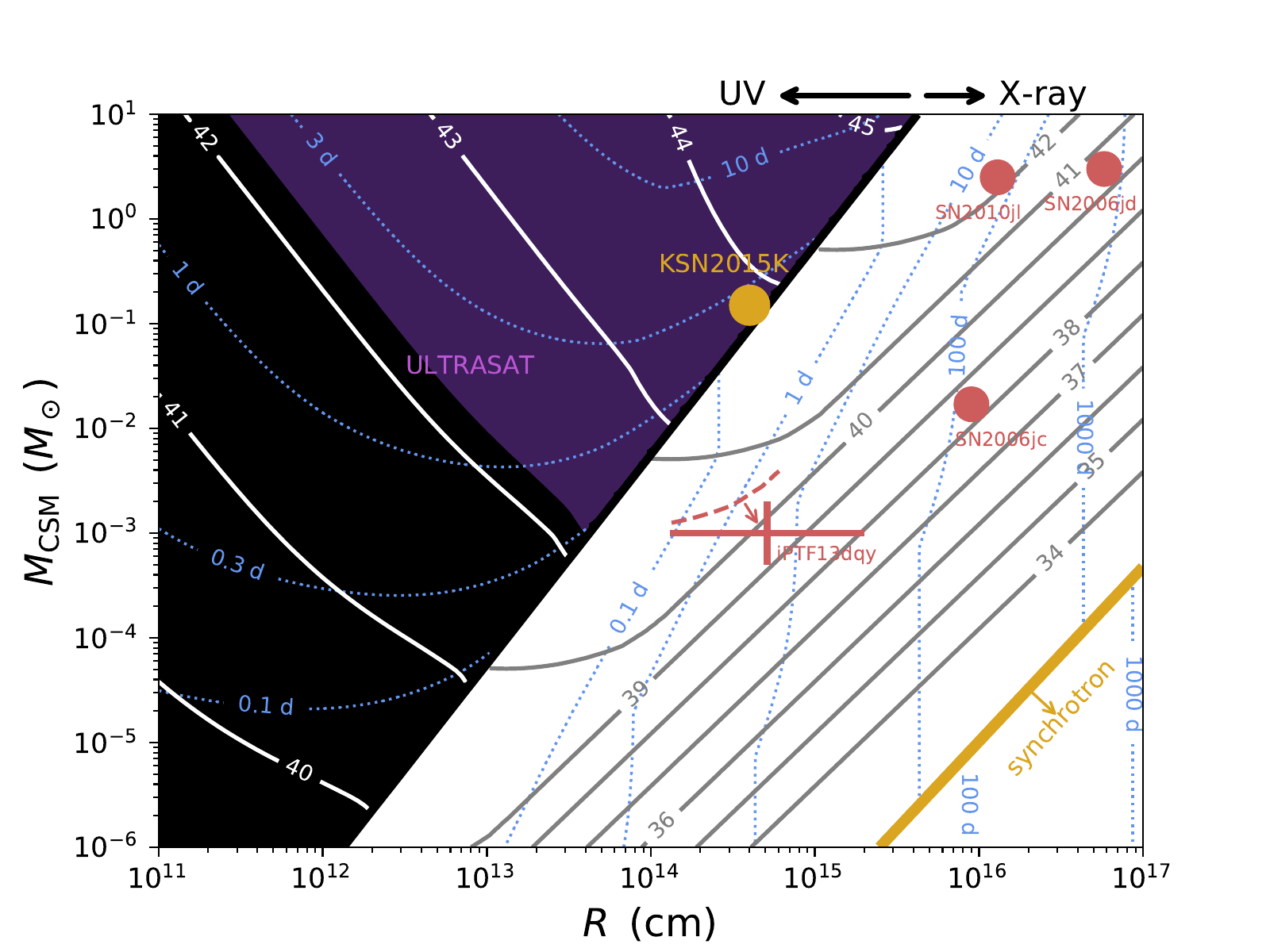}
    \caption{
    Same as Fig.~\ref{fig:MR}, but for the near-UV instead of optical band. The white (blue) curves within the dark shaded region to the left (where $\tau_T > c/v$) show contours of peak luminosity (and corresponding duration) at $2500 \Angstrom$, calculated as described in \S\ref{sec:UVOptical}. Curves within the right-side region show luminosity/timescale contours at 1\,keV, with symbols and notation following Fig.~\ref{fig:MR}. Dense CSM with $\tau_T > c/v$ produces bight UV transients that would be prime targets for upcoming UV missions. The purple shaded region (a subset of the black region) shows the parameter space in which the ULTRASAT detection rate is $\dot{N}_{\rm det} > 1 \, {\rm yr}^{-1}$ (eq.~\ref{eq:Ndet_ULTRASAT}; \S\ref{sec:UV}).
    }
    \label{fig:MR_UV}
\end{figure}

We conclude this section by highlighting the importance of wide-field UV instruments in constraining the CSM phase space. Eq.~(\ref{eq:T_RMS}) shows that transients powered by CSM with $\tau_T \gtrsim c/v$ emit most of their energy in the UV. This is empirically supported by observations of fast blue optical transients. Future wide-field high-cadence UV missions are therefore critical to further explore such CSM interaction. The Ultraviolet Transient Astronomy Satellite (ULTRASAT) is a near-UV mission currently under development that is particularly important in this context \citep{Sagiv+14}. Indeed, one of the primary science goals of ULTRASAT is detecting shock-breakout from SNe, which is similar to the dense CSM interaction discussed in \S\ref{sec:UVOptical}. 
In Fig.~\ref{fig:MR_UV} we show the luminosity and timescale associated with such transients in the near-UV band. This is the same as Fig.~\ref{fig:MR}, except that the luminosity/duration contours within the $\tau_T > c/v$ region (dark shaded region) are calculated at $2500 \Angstrom$ where the ULTRASAT sensitivity peaks.
We can roughly estimate the number of such events detectable by ULTRASAT by assuming that any transient whose flux exceeds $F_{\rm lim} \approx 6.9 \times 10^{-14} \, {\rm erg \, cm}^{-2} \, {\rm s}^{-1}$ is correctly identified. This flux corresponds to an effective limiting AB magnitude of $22$ (E. Ofek, private communication).
Taking the instrument's field of view to be $\Omega = 180 \, \degsq$, we find that the rate at which ULTRASAT will detect such events is
\begin{align}
\label{eq:Ndet_ULTRASAT}
    \dot{N}_\mathrm{det} 
    &= \frac{\Omega}{4\pi} \mathcal{R} V_{\rm lim} 
    \\ \nonumber
    &\approx 17 \, {\rm yr}^{-1} \, \left(\frac{\mathcal{R}}{700 \, {\rm Gpc}^{-3} \, {\rm yr}^{-1}}\right) \left(\frac{L_{\rm NUV}}{10^{43}\,{\rm erg \, s}^{-1}}\right)^{3/2}
    .
\end{align}
This is qualitatively consistent with the estimates of \cite{Ganot+16} (see their Table 2), with the primary quantitative difference arising from our normalization to a lower volumetric rate $\mathcal{R}$ (and additional minor differences in the assumed sensitivity threshold $F_{\rm lim}$ and model light-curves).
The purple shaded region in Fig.~\ref{fig:MR_UV} shows the CSM parameter space where $\dot{N}_{\rm det}$ exceeds one-per-year. Clearly, ULTRASAT would probe a significant region in CSM mass-radius parameter space that is currently under-explored, even with our conservative (low) fiducial volumetric rate.

\section{Inferring CSM Properties from X-ray Detections}
\label{sec:inferring_CSM}

In the previous section we presented the phase space of CSM-interaction as a function of physical properties such as CSM mass, radius, and shock velocity (e.g. Fig.~\ref{fig:MR}). This is a natural approach based on the forward modeling derived in \S\ref{sec:GeneralConsiderations}--\ref{sec:ExpansionPhase}.
Here we turn the problem around and ask---can the CSM properties be inferred from observed X-ray data?

Based on Figs.~\ref{fig:Results}--\ref{fig:Xray_both_sides} we find that the X-ray transient duration $t_X$ is typically set by the post-interaction phase, and that emission during this phase is mostly governed by un-Comptonized bremsstrahlung unless the shock speed is larger than a typical SN shock speed of $\sim 10,000 \ {\rm km \, s}^{-1}$. This implies that $t_X \sim t_{\rm dyn}$ for an adiabatic shock and $t_X \sim t_{\rm ff}$ in the radiative regime.
To enforce continuity between the two cases, we assume that $t_X = \min\left(t_{\rm dyn},t_{\rm ff}/\eta\right)$ as implied by eq.~(\ref{eq:L_ff}).
Inverting eqs.~(\ref{eq:t_dyn},\ref{eq:t_ff},\ref{eq:LX_unComptonized}) we find that the CSM mass and radius can be expressed as a function of the duration $t_X$ and peak luminosity $L_X$ of the X-ray transient,
\begin{equation}
\label{eq:M_long}
    M \approx
    \max
    \begin{cases}
    0.39 \,M_\odot \, \left(\frac{L_X/\nu_{\rm keV}}{10^{41}\,{\rm erg \,s}^{-1}}\right)^{1/2} \left(\frac{t_X}{100\,{\rm d}}\right)^{3/2} v_9^2
    \\
    0.029\,M_\odot \, \left(\frac{L_X/\nu_{\rm keV} }{10^{41}\,{\rm erg \,s}^{-1}}\right)^{3/2} \left(\frac{t_X}{100\,{\rm d}}\right)^{1/2} v_9^{-2} 
    \end{cases}
\end{equation}
and
\begin{equation}
\label{eq:R_long}
    R \approx
    \max
    \begin{cases}
    8.6 \times 10^{15}\,{\rm cm}\, \left(\frac{t_X}{100\,{\rm d}}\right) v_9
    \\
    2.3\times 10^{15}\,{\rm cm}\, \left(\frac{L_X/\nu_{\rm keV}}{10^{41}\,{\rm erg \,s}^{-1}}\right)^{1/2} \left(\frac{t_X}{100\,{\rm d}}\right)^{1/2} v_9^{-1} 
    \end{cases}
    .
\end{equation}
In both equations above the top (bottom) case corresponds to the adiabatic (radiative) regime.

The adiabatic and radiative expressions in eqs.~(\ref{eq:M_long},\ref{eq:R_long}) equal one another at a critical value of the shock velocity,
\begin{equation}
\label{eq:v_rad}
    v_{\rm rad} \approx 5,200\,{\rm km \,s}^{-1}\,
    \left(\frac{L_X /\nu_{\rm keV}}{10^{41}\,{\rm erg \,s}^{-1}}\right)^{1/4} \left(\frac{t_X}{100\,{\rm d}}\right)^{-1/4} 
    .
\end{equation}
For $v>v_{\rm rad}$ the shock is adiabatic, while at $v<v_{\rm rad}$ it is radiative (see e.g. Fig.~\ref{fig:PhaseSpace}).
It is convenient to define a dimensionless variable
\begin{equation}
\label{eq:v_tilde}
    \tilde{v} \equiv 
    \max
    \begin{cases}
    v/v_{\rm rad}
    \\
    \left(v/v_{\rm rad}\right)^{-1}
    \end{cases}
\end{equation}
such that $\tilde{v} \geq 1$.
With this new variable, we can write down the CSM mass and radius (eqs.~\ref{eq:M_long},\ref{eq:R_long}) in a simple form that is applicable in both adiabatic and radiative regimes,
\begin{equation}
\label{eq:M}
    M \approx 
    0.11 \,M_\odot \,
    \left(\frac{\tilde{v}L_X/\nu_{\rm keV}}{10^{41}\,{\rm erg \,s}^{-1}}\right) \left(\frac{\tilde{v}t_X}{100\,{\rm d}}\right)
\end{equation}
and
\begin{equation}
\label{eq:R}
    R \approx 
    4.5 \times 10^{15} \,{\rm cm} \,
    \left(\frac{\tilde{v}L_X/\nu_{\rm keV}}{10^{41}\,{\rm erg \,s}^{-1}}\right)^{1/4} \left(\frac{\tilde{v}t_X}{100\,{\rm d}}\right)^{3/4} 
    .
\end{equation}

\begin{figure}
    \includegraphics[width=0.5\textwidth]{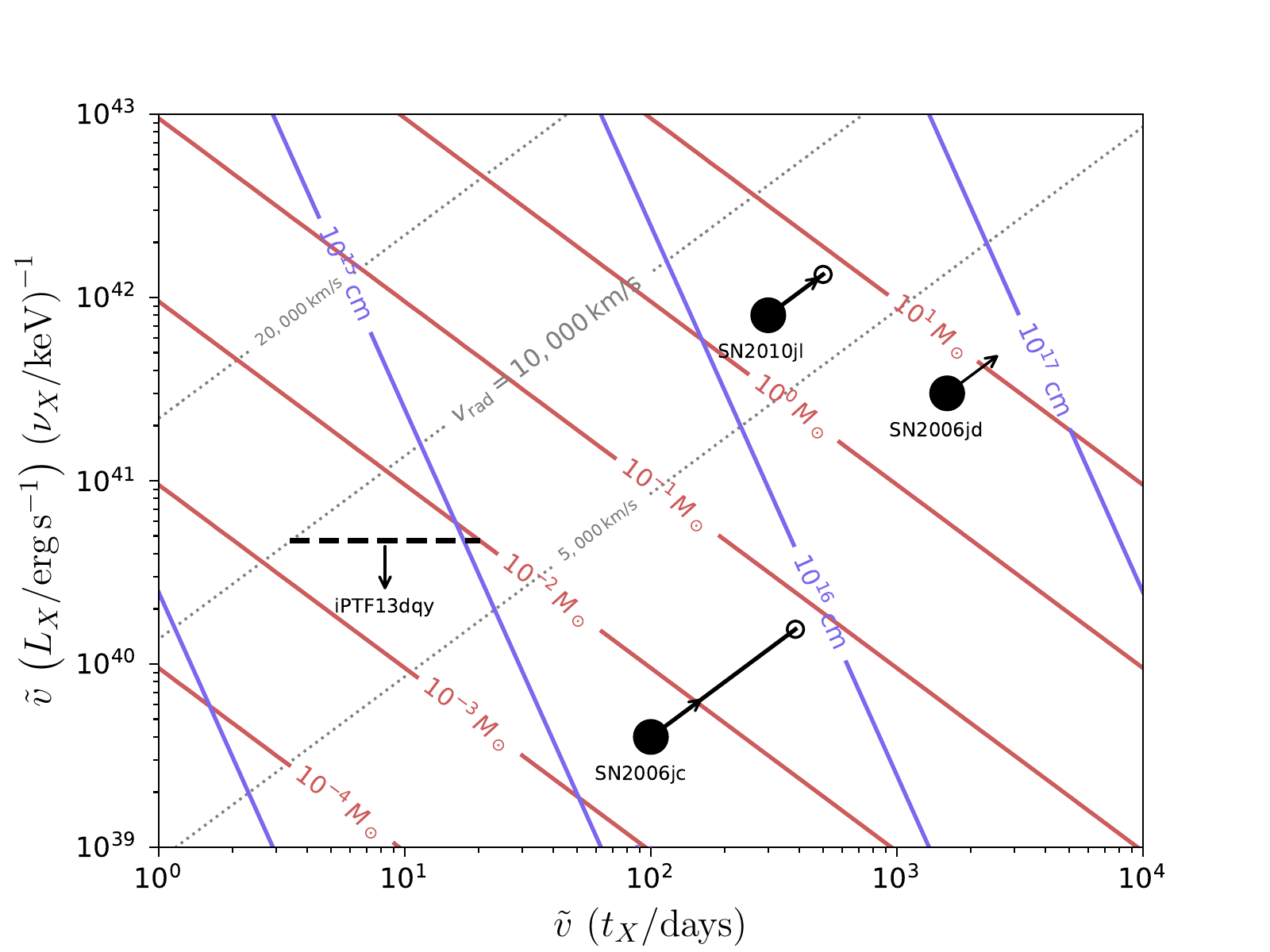}
    \caption{
    Luminosity--duration phase-space of CSM-powered thermal (bremsstrahlung) X-ray transients neglecting Comptonization: red (blue) contours show the CSM mass (radius) required to produce an X-ray transient with peak luminosity $L_X$ at frequency $\nu_X$ and duration $t_X$ (eqs.~\ref{eq:M},\ref{eq:R}). The dimensionless variable $\tilde{v}\left(v,L_X,t_X\right) \gtrsim 1$ that multiplies both axes can be calculated from eqs.~(\ref{eq:v_rad},\ref{eq:v_tilde}) if the shock velocity $v$ is known. Otherwise, a conservative assumption of $\tilde{v}=1$ yields minimum values of mass and radius. 
    As in Figs.~\ref{fig:MR}--\ref{fig:MR_UV} we show the events: SN2010jl \citep{Ofek+14,Chandra+15}, SN2006jd \citep{Chandra+12}, SN2006jc \citep{Immler+08}, and iPTF13dqy \citep{Yaron+17}.
    Dotted grey contours show $v_{\rm rad}$ (eq.~\ref{eq:v_rad}). Events that fall below the curve corresponding to $v=v_{\rm rad}$ (shock velocity equals the labeled contour value) are within the radiative regime.
    This diagram provides a convenient framework for inferring CSM properties of X-ray transients.
    See \S\ref{sec:inferring_CSM} for further details.
    }
    \label{fig:LX_tX}
\end{figure}

Equations~(\ref{eq:M},\ref{eq:R}) can be used to infer the CSM mass and radius of observed thermal X-ray transients (we note again that eqs.~\ref{eq:M},\ref{eq:R} assume  Compton $y \lesssim 1$, which is a consistency check that should be made when using these results). This is shown in Fig.~\ref{fig:LX_tX}, where contours of constant CSM mass (red) and radius (blue) are plotted as a function of $\tilde{v} t_X$ and $\tilde{v} L_X / \nu_{\rm keV}$.
If the shock velocity of a given event is measured by other means, then $\tilde{v}$ can be calculated from eqs.~(\ref{eq:v_rad},\ref{eq:v_tilde}) and the source can be unambiguously placed within this diagram. In practice, the shock velocity is usually unknown. In this case, the CSM mass/radius cannot be uniquely determined. A conservative assumption is to adopt $\tilde{v}=1$ when placing events on the luminosity--duration diagram~\ref{fig:LX_tX}. This corresponds to the assumption that the shock is marginally radiative ($v = v_{\rm rad}$) and yields minimum values of $M$ and $R$ that are consistent with the data. If the shock velocity is either greater than or lower than $v_{\rm rad}$ then $\tilde{v}>1$ and the source would move along an upwards diagonal trajectory in Fig.~\ref{fig:LX_tX}. 
This is illustrated by the black arrows in the figure.

We illustrate the method by placing a handful of SNe with observed X-ray emission within this luminosity--duration phase space, as described above (Fig.~\ref{fig:LX_tX}).
For SN2006jc and SN2006jd we adopt $L_X \approx 4 \times 10^{39}\,{\rm erg\,s}^{-1}$, $t_X \approx 100\,{\rm d}$ and $L_X \approx 3 \times 10^{41}\,{\rm erg\,s}^{-1}$, $t_X \gtrsim 1,600\,{\rm d}$, respectively \citep{Immler+08,Chandra+12}.
These quoted luminosities are in the 0.2--10\,keV band, and depend on the assumed photon index. Here we adopt $\nu_X = 1\,{\rm keV}$ as a characteristic frequency, however we note that for electron temperatures above $\geq 10 \,{\rm keV}$ (eq.~\ref{eq:Te}) the luminosity should be dominated by the top of the band $\sim 10\,{\rm keV}$ (for the bremsstrahlung emission spectrum relevant here the photon index would be $\Gamma = 1$).
Lacking a direct constraint on the shock velocity for these events (from which $\tilde{v}$ could be calculated; eq.~\ref{eq:v_tilde}), we conservatively assume $\tilde{v}=1$. 
This amounts to assuming that the shock velocity is $v = v_{\rm rad} \approx 2,300\,{\rm km \,s}^{-1}$ for SN2006jc and $\approx 3,400\,{\rm km \,s}^{-1}$ for SN2006jd.
If the actual shock velocity of either event is larger/smaller then $\tilde{v}$ would be $>1$ and the events would move in the direction of the black arrows.
For example, \cite{Immler+08} suggested that $v \approx 9,000\,{\rm km\,s}^{-1}$ for SN2006jc, which would imply $\tilde{v} \approx 3.9$ and a larger CSM mass and radius (marked with a connected open circle).
For this velocity, SN2006jc would be in the radiative regime. This may explain why the observed X-ray rise/fall time is shorter than the time elapsed since the SN explosion ($\sim t_{\rm dyn}$).

For SN2010jl we similarly adopt $L_X \approx 8 \times 10^{41}\,{\rm erg\,s}^{-1}$ at $\nu_{\rm keV}=1$ and $t_X \approx 300\,{\rm d}$ based on \cite{Chandra+15}. This event was particularly well-observed, allowing direct estimates of the peak temperature (from the SED), and thus the shock velocity \citep{Ofek+14}. Here we adopt the value $k_B T_e \approx 19\,{\rm keV}$ from \cite{Chandra+15}, which implies a shock velocity $v \approx 4,000 \,{\rm km \,s}^{-1}$ (eq.~\ref{eq:Te}). Note that the values above (and especially $L_X$) may be uncertain by factors of a few, e.g. due to the intervening neutral column density $N_{\rm H}$ (compare \citealt{Ofek+14} with \citealt{Chandra+15}).
Similar to the other events, we show the conservative $\tilde{v}=1$ location of SN2010jl with a filled black circle in Fig.~\ref{fig:LX_tX}. The more realistic location, utilizing the inferred shock velocity is shown with a connected open circle. This implies $\tilde{v} \approx 1.7$ ($v_{\rm rad} \approx 6,600 \,{\rm km\,s}^{-1}$) and that SN2010jl is radiative. 
Finally, we show the {\it Swift} upper limits on X-ray emission from iPTF13dqy with a dashed black curve \citep{Yaron+17}. This limit is marginally consistent with the inferred CSM mass and radius of this event based on flash-ionization spectroscopy.
Future detected events may be similarly placed on this diagram, and help unveil CSM properties of different stellar populations.

\section{Conclusions}
\label{sec:conclusions}

Dense CSM interaction may produce bright electromagnetic emission that manifests in myriad ways depending on the CSM and shock properties. In this work we have described the optical to X-ray signatures that arise from such interaction. The shock--CSM parameter space is divided into distinct regions based on the shock velocity $v$ and CSM  column density, e.g. parameterized by the Thompson optical depth $\tau_T$ (Fig.~\ref{fig:PhaseSpace}): at low optical depths $\tau_T < c/v$ the (collisionless) shock heats the CSM to $\gg$keV temperatures (eq.~\ref{eq:Te}) and a hard X-ray thermal transient with a (potentially Comptonized) bremsstrahlung spectrum is produced;
at high optical-depths $\tau_T > c/v$ a radiation-mediated shock that heats the CSM to significantly lower temperatures (eq.~\ref{eq:T_RMS}) is formed instead, and the resulting signature is an optical/UV thermal blackbody transient.
The latter is often termed `shock-breakout' and `cooling-envelope' emission and has been discussed extensively as a mechanism for producing fast-optical transients \citep[e.g.][]{Ofek+10,Nakar&Sari10,Chevalier&Irwin11,Nakar&Piro14,Piro15,Rest+18,Ho+19,Piro+21,Margalit21}.
Here we have instead focused primarily on aspects of the X-ray transient produced within the $\tau_T < c/v$ regime, a problem that was first addressed by \cite{Chevalier&Irwin12} but has received far less attention
(though see \citealt{Pan+13,Svirski+12,Tsuna+21}).
In particular, we treat the case where the CSM has an outer truncation radius, as motivated by observations of fast optical transients (e.g. \citealt{Rest+18}) and enhanced mass-loss in late stages of stellar evolution (e.g. \citealt{Quataert2012}).

Properties of the X-ray transient depend on the thermal state of the shock and on photon propagation effects. In \S\ref{sec:ParameterSpace} we showed that the shock--CSM parameter space can be divided into several regions depending on whether the shock is radiative or adiabatic, and whether Comptonization is important (see Fig.~\ref{fig:PhaseSpace}).
Additionally, the X-ray light-curve can be separated into two phases: (i) the `interaction' phase in which the shock propagates within the CSM, and (ii) the subsequent `post-interaction' phase that describes expansion and cooling of the CSM after it has been fully shocked. These are analogous to the shock-breakout and cooling-envelope emission phases discussed in the context of radiation-mediated shocks.

In \S\ref{sec:InteractionPhase} we discussed the X-ray signature produced during the interaction phase, paying special attention to Comptonization of low-frequency bremsstrahlung photons by hot post-shock electrons, a process that becomes important in shaping both the emergent spectrum and total energetics at high shock velocities. We carefully treat Compton scattering in Appendix~\ref{sec:Appendix_Comptonization} and show that the commonly used expression for the inverse-Compton cooling rate
$\Lambda_{\rm C} = n_e \sigma_T c \left( 4 k_B T_e / m_e c^2 \right) U_{\rm rad}$ is not valid (in a global sense) in regimes where the Compton-y parameter is large, $y \gtrsim 1$.

As first pointed out by \cite{Chevalier&Irwin12} and \cite{Svirski+12} and discussed in \S\ref{sec:Propagation}, propagation effects through the upstream unshocked-CSM can severely inhibit the emergent X-ray luminosity during the interaction phase. 
In particular, photoelectric absorption of $\sim$keV photons can quench the X-ray signature within this band until the shock reaches very near to the CSM outer edge (where the unshocked CSM column density is small). This implies a short duration $\ll t_{\rm dyn}$ of the interaction-phase X-ray light curve (eq.~\ref{eq:t_int}), unless X-rays produced by the shock manage to photoionize the upstream CSM. As shown by eq.~(\ref{eq:v_ionization_bo}; see also Appendix~\ref{sec:Appendix_IonizationBreakout}), this is only possible for extremely high shock velocities and/or low density CSM, and is therefore irrelevant throughout most of the parameter space.

Because X-ray emission is likely severely inhibited by bound-free absorption during the bulk of the interaction phase, we were motivated to consider the novel regime of the post-interaction phase (\S\ref{sec:ExpansionPhase}). Accounting for adiabatic expansion, Comptonization, and radiative cooling, we derived the range of possible X-ray light-curves within this phase. These are summarized in Fig.~\ref{fig:lightcurves}.

Transitioning to the $\tau_T > c/v$ parameter space, we briefly reviewed and discussed the (primarily) optical/UV emission that is expected within this radiation-mediated shock regime (\S\ref{sec:UVOptical}). Details of these processes are derived and discussed in greater detail elsewhere \citep[e.g.][]{Chevalier&Irwin11,Ginzburg&Balberg12,Piro15,Margalit21}, and we here mainly recapitulated a few pertinent points for completeness.

In \S\ref{sec:results} we described and discussed observable implications of our results. We plot the luminosity and duration of CSM-powered X-ray transients within the phase-space of shock velocity and CSM mass/radius (Figs.~\ref{fig:Results}--\ref{fig:MR_UV}). The number of these transients that would have been detectable by the ROSAT all-sky survey (RASS) and that may be found with eROSITA are estimated in \S\ref{sec:detectability} and shown in Fig.~\ref{fig:Xray_both_sides} (see also Table~\ref{tab:rates}). For a volumetric event-rate of $\mathcal{R} = 700 \, {\rm Gpc}^{-3} \, {\rm yr}^{-1}$ ($\sim 1\%$ of the CCSN rate; motivated by the rate of FBOTs) we find that RASS most likely would not have been sensitive to such transients, consistent with the lack of candidate events in previous searches \citep{Donley2002}.
In contrast, eROSITA is expected to discover $N_{\rm det}>1$ events with $M \gtrsim 10^{-1} \, M_\odot$ and $10^{14} \, {\rm cm} \lesssim R \lesssim 10^{16} \, {\rm cm}$ and potentially many more short-duration ($\sim 10^2 \, {\rm s}$) shock-breakout X-ray flashes.
The latter would be associated with particularly compact CSM shells ($R \sim 10^{13}\,{\rm cm}$). One concern is that X-ray emission from such shells might be bound-free absorbed by even low-density material that may enshroud the dense CSM shell (e.g., a standard stellar wind).  We show in Appendix~\ref{sec:Appendix_Wind_IonizationBreakout}, however, that radiation is typically capable of photoionizing its way out of such material so that this should not in fact affect detectability prospects (see Fig.~\ref{fig:Xray_both_sides}).

We additionally discussed the dichotomy between observable manifestations of collisionless and radiation-mediated shocks, as illustrated in Fig.~\ref{fig:MR}. Collisionless shocks ($\tau_T < c/v$) produce hard X-ray emission, whereas radiation-mediated shocks with $\tau_T \gtrsim c/v$ manifest as bright optical/UV transients. Future wide-field UV missions would be especially sensitive to such events (\S\ref{sec:UV}). In particular, we estimate that the planned ULTRASAT mission \citep{Sagiv+14} may detect $\gtrsim 10^2$ events per year (eq.~\ref{eq:Ndet_ULTRASAT}; Fig.~\ref{fig:MR_UV}). This would revolutionize our ability to probe confined dense CSM and would help improve our understanding of stellar mass loss during the final months-to-years of a star's life.

Finally, in \S\ref{sec:inferring_CSM} we showed how X-ray observations may be used to infer properties of the underlying CSM. The CSM mass and radius can be found using the peak X-ray luminosity and duration of observed events, provided that the shock velocity is known (e.g. by constraining the X-ray SED). A convenient parameterization that is applicable in both the adiabatic and radiative regimes is given by eqs.~(\ref{eq:M},\ref{eq:R}) and a dimensionless variable $\tilde{v} \gtrsim 1$ (eq.~\ref{eq:v_rad},\ref{eq:v_tilde}). Fig.~\ref{fig:LX_tX} illustrates this luminosity--duration phase-space, showing a handful of SNe with X-ray detections. This figure provides a convenient framework for inferring CSM properties and comparing X-ray transients.
This can be viewed as an analog to the luminosity--duration phase-space of synchrotron self-absorbed radio transients \citep{Chevalier98} for the case of optically-thin thermal (bremsstrahlung) X-ray transients.
Systematic sensitive X-ray follow-up of nearby SNe on timescales of months--decades will be necessary to further fill in this phase-space \citep[e.g.][]{Ofek+13}.

We conclude by commenting on our choice of density profile. Throughout this work we have considered the CSM to be a constant density (top-hat) shell. An outer truncation radius is motivated by observations and modeling of FBOTs; however, our assumption of a constant density medium within this radius is somewhat ad-hoc and chosen for simplicity.
We note however that our main results do not depend strongly on this assumption. In particular, our results can easily be applied also to a wind density profile 
$\rho_{\rm w} = \dot{M}/4\pi v_{\rm w} r^2$ (where $\dot{M}$ is the mass-loss and $v_{\rm w}$ the wind velocity)
under the simple transformation
$M \to R \dot{M}/v_{\rm w}$
in every equation (where $R$ is the outer wind truncation radius).

\acknowledgements
We thank Eran Ofek and the anonymous referee for a careful reading and helpful comments that helped improve this work.
A.Y.Q.H. would like to thank Jenn Donley, Michael Eracleous, and William Brandt for answering questions about their RASS transient search.
BM is supported by NASA through the NASA Hubble Fellowship grant \#HST-HF2-51412.001-A awarded by the Space Telescope Science Institute, which is operated by the Association of Universities for Research in Astronomy, Inc., for NASA, under contract NAS5-26555.   EQ was supported in part by a Simons Investigator Grant from the Simons Foundation.
This work benefited from workshops and collaborative interaction supported by the Gordon and Betty Moore Foundation through grant GBMF5076.

\appendix
\section{Inverse Compton Scattering}
\label{sec:Appendix_Comptonization}

In the following we derive pertinent results regarding Componization of bremsstrahlung emission for arbitrary Compton-y parameter. These topics were first discussed by \cite{Illarionov&Sunyaev72} and \cite{Felten&Rees72}. Below we re-derive and extend some of these results, and identify a minor point of contention. We also point out important differences between our treatment of inverse Compton cooling and previous treatments in the context of CSM X-ray emission studies (e.g. \citealt{Chevalier&Irwin12}, \citealt{Svirski+12}).

\subsection{Inverse Compton Cooling}

Inverse Compton scattering increases the energy of photons at the expense of scattering electrons so long as $h \left\langle \nu \right\rangle < 4 k_B T_e$.
In the the small Compton-y regime, $y \ll 1$, the inverse Compton cooling rate can be calculated from the perspective of an electron scattering off the radiation field, resulting in the well-known expression
$\Lambda_{\rm C} = n_e \sigma_T c \left(\frac{4 k_B T_e}{m_e c^2}\right) U_\gamma$, where $U_\gamma$ is the radiation energy density.
An important point is that this expression is not applicable in the $y \gtrsim 1$ regime, where photons are upscattered appreciably. In this regime, the radiation energy density is no longer an appropriate parameter since it is not conserved in the scattering process (i.e. the emergent $U_\gamma$ differs significantly from its initial value, before scattering). 
For the same reason, it is also more convenient to consider the scattering process from the point of view of photons, assuming that the electron distribution function and temperature remain fixed.

A photon within the scattering medium is upscattered in frequency by a factor $\sim e^y$, and up to a maximum of $h\nu \sim 3 k_B T_e$.
The energy gain for a given photon of initial frequency $\nu$ is therefore
\begin{equation}
\delta E_\gamma(\nu) 
= h\nu \left[ \min \left( e^y , \frac{3 k_B T_e}{h\nu} \right) - 1 \right]
.
\end{equation}
The electron cooling rate is equal to the energy gain of photons per unit time,
\begin{align}
\Lambda_{\rm C} 
&= \frac{dE_\gamma}{dV dt}
\approx \int d\nu \frac{8\pi \nu^2}{c^3} \dot{n}(\nu) \delta E_\gamma(\nu) 
\\ \nonumber
&= \left( e^y -1 \right) \int^{\nu_y} \frac{8\pi \nu^2}{c^3} \dot{n}(\nu) h\nu d\nu
\\ \nonumber
&+
\int_{\nu_y} \frac{8\pi \nu^2}{c^3} \dot{n}(\nu) \left( 3k_B T_e - h\nu \right) d\nu
\\ \nonumber
&= \left( e^y -1 \right) \dot{U}_\gamma - e^y \dot{U}_\gamma(>\nu_y) + 3 k_B T_e \dot{n}_\gamma(>\nu_y)
\end{align}
where $\nu_y = e^{-y} k_B T_e/h$ is the frequency above which all emitted photons saturate in Compton scattering up to the thermal peak.

For small Compton-y $\lesssim 1$, $h \nu_y \approx k_B T_e$. If the injected spectrum is sufficiently ``soft'', then both $\dot{U}_\gamma(>\nu_y)$ and $\dot{n}_\gamma(>\nu_y)$ are small, and the cooling rate reduces to the familiar expression $\Lambda_{\rm C} \approx y \dot{U}_\gamma = n_e \sigma_T c \left(\frac{4 k_B T_e}{m_e c^2}\right) U_\gamma$ (where we used the fact that $y = 4 t_{\rm esc}/t_{\rm IC}$ and $U_\gamma = \dot{U}_\gamma t_{\rm esc}$).
In the opposite, large Compton-y, regime we instead have that the cooling rate is proportional to the volumetric rate at which soft photons are produced.

For the specific case where photon-production is governed by bremsstrahlung emission, we note that the free-free cooling rate is simply $\Lambda_{\rm ff} = \dot{U}_\gamma$.
We then express the Compton cooling rate in terms of an enhancement factor 
$\mathcal{C} \equiv \left( \Lambda_{\rm ff} + \Lambda_{\rm C} \right) / \Lambda_{\rm ff}$ 
to the free-free cooling rate, so that
\begin{equation}
\label{eq:Appendix_Cy}
\mathcal{C}(y) 
= e^y \left[ 1 - \frac{\dot{U}_\gamma(>\nu_y)}{\dot{U}_\gamma} \right] + 3 k_B T_e \frac{\dot{n}_\gamma(>\nu_y)}{\dot{U}_\gamma}
.
\end{equation}
For bremsstrahlung emission, the source function is proportional to $\dot{n}(x) \propto g(x) x^{-3} e^{-x}$ where $g(x) \approx \ln \left( 2.2 / x \right)$ is the Gaunt factor, and $x = h\nu / k_B T_e$ is a normalized frequency ordinate.
In this case, the second term on the RHS of eq.~(\ref{eq:Appendix_Cy}) can be estimated as
\begin{equation}
3 \frac{\int_{x_y} g(x) x^{-1} e^{-x} dx}{\int_{x_{\rm coh}} g(x) e^{-x} dx}
\approx 
3y \left( \frac{1}{2}y + \ln (2.2) \right)
\end{equation}
for $y < y_{\rm sat}$, where
we have approximated the integral by taking $e^{-x} \sim \Theta\left(x\leq1\right)$,
and where 
the saturation Compton-y parameter is defined as
\begin{equation}
\label{eq:Appendix_ysat}
y_{\rm sat} \equiv \ln (x_{\rm coh}^{-1}) .
\end{equation}
The frequency $x_{\rm coh}$ above plays an important role in Comptonization of bremsstrahlung radiation---it represents the minimal frequency of photons that can effectively inverse Compton upscatter in the medium (\citealt{Kompaneets57,Illarionov&Sunyaev72})
\begin{align}
\label{eq:x_coh}
\frac{x_{\rm coh}}{\sqrt{\ln(2.2/x_{\rm coh})}}
&\simeq 3 \times 10^5 \, n_e^{1/2} T_e^{-9/4} 
\\ \nonumber
&\approx 1.4 \times 10^{-10} \, 
M_{-1}^{1/2} R_{15}^{-3/2} \epsilon_T^{-9/4} v_9^{-9/2} .
\end{align}
Below this frequency free-free absorption operates faster than inverse-Compton scattering and photons are destroyed (absorbed) before they can Compton-upscatter in frequency \citep{Kompaneets57}.

The first term on the RHS of eq.~(\ref{eq:Appendix_Cy}) can similarly be estimated as
\begin{equation}
e^y \left[ 1 - \frac{\int_{x_y} g(x) e^{-x} dx}{\int_{x_{\rm coh}} g(x) e^{-x} dx} \right]
\approx 1 + y - y_{\rm sat} 
e^{y-y_{\rm sat}}
,
\end{equation}
where above we assumed that $x_{\rm coh} \ll 1$.
Overall, we find that
\begin{align}
\label{eq:Appendix_Cy_final}
\mathcal{C}(y)
&\approx
\begin{cases}
1 + y - y_{\rm sat} e^{y-y_{\rm sat}} +
3y \left( \frac{1}{2}y + \ln (2.2) \right)
&, y < y_{\rm sat}
\\
1 + 
3y_{\rm sat} \left( \frac{1}{2}y_{\rm sat} + \ln (2.2) \right)
&, y \geq y_{\rm sat}
\end{cases}
\nonumber \\
&\propto 
\begin{cases}
1 + y &, y < 1
\\
y^2 &, 1 \lesssim y < y_{\rm sat}
\\
const. &, y \geq y_{\rm sat}
\end{cases}
\end{align}
and $\mathcal{C} \approx \frac{3}{2} \ln \left( x_{\rm coh}^{-1} \right)^2$ in the saturated Comptonized regime ($y \gtrsim y_{\rm sat}$), in agreement with the result first derived by Kompaneets \citep{Kompaneets57}. Equation~(\ref{eq:Appendix_Cy_final}) can be seen as an extension of that result to arbitrary Compton-y.

A final point---the Compton cooling correction in eq.~(\ref{eq:Appendix_Cy_final}) is derived without reference to the spatial distribution of electrons and photons. This is akin to the infinite homogeneous medium approach commonly adopted in solving the Kompaneets equation. A physical (finite size) cloud from which photons escape will be influenced by edge effects (e.g. photons emitted near the surface of the medium will experience less scatterings and lower Comptonization), and the cooling correction will be somewhat modified. In the following subsection we treat the spatial Comptonization problem in detail. The effective volume-averaged Compton cooling correction that arises from this treatment, $\mathcal{C} \sim \int \left( L_\nu + L_\nu^{\rm w} \right) d\nu / L_{\rm ff}$ (where $L_\nu + L_\nu^{\rm w}$ is the emergent specific luminosity after Comptonization; eqs.~\ref{eq:Appendix_Lnu_flat},\ref{eq:Appendix_Lnu_w}), is broadly in agreement with eq.~(\ref{eq:Appendix_Cy_final}). The primary difference is a smoother transition into the saturated regime, around $y \sim y_{\rm sat}$.

\subsection{Comptonized Spectrum}

The emergent spectrum of Comptonized bremsstrahlung radiation was first discussed by \cite{Illarionov&Sunyaev72}, and independently by \cite{Felten&Rees72}.
In the following, we review these results and extend them to a broader range of parameters.

The treatment of Comptonization and solution of the Kompaneets equation can be separated into two distinct cases: that of an infinite homogeneous medium; and the more physically-relevant scenario of a finite medium.
The emergent spectrum, though qualitatively similar in both regimes, scales differently with the Compton-y parameter, leading to potential confusion.
Indeed, the Compton-y label should be interpreted quite differently in the two scenarios: it corresponds to a time coordinate in the infinite medium scenario; but represents physical depth (or column density) within the scattering region for the finite-medium scenario.
In the following we treat the finite medium case.

Consider a scattering cloud of size $R$ that is characterized by a constant electron density $n_e$ and temperature $T_e$.
We consider the physical setting in which photons that undergo scattering by the cloud are also produced within cloud itself. In this scenario, photons emitted from different depths within the scattering medium will experience a different number of scatterings (on average) before escaping to an external observer. 
One way to view this is that photons produced at different radii $r$ within the cloud have a different effective $y$ parameter
\begin{equation}
\label{eq:y_eff}
y^\prime(r) = 
4 n_e^2 \sigma_T^2 \left( R-r \right)^2 \left(\frac{k_B T_e}{m_e c^2}\right)
= y \left( 1 - \frac{r}{R} \right)^2
,
\end{equation}
where above we assume that $\tau_T > 1$.
Photons that are produced at frequency $\nu^\prime$ and depth $r$ will be upscattered to $\nu \sim \nu^\prime e^{y^\prime(r)}$ (and up to a maximum of $h \nu \sim 3 k_B T_e$). There is thus a correspondence between the frequency of a photon escaping the scattering medium and the radius $r$ at which it was produced.

The emergent luminosity (assuming steady-state) depends only on the rate of photons produced at different frequencies and depths within the cloud.
The differential number of photons that are produced within the scattering medium and that escape with frequency $\nu$ is thus a convolution of the photon emission frequency $\nu^\prime$ and the depth within the scattering medium $r$,
\begin{equation}
\label{eq:Appendix_dNdot}
d\dot{N}_\gamma(\nu) \sim \int d \nu^\prime 
\frac{4\pi j_\nu(r,\nu^\prime) }{h\nu^\prime}
4\pi r(\nu,\nu^\prime)^2 dr(\nu,\nu^\prime)
\end{equation}
where $j_\nu$ is the emissivity, and
$r(\nu,\nu^\prime) = r\left(y^\prime\right)$ following eq.~(\ref{eq:y_eff}) such that
\begin{equation}
\label{eq:y_eff_nu}
y^\prime(\nu,\nu^\prime) = \min \left[ \ln \left(\frac{\nu}{\nu^\prime}\right) , \ln \left(\frac{3 k_B T_e}{h \nu^\prime}\right) \right] 
\end{equation}
relates a photon's initial frequency to its value upon escape from the scattering cloud. These expressions are correct for an idealized situation in which the number of scattering events ($\propto y^\prime$) is fully determined by $r$. In reality this process is stochastic, so that different photons emitted from the same location $r$ may undergo a somewhat different number of scatterings prior to their escape. This will broaden sharp peaks in the source function spectrum, and can be incorporated into the formalism above by integrating over a (geometry dependent) probability distribution function for $y^\prime(r)$. Here we neglect this complication for sake of simplicity, and note that because the probability distribution function of $y^\prime$ is peaked at a value comparable to eq.~(\ref{eq:y_eff}), and since the source function we consider is weakly dependent on frequency, our simplified treatment will suffice.
This treatment will also allow derivation of results in closed analytic form.

From eqs.~(\ref{eq:y_eff},\ref{eq:y_eff_nu}) we find that
\begin{align}
\label{eq:Appendix_drdnu}
\frac{dr}{d\nu}
&= \left(\frac{dr}{d y^\prime}\right) \left(\frac{d y^\prime}{d\nu}\right)
\\ \nonumber
&= -\frac{R}{2 y^{1/2}}
\begin{cases}
\frac{1}{\nu} \left[ \ln \left(\frac{\nu}{\nu^\prime}\right) \right]^{-1/2} &,~\nu \lesssim \frac{3 k_B T_e }{ h }
\\
0 &,~{\rm else}
\end{cases}
.
\end{align}
This result is of significant importance because it implies that the specific luminosity $L_\nu \propto d\dot{N}_\gamma /d\nu \propto dr/d\nu$ scales as $\propto y^{-1/2}$ in the limit of large $y$ (eqs.~\ref{eq:Appendix_dNdot},\ref{eq:Appendix_drdnu}).
More precisely, 
we can write the specific luminosity $L_\nu$ at frequencies $\nu < 3 k_B T_e$ and below the Wien peak as (eqs.~\ref{eq:Appendix_dNdot},\ref{eq:Appendix_drdnu})
\begin{align}
\label{eq:Appendix_Lnu_flat}
L_\nu &\sim h \nu \left(\frac{d\dot{N}_\gamma}{d\nu}\right)
\approx h \nu \int d \nu^\prime \frac{4\pi j_\nu(r,\nu^\prime)}{h \nu^\prime} 4\pi r^2 \left(\frac{dr}{d\nu}\right)
\nonumber \\
&= 
8\pi^2 R^3 y^{-1/2}
\int_{\nu e^{-y}}^{\nu} 
\frac{d\nu^\prime}{\nu^\prime} 
j_\nu(\nu^\prime)
\left[ \ln \left(\frac{\nu}{\nu^\prime}\right) \right]^{-1/2}
\\ \nonumber
&\times 
\left[ 1 - \left(\frac{\ln(\nu/\nu^\prime)}{y}\right)^{1/2} \right]^2 
\Theta \left( \nu < 3 k_B T_e/h \right) ,
\end{align}
where in the final line we have assumed that the source function $j_\nu$ does not vary with $r$ (which follows from our treatment of constant density, constant temperature media).
One can formally show that this reduces to $L_\nu = 16\pi^2 R^3 j_\nu / 3$ in the limit $y \to 0$, as expected when Comptonization is negligible.

Photons that manage to upscatter all the way to the thermal peak will accumulate at a mean frequency $h\nu = 3 k_B T_e$ and form a Wien spectrum.
The luminosity of this Wien component is
\begin{equation}
\label{eq:Appendix_Lnu_w}
L_\nu^{\rm w} = \frac{1}{2} h \dot{N}_\gamma^{\rm w} x^3 e^{-x}
\end{equation}
where $x \equiv h\nu / k_B T_e$ is the normalized frequency, and the number of photons in the Wien peak is
\begin{align}
\label{eq:Appendix_Ndot_w}
\dot{N}_\gamma^{\rm w} 
&= 
\int d\dot{N}_\gamma
\approx
\int d\nu^\prime \frac{4\pi j_\nu(\nu^\prime)}{h\nu^\prime} \frac{4\pi}{3} 
r_{\rm crit}(\nu^\prime)^3
\\ \nonumber
&=
\frac{16\pi^2 R^3}{3 h}
\int_{e^{-y}}^{3}
\frac{dx^\prime}{x^\prime} j_\nu(x^\prime)
\left[ 1 - \left(\frac{\ln( 3/x^\prime )}{y}\right)^{1/2} \right]^3
\end{align}
where $r_{\rm crit}(\nu^\prime)$ is the minimal depth from which a photon of initial frequency $\nu^\prime$ would be upscattered into the Wien peak. This radius is determined by eq.~(\ref{eq:y_eff}) and the condition $y^\prime(r_{\rm crit}) = \ln (3/x^\prime)$.

For the specific case of bremsstrahlung emission we have $j_\nu(x) \propto g(x) e^{-x}$, where $g(x) \approx \ln (2.2/x)$ is the Gaunt factor.
We can approximate the integral in eq.~(\ref{eq:Appendix_Lnu_flat}) by taking the exponential factor in $j_\nu$ to be a step function that terminates at $x^\prime = 1$. This yields
\begin{align}
\label{eq:Appendix_I_definition}
I(x,y) 
&\equiv 
y^{-1/2}
\int_{\max(x_{\rm coh},xe^{-y})}^x dx^\prime \ln\left(\frac{2.2}{x^\prime}\right) 
\\ \nonumber
&\times
\frac{1}{x^\prime} 
\left[ \ln \left(\frac{1}{x^\prime}\right) \right]^{-1/2}
\left[ 1 - \left(\frac{\ln (1/x^\prime)}{y}\right)^{1/2} \right]^2
\end{align}
where the lower integration limit cannot be below the frequency $x_{\rm coh}$ at which free-free absorption destroys photons faster than inverse-Compton scattering. This frequency defines a critical Compton-y parameter $\ln ( x/x_{\rm coh} ) \equiv \zeta(x)$ through which the solution to the integral (\ref{eq:Appendix_I_definition}) can be expressed as
\begin{align}
\label{eq:Appendix_I1}
I &= 
\frac{1}{15}y + \frac{2}{3} \ln\left(\frac{2.2}{x}\right)
~;~~~~~ y<\zeta 
\equiv \ln \left(\frac{x}{x_{\rm coh}}\right)
\\ 
\label{eq:Appendix_I2}
I &= 
2 \left(\frac{y}{\zeta}\right)^{-1/2}
\left\{
\frac{1}{3}\zeta + \ln\left(\frac{2.2}{x}\right)
\right.
&
\\ \nonumber
&-
\left[ \frac{1}{2}\zeta + \ln\left(\frac{2.2}{x}\right) \right] \left(\frac{y}{\zeta}\right)^{-1/2}
&
\\ \nonumber
&+ \left.
\frac{1}{3} \left[ \frac{3}{5}\zeta + \ln\left(\frac{2.2}{x}\right) \right] \left(\frac{y}{\zeta}\right)^{-1}
\right\}
~;~~~~ y>\zeta(x)
.
\end{align}
These expressions enter the Comptonized specific luminosity and are valid except at frequencies $x \gtrsim 1$ where the step-function approximation to $e^{-x}$ in the bremsstrahlung source function is no longer applicable. Furthermore, at frequencies $x \sim 1$ and sufficiently large Compton-y, the Wien component (eq.~\ref{eq:Appendix_Lnu_w}) may dominate that given by eq.~(\ref{eq:Appendix_Lnu_flat}; this occurs for $x>x_{\rm w}$, see eq.~\ref{eq:Appendix_xw}).

It is convenient to express the Comptonized emission in terms of a multiplicative correction factor to the intrinsic bremsstrahlung emission, $\Psi\left(x,y\right) \equiv L_{\nu}^{\rm C}/L_{\nu}^{\rm ff}$,
\begin{align}
\label{eq:Appendix_Psi}
\Psi\left(x,y\right) 
&\underset{x<1}{\approx} 
\frac{I(x,y)}{I(x,0)}
= 
\frac{3}{2} 
\left[ \ln\left(\frac{2.2}{x}\right) \right]^{-1} 
I(x,y)
\\ \nonumber
&\sim
\begin{cases}
1 
&;~~ y \leq y_{\rm crit}
\\
\left( \frac{y}{y_{\rm crit}} \right)^{-1/2}
&;~~ y > y_{\rm crit}
\end{cases}
\end{align}
where the final line is a crude approximation to eqs.~(\ref{eq:Appendix_I1},\ref{eq:Appendix_I2}) which increases in accuracy at low frequencies $x \ll 1$, and where
\begin{equation}
\label{eq:Appendix_y_crit}
y_{\rm crit}(x) 
\equiv 
\left[ 3 + \frac{ \ln\left({x}/{x_{\rm coh}}\right) }{ \ln\left({2.2}/{x}\right) } \right]^2 \ln\left(\frac{x}{x_{\rm coh}}\right)
.
\end{equation}

\begin{figure*}
    \includegraphics[width=0.5\textwidth]{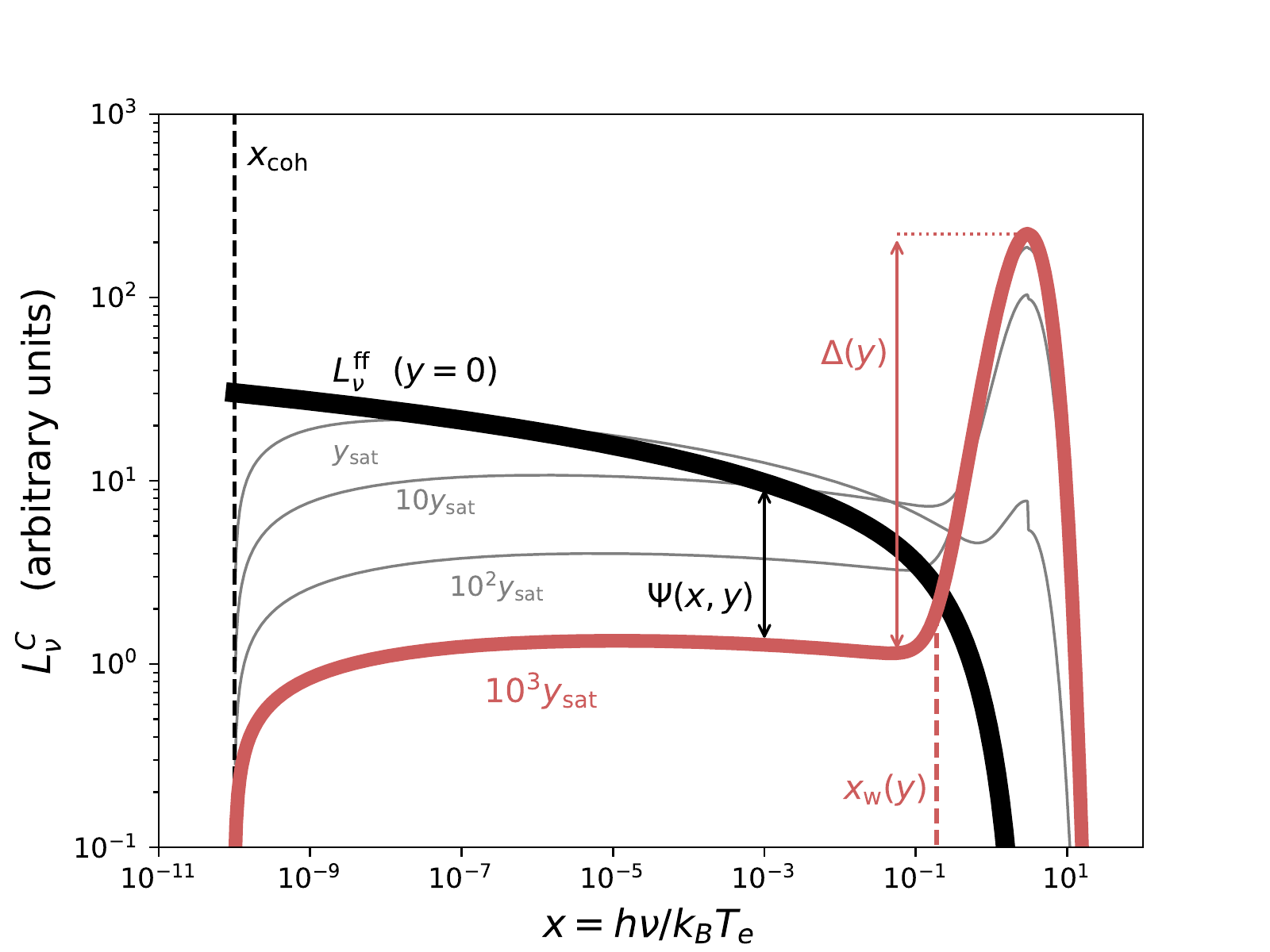}
    \includegraphics[width=0.5\textwidth]{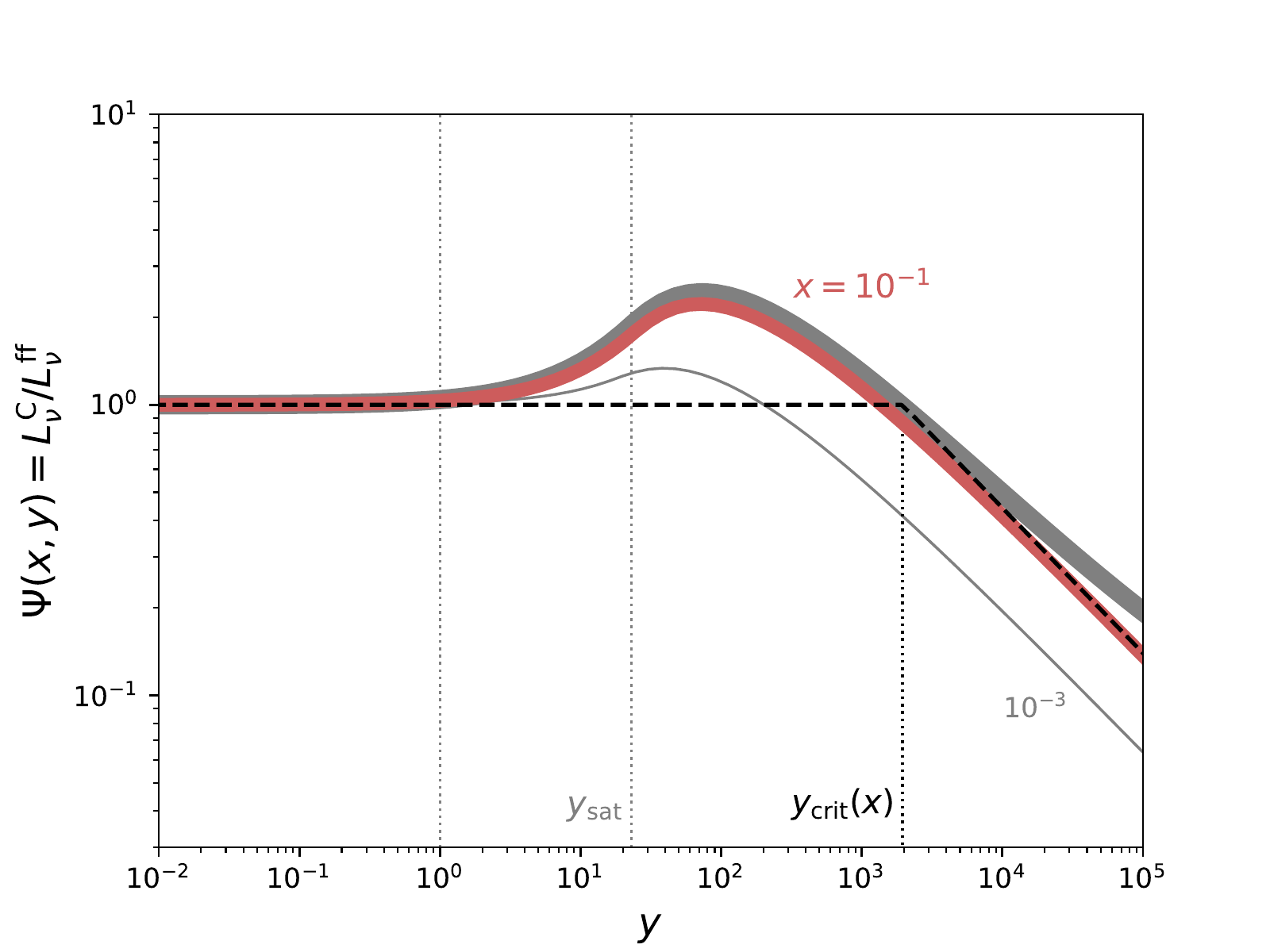}
    \caption{{\it Left:} Spectrum of Comptonized bremsstrahlung emission
    from a finite-size slab with uniform density and temperature. The spectrum is shown for varying
    Compton-y parameters (with $L_\nu^{\rm ff} \propto R^3 j_\nu$ kept fixed; eqs.~\ref{eq:Appendix_Lnu_flat},\ref{eq:Appendix_Lnu_w}). 
    We assume $x_{\rm coh} = 10^{-10}$, implying $y_{\rm sat} \approx 23$ (eq.~\ref{eq:Appendix_ysat}).
    The solid black curve shows the un-Comptonized spectrum ($y=0$), solid red for $y=10^3 y_{\rm sat}$, and grey for intermediate Compton-y (as labeled). In the strongly Comptonized regime ($y \gg y_{\rm sat}$) a Wien spectrum $\propto x^3 e^{-x}$ dominates at frequencies $x > x_{\rm w}$ (eq.~\ref{eq:Appendix_xw}), and the ratio between this peak and the $\sim$flat spectrum at $x<x_{\rm w}$ is $\Delta(y)$ (eq.~\ref{eq:Appendix_Delta}).
    {\it Right:} The Compton ``correction factor'' $\Psi(x,y) \equiv L_\nu^{\rm C} / L_\nu^{\rm ff}$ (eq.~\ref{eq:Appendix_Psi}) as a function of Compton-y and for frequencies $x=10^{-1}$ (solid red) and $10^{-3}$ (thin grey). In the former case, the thick-grey curve shows the result of numerically integrating the source function (as shown in the left panel) while the red curve shows the analytic approximation using eq.~(\ref{eq:Appendix_I1}). The two begin to deviate at very large $y$, once $x_{\rm w}(y) \lesssim x$ and the Wien bump dominates at the observed frequency (eq.~\ref{eq:Appendix_xw}).
    The dashed black curve shows the broken power-law approximation given in the bottom row of eq.~(\ref{eq:Appendix_Psi}). This rough approximation is correct to within a factor of $\sim 3$ for any $x<10^{-1}$ (and improved accuracy for lower frequencies).
    }
    \label{fig:Appendix_spectrum}
\end{figure*}

Equation~(\ref{eq:Appendix_Psi}) shows that Comptonization does not appreciably change the specific luminosity of bremsstrahlung emission at low frequencies of interest, $x_{\rm coh} \ll x \ll 1$, except for very high Compton-y parameters $\gg y_{\rm crit} \sim \mathcal{O}(10^3)$. This is illustrated in Fig.~\ref{fig:Appendix_spectrum},
which shows the emergent spectrum as a function of $y$.
For $y \ll 1$, the spectrum and luminosity are not appreciably altered by inverse Compton scattering (the luminosity increases as $1+y$ in this regime; see eq.~\ref{eq:Appendix_Cy_final}). For $1 \lesssim y \lesssim y_{\rm sat}$, the total luminosity increases as $\propto y^2$, and the spectrum---now regulated by inverse-Compton scattering---is relatively flat, $L_\nu \propto \ln(\nu)^{-1/2}$. The fact that the emergent spectrum in this regime is not dramatically changed from the un-Comptonized case is a coincidence due to the similarly $\sim$flat free-free source function. 
Finally, at $y > y_{\rm sat}$, the Wien peak becomes apparent in the spectrum. In this saturated regime, the total luminosity does not continue to grow with increasing Compton-y (eq.~\ref{eq:Appendix_Cy_final}). Instead the Wien peak remains fixed while the flat portion of the spectrum decreases in luminosity as $\propto y^{-1/2}$ (eq.~\ref{eq:Appendix_Psi}).

We note that for practical purposes, the Compton-y parameter of many astrophysical systems will be $y < \ln (x/x_{\rm coh}) \sim \mathcal{O}(10)$. In this regime the Compton cooling correction (eq.~\ref{eq:Appendix_Psi}) attains a particularly simple form,
\begin{equation}
\label{eq:Appendix_Psi_lowy}
    y<\ln\left({x}/{x_{\rm coh}}\right) :~
    \Psi \left(x,y\right)
    \underset{
    x<1 
    }{=} 
    1 + \frac{y}{10 \ln\left(2.2/x\right)} .
\end{equation}
Note that eq.~(\ref{eq:Appendix_Psi_lowy}) implies $\Psi \sim 1$ for low $y$, consistent with the bottom row of eq.~(\ref{eq:Appendix_Psi}).

We conclude with an estimate of the ratio $\Delta$ between the peak (Wien) and flat spectral luminosities.
We define a second integral term,
\begin{align}
\label{eq:Appendix_Iw}
I_{\rm w}(y) 
&\equiv 
\int_{\max(x_{\rm coh},e^{-y})}^1 dx^\prime \ln\left(\frac{2.2}{x^\prime}\right)
\\ \nonumber 
&\times 
\frac{1}{x^\prime} \left[ 1 - \left(\frac{\ln (1/x^\prime)}{y}\right)^{1/2} \right]^2
\\ \nonumber
&\underset{y \gg y_{\rm sat}}{\approx}
y_{\rm sat} \left( \frac{1}{2} y_{\rm sat} + \ln 2.2 \right)
,
\end{align}
that enters eq.~(\ref{eq:Appendix_Ndot_w}).
With eqs~(\ref{eq:Appendix_Lnu_flat},\ref{eq:Appendix_Lnu_w},\ref{eq:Appendix_I2},\ref{eq:Appendix_Iw}) we find that
\begin{align}
\label{eq:Appendix_Delta}
\Delta(y) 
&\equiv \frac{L_\nu^{\rm w}(x=3)}{L_\nu(x=1)} 
\approx \frac{2}{3} \frac{ \frac{1}{2}\left(\frac{3}{e}\right)^3 I_{\rm w}(y)}{I(1,y)}
\\ \nonumber
&\underset{y \gg y_{\rm sat}}{\approx}
\frac{1}{4} \left(\frac{3}{e}\right)^3 y_{\rm sat}^{1/2} y^{1/2}
.
\end{align}
\cite{Illarionov&Sunyaev72} considered only the saturated Comptonized regime and comment that $\Delta = \ln ( x_{\rm coh}^{-1} ) y^{1/2} = y_{\rm sat} y^{1/2}$, but do not provide a derivation for this (see also \citealt{Felten&Rees72}).
Equation~(\ref{eq:Appendix_Delta}) recovers the same $\propto y^{1/2}$ scaling in the saturated regime, albeit with a prefactor $\simeq 3 y_{\rm sat}^{1/2} \sim 10$ smaller than quoted by \cite{Illarionov&Sunyaev72}.

A final parameter of relevance in the Comptonized regime is the transition frequency $x_{\rm w}$ between the $\sim$flat Comptonized spectrum (eq.~\ref{eq:Appendix_Lnu_flat}) and the Wien peak (eq.~\ref{eq:Appendix_Lnu_w}).
We focus on the saturated Comptonized regime, $y>y_{\rm sat}$, in which the Wien peak is pronounced and of greater relevance (at low $y$, the Wien portion of the spectrum becomes buried below the flat spectrum). In this regime, $x_{\rm w}$ can be found as a solution to the transcendental equation
\begin{equation}
L_\nu(x_{\rm w}) = L_\nu^{\rm w}(x_{\rm w}) 
~~\leftrightarrow~~
\Delta\left(y\right) \frac{x_{\rm w}^3 e^{-x_{\rm w}}}{3^3 e^{-3}} \approx 1
.
\end{equation}
Given that $x_{\rm w} \lesssim 1$, the exponential term in the equation above can be Taylor expanded to obtain a closed-form analytic solution for $x_{\rm w}$, 
\begin{equation}
\label{eq:Appendix_xw}
x_{\rm w}(y) \underset{y \gg y_{\rm sat}}{\approx}
\left(\frac{y_{\rm sat}y}{16}\right)^{-1/6} + \frac{1}{3} \left(\frac{y_{\rm sat}y}{16}\right)^{-1/3} + ...
\end{equation}
where the second term is typically only a small correction, and we have made use of eq.~(\ref{eq:Appendix_Delta}).

\subsection{Upscattering of Soft Synchrotron Photons}
\label{sec:Appendix_Synch_Comptonization}

Above we have considered IC scattering of soft photons by thermal electrons, where the source of soft photons was assumed to be bremsstrahlung emission by the same thermal electrons. Here we consider an alternative possibility, that the soft photon field is dominated by low-frequency synchrotron emission from non-thermal electrons.

The absorption coefficient of synchrotron radiation for a power-law electron distribution with $p=3$ is given by \citep{Rybicki&Lightman}
\begin{equation}
    \alpha_\nu = 
    \frac{9e^2 m_p}{32\pi m_e^2 c^2} \Gamma\left(\frac{11}{12}\right) \Gamma\left(\frac{31}{12}\right) \epsilon_e n v^2 \left(\frac{eB}{2\pi m_e c}\right)^{5/2} \nu^{-7/2} 
    .
\end{equation}
Taking the magnetic field to be $B = \sqrt{16\pi \epsilon_B n m_p v^2}$, the timescale for synchrotron self-absorption (SSA) of a photon at frequency $x$ ($=h\nu/k_B T_e$) is 
\begin{align}
    t_{\rm abs}^{\rm syn}(x) = \frac{1}{\alpha_\nu(x) c}
    \approx 4 \times 10^{18} \,
    &{\rm s} \,
    \epsilon_{e,-1}^{-1} \epsilon_{B,-1}^{-5/4} \epsilon_T^{7/2} 
    \\ \nonumber
    &\times M_{-1}^{-9/4} R_{15}^{27/4} v_9^{5/2} x^{7/2}
    .
\end{align}
Clearly, SSA is irrelevant for hard X-ray photons (where $x \sim 1$), however the strong frequency dependence $t_{\rm abs}^{\rm syn} \propto x^{7/2}$ makes synchrotron absorption important at lower frequencies.
Equating $t_{\rm abs}^{\rm syn}$ to the IC scattering timescale $t_{\rm IC}$ (eq.~\ref{eq:t_IC}) yields the frequency $x_{\rm coh}^{\rm syn}$ above which IC scattering of synchrotron photons becomes effective (that is, at frequencies $x>x_{\rm coh}^{\rm syn}$ photons can be IC scattered before they are absorbed, $t_{\rm IC} < t_{\rm abs}^{\rm syn}$). It is
\begin{equation}
    x_{\rm coh}^{\rm syn} \approx
    4 \times 10^{-5} \epsilon_{e,-1}^{2/7} \epsilon_{B,-1}^{5/14} \epsilon_T^{-9/7} M_{-1}^{5/14} R_{15}^{-15/14} v_9^{-9/7} .
\end{equation}
This is the direct analog of $x_{\rm coh}$ for bremsstrahlung emission that was previously discussed (eq.~\ref{eq:x_coh}).

The above estimates assume that the non-thermal electrons are slow-cooling, that is, that $t_{\rm syn}(x) > t_{\rm dyn}$. However this is only correct at frequencies less than the synchrotron cooling frequency
\begin{equation}
    x_{\rm cool} 
    \approx 2.5 \times 10^{-15} \, \epsilon_{B,-1}^{-3/2} M_{-1}^{-3/2} R_{15}^{5/2} v_9^{-3} \epsilon_T^{-1}
    .
\end{equation}
The frequencies of relevance in our current problem lie well above $x_{\rm cool}$, and therefore the above estimates should be modified to account for the fast-cooling regime.

In the fast-cooling regime, the absorption  coefficient above should be replaced with $\alpha_\nu \to \alpha_\nu t_{\rm syn}(x)/t_{\rm dyn} = \alpha_\nu (x / x_{\rm cool})^{-1/2}$. Therefore the SSA timescale is $t_{\rm abs}^{\rm syn} \to t_{\rm abs}^{\rm syn} (x/x_{\rm cool})^{1/2}$,
\begin{equation}
    t_{\rm abs}^{\rm syn}(x>x_{\rm cool}) 
    \approx 8 \times 10^{25} \, {\rm s} \,
    \epsilon_{e,-1}^{-1} \epsilon_{B,-1}^{-1/2} \epsilon_T^{4} 
    M_{-1}^{-3/2} R_{15}^{11/2} v_9^{4} x^{4}
    .
\end{equation}
This implies that
\begin{equation}
    x_{\rm coh}^{\rm syn}(>x_{\rm cool}) \approx
    2 \times 10^{-6} \epsilon_{e,-1}^{1/4} \epsilon_{B,-1}^{1/8} \epsilon_T^{-5/4} M_{-1}^{1/8} R_{15}^{-5/8} v_9^{-3/2} .
\end{equation}

We can now compare the production rate of soft synchrotron photons vs soft bremsstrahlung photons. It is
\begin{align}
    &\left.\frac{\dot{n}_{\rm syn}(x)}{\dot{n}_{\rm ff}(x)}\right\vert_{x_{\rm coh}^{\rm syn}}
    \approx
    \\ \nonumber
    &\max
    \begin{cases}
    1.3 \times 10^5 \, \epsilon_{e,-1}^{5/8} \epsilon_{B,-1}^{1/16} \epsilon_T^{31/8} M_{-1}^{-15/16} R_{15}^{35/16} v_9^{39/4}
    \\
    2.3 \times 10^2 \, \epsilon_{e,-1}^{5/8} \epsilon_{B,-1}^{1/16} \epsilon_T^{27/8} M_{-1}^{1/16} R_{15}^{3/16} v_9^{31/4}
    \end{cases}
    ,
\end{align}
where the top (bottom) case corresponds to an adiabatic (radiative) shock.
The above estimate shows that synchrotron emission produces far more photons than free-free emission does at frequency $x_{\rm coh}^{\rm syn}$.
Naively, this suggests that the IC cooling rate of thermal electrons should be set by soft synchrotron photons rather than---as estimated in the previous subsections---by bremsstrahlung photons.
However, because the (non-thermal) synchrotron-emitting electrons are fast-cooling at frequencies of interest, photon production at these frequencies is confined to within a thin shell of relative width $\sim t_{\rm syn}(x_{\rm coh}^{\rm syn})/t_{\rm dyn} = \left(x_{\rm coh}^{\rm syn}/x_{\rm cool}\right)^{-1/2} \sim 4 \times 10^{-5}$ behind the shock front.
The Compton-y parameter relevant for upscattering of these photons is therefore $\sim 4 \times 10^{-5}$ times smaller than the Compton-y throughout the full CSM shell (as relevant in the case of bremsstrahlung emission). This implies that the effective $y$ of synchrotron photons is always $\ll 1$, and that therefore---IC scattering of soft synchrotron photons is irrelevant in our current scenario (despite the fact that synchrotron emission produces more soft photons than bremsstrahlung). Our estimates from the previous subsections are therefore appropriate, even in the presence of non-thermal synchrotron emission.

\section{Ionization Breakout}
\label{sec:Appendix_IonizationBreakout}

In the following we derive conditions under which X-rays emitted by the shock can escape without undergoing significant bound-free absorption by upstream matter. This can differ from the simple condition $\tau_{\rm bf} \lesssim 1$ (eq.~\ref{eq:tau_bf}) because the shock-produced X-rays can feed back on their environment by photoionizing the upstream and reducing the column density of matter that contributes to bound-free absorption.
Such `ionization breakout' has been discussed in previous work (e.g. \citealt{Metzger+14,Margalit+18} in the context of SLSNe). Here we present and extend key results of these papers, applying them to the shock-powered scenario of interest in this work.

The main properties of ionization breakout can be understood using a Stromgren-sphere-like analysis.
In this treatment, an infinite homogeneous medium is considered, and it is assumed that a photon that is emitted by a recombining ion is immediately re-absorbed by photoionizing a neighboring ion (the so-called `on-the-spot' approximation).
This approximation dictates that, in photoionization equilibrium (steady-state), the number of free electrons is equal to the number of ionizing photons emitted over a recombination timescale, $N_e \approx Q_0 t_{\rm rec}$, where $t_{\rm rec} \sim \left( \alpha_{\rm rr} n_e \right)^{-1}$ is the recombination time ($\alpha_{\rm rr}$ is the radiative recombination rate) 
and $Q_0 = \int \left(L_\nu / h\nu \right) d\nu$ is the emission rate of ionizing photons.

The classic Stromgren analysis determines the size of the photoionized region $d_{\rm s}$ by equating $N_e \sim Q_0 t_{\rm rec}$ to the number of free electrons within a volume enclosed by the Stromgren region. For a homogeneous medium this is $N_e \sim 4\pi R^2 d_{\rm s} n_e$ in the planar regime ($d_{\rm s} \lesssim R$) that is relevant for the compact CSM shells considered in this work ($\Delta R \lesssim R$).
This yields 
\begin{equation}
\label{eq:Appendix_ds}
d_{\rm s} = 
\frac{Q_0}{4\pi R^2 X_A \alpha_{\rm rr} n_e^2}
,
\end{equation}
which is an analog of the Stromgren-sphere result for planar geometry.\footnote{In the spherical regime ($d_{\rm s} \gg R$) we instead have $N_e \sim 4\pi d_{\rm s}^3 n_e/3$ and the result reduces to the standard Stromgren-sphere, $d_{\rm s} = \left({3 Q_0}/{4\pi \alpha_{\rm rr} n_e^2}\right)^{1/3}$.} In the above, $X_A$ is the fractional number density of ion $A$ of interest (so that $d_{\rm s}$ is the ionization depth of this ion).

The result above gives the steady-state solution, achieved at $t \gtrsim t_{\rm rec}$. How does the ionization front expand towards this equilibrium?
At times $t \ll t_{\rm rec}$ every ionized atom has not yet had time to recombine, therefore the number of ionized electrons is simply set by the number of ionizing photons emitted during this time, $N_e(t) \sim Q_0 t$, 
which implies $d_{\rm s} \propto t$ at $t \lesssim t_{\rm rec}$ 
and that the ionization front phase velocity is $\sim d_{\rm s}/t_{\rm rec}$
(in the spherical regime $d_{\rm s} \propto t^{1/3}$ instead).

Ionization breakout through the full CSM shell requires that both $d_{\rm s} \geq \Delta R$ and $d_{\rm s}/t_{\rm rec} \gg v$. The second condition ensures that the ionization front phase velocity exceed the shock velocity. Typically $t_{\rm rec} \ll t_{\rm dyn}$ so that this condition is automatically satisfied if $d_{\rm s} \geq \Delta R$.

Because $d_{\rm s} \propto Q_0 \propto L_X$, ionization breakout requires high ionizing luminosity.
In the present scenario, the breakout conditions can be cast instead as a requirement on the shock velocity. We express the ionizing luminosity as a fraction $\epsilon_X$ of the kinetic shock power $L_{\rm sh} \propto v^3$ (eq.~\ref{eq:Lsh}), which gives $Q_0 \sim \epsilon_X L_{\rm sh} / h\nu$ and yields the ionization breakout condition
\begin{equation}
v 
> 
\max \left[ 
\left(\frac{2 h\nu R n_e X_A \alpha_{\rm rr}}{\epsilon_X \mu_e m_p}\right)^{1/3}
,
\left(\frac{2 h\nu}{\epsilon_X \mu_e m_p}\right)^{1/2}
\right]
\end{equation}
or quantitatively,
\begin{equation}
\label{eq:Appendix_v_breakout}
v_9 > 
\max
\begin{cases}
5.1
\, 
M_{-1}^{1/3} R_{15}^{-2/3} X_{A,-2}^{1/3} \alpha_{{\rm rr},-12}^{1/3} \nu_{\rm keV}^{1/3} \epsilon_{X,-3}^{-1/3}
\\
1.3
\, \epsilon_{X,-3}^{-1/2} \nu_{\rm keV}^{1/2}
\end{cases}
.
\end{equation}
This is an implicit relation for $v$ because $\epsilon_X$ (normalized above to a fiducial $\epsilon_{X,-3} = \epsilon_X/10^{-3}$) depends on the emission process and is in general also a function of velocity $v$ (eqs.~\ref{eq:epsilon_syn},\ref{eq:epsilon_ff}).
Above we have normalized $X_{A,-2} = X_A/10^{-2}$ and $\alpha_{{\rm rr},-12} = \alpha_{\rm rr} / 10^{-12} \, {\rm cm}^3 \, {\rm s}^{-1}$, roughly appropriate to $^{20}$Ne, whose K-shell edge falls immediately below $1 \, {\rm keV}$ \citep[e.g.][]{Wilms+00}.

The Stromgren analysis described above neglects optical depth effects, in essence assuming that the bound-free optical depth $\tau_{\rm bf} \to \infty$ outside the ionization front and $\tau_{\rm bf} \approx 0$ interior to it (i.e. that this region is fully ionized).
Clearly, if $\tau_{\rm bf} \lesssim 1$ through the CSM shell then X-rays freely escape regardless of photoionization/recombination, and the Stromgren analysis is irrelevant.
To connect these two regimes we proceed by calculating the penetration depth of ionizing photons accounting for the bound-free optical depth, i.e. relaxing the previous implicit assumption that the neutral fraction is $f_{\rm n} = 0$ interior to the ionization front.
This closely follows \cite{Metzger+14}.

In {\it local} ionization-recombination equilibrium, the neutral fraction of the CSM at a depth $r$ is
\begin{equation}
\label{eq:Appendix_f_n}
f_{\rm n}(r) = \left[ 1 + \frac{4\pi}{\alpha_{\rm rr} n_e} \int_{\nu_{\rm min}(r)} \frac{J_\nu(r) \sigma_{\rm bf}(\nu)}{h\nu} \right]^{-1} ,
\end{equation} 
where $J_\nu$ is the specific intensity of ionizing radiation, $\sigma_{\rm bf}$ is the bound-free cross section, and $\nu_{\rm min}(r)$ the minimal frequency of photons that can penetrate a depth $r$.
The effective optical-depth of the CSM to photons of frequency $\nu$ is 
\begin{equation}
\tau_{\rm eff}(r) = \int_0^r n_A \sigma_{\rm bf}(\nu) f_{\rm n}(r^\prime) \left( 1 + n_e \sigma_T r^\prime \right) \, dr^\prime .
\end{equation}
This accounts for the additional path-length traveled by photons in the large scattering optical-depth regime, $\tau_{\rm T} \gtrsim 1$. Following Appendix~B of \cite{Metzger+14} but retaining the first term in brackets in eq.~(\ref{eq:Appendix_f_n}) above (which is neglected in \citealt{Metzger+14}; c.f. their eq.~B3), we find that
\begin{align}
\label{eq:Appendix_local_equilibrium}
\left( 1 + n_e \sigma_T r \right) \, dr
=
3 &\left[
\frac{1}{n_A \sigma_{\rm bf}(\nu_{\rm min})}
\right.
\\ \nonumber
&+
\left.
\frac{4\pi J_{\nu_{\rm min}}}{(\Gamma+2) h \alpha_{\rm rr} n_e n_A}
\right] \, d\ln\nu_{\rm min}
,
\end{align}
where $\Gamma$ is the photon index ($J_\nu \propto \nu^{1-\Gamma}$; $\Gamma=1$ for the typical case of bremsstrahlung emission) and we have assumed that $\sigma_{\rm bf}(\nu) \propto \nu^{-3}$. This is analogous to eq.~B5 in \cite{Metzger+14}.
At $\nu_{\rm min} = \nu$, the ionizing photon penetration depth $d_{\rm p.i.}$ is defined such that $\tau_{\rm eff}(d_{\rm p.i.}) = 1$. This can be calculated by integrating eq.~(\ref{eq:Appendix_local_equilibrium}) from $r=0$ to $r=d_{\rm p.i.}$, resulting in
\begin{align}
\label{eq:Appendix_taupi}
\tau_{\rm p.i.}^{-1} \equiv
\frac{d_{\rm p.i.}}{\Delta R}
&= \frac{\sqrt{1+2\tau_{\rm T} \left( \tau_{\rm bf}^{-1} + \tau_{\rm s}^{-1} \right)} - 1}{\tau_{\rm T}}
\\ \nonumber
&\approx
\begin{cases}
\tau_{\rm bf}^{-1} + \tau_{\rm s}^{-1} &;~ \tau_{\rm T} \ll 1
\\
\sqrt{ 2 \tau_{\rm T} \left( \tau_{\rm bf}^{-1} + \tau_{\rm s}^{-1} \right) } &;~ \tau_{\rm T} \gg 1
\end{cases}
\end{align}
where
\begin{align}
\tau_{\rm s}^{-1} 
\equiv 
\frac{\Gamma+2}{4\pi} \frac{d_{\rm s}}{\Delta R} 
&\approx 
1.9 \times 10^{-3} \,
\left(\frac{\Gamma+2}{3}\right) \epsilon_{X,-3}
\\ \nonumber
&\times
X_{A,-2}^{-1} \alpha_{{\rm rr},-12}^{-1} \nu_{\rm keV}^{-1} M_{-1}^{-1} R_{15}^2 v_9^3
\end{align}
is an effective Stromgren optical depth (see eq.~\ref{eq:Appendix_ds}), and $\tau_{\rm bf}$ the usual bound-free optical depth (eq.~\ref{eq:tau_bf}).

In the limit $\tau_{\rm bf} \gg 1$ and small scattering opacity ($\tau_{\rm T} \ll 1$), we recover the Stromgren result, namely $d_{\rm p.i.} \sim d_{\rm s}$.
However, in the regime where $\tau_{\rm bf} \lesssim \tau_{\rm s}$ then the optical-depth, rather than photoionization, is what sets the photon penetration depth. An alternative way to understand this regime is to note that $\tau_{\rm bf} \lesssim \tau_{\rm s}$ is satisfied when the photon mean-free path $\lambda_{\rm mfp} \sim 1/n_A \sigma_{\rm bf}$ exceeds the Stromgren distance $d_{\rm s}$. Clearly, in this case photons will not be photoelectrically absorbed within the region $r \lesssim d_{\rm s} < \lambda_{\rm mfp}$ so that the assumptions on which the Stromgren analysis are derived are clearly violated.
Indeed, in the standard Stromgren regime the neutral fraction within the ionized nebula can be shown to be $f_{\rm n} \approx (n_A \sigma_{\rm bf} d_{\rm s} )^{-1} = (d_{\rm s}/\lambda_{\rm mfp})^{-1} \ll 1$ because $\lambda_{\rm mfp} \ll d_{\rm s}$.

\subsection{Ionization Breakout from a Wind}
\label{sec:Appendix_Wind_IonizationBreakout}

In the following we consider the necessary conditions for X-rays to escape unattenuated from a wind density profile. Our primary motivation is considering whether a low-density stellar wind, that may surround the dense CSM shell, could inhibit X-ray emission due to photoelectric absorption. 
From a physics standpoint, this is an extension of our previous discussion in this Appendix to the case of a density profile that scales as $\rho \propto r^{-2}$.
Specifically, we take the wind density to be $\rho_{\rm w} = \dot{M}/4\pi v_{\rm w} r^2$.

There are three characteristic radii in this problem ($R_{\rm bf}$, $R_{\rm s}$, and $R_{\rm rec}$), which we consider below. The first is the radius at which the bound-free optical depth $\tau_{\rm bf}^{\rm w} = \int \kappa_{\rm bf} \rho_{\rm w} dr$ equals one,
\begin{align}
\label{eq:Appendix_Rbf}
    R_{\rm bf} \equiv \frac{\kappa_{\rm bf} \dot{M}}{4\pi v_{\rm w}}
    &\approx 5.3 \times 10^{14}\,{\rm cm}\,
    \left(\frac{\dot{M}}{10^{-5}\,M_\odot\,{\rm yr}^{-1}}\right) 
    \nonumber \\
    &\times 
    \left(\frac{v_{\rm w}}{100\,{\rm km \,s}^{-1}}\right)^{-1}
    \nu_{\rm keV}^{-8/3}
    .
\end{align}
If the CSM shell is located at $R > R_{\rm bf}$ then the bound-free optical depth of the surrounding wind is $< 1$ and X-rays can freely escape to an observer.
Alternatively, if $R < R_{\rm bf}$ then photoelectric absorption plays an important role. However X-rays may still be able to burrow out if the emission is luminous enough to photoionize the wind. We now consider this possibility.

We adopt a Stromgren-sphere approach, as discussed earlier in this Appendix. This is a steady-state approach that assumes that photoionization and recombination are in equilibrium. In this limit, the extent of the photoionized region is determined by equating the emission rate of ionizing photons $Q_0$ with the volume-integrated recombination rate within this region,
$\int_R^{R_{\rm s}} (4\pi r^2 n_e / t_{\rm rec}) \,dr$.
This yields the `Stromgren radius' $R_{\rm s}$ out to which the wind would be ionized,
\begin{equation}
\label{eq:Appendix_Rs_w_1}
    R_{\rm s} = \left( 1 - \frac{Q_0}{Q_{\rm crit}} \right)^{-1} R
\end{equation}
where 
\begin{align}
\label{eq:Appendix_Qcrit}
    Q_{\rm crit} 
    &\equiv \left(\frac{\dot{M}}{\mu_e m_p v_{\rm w}}\right)^2 \frac{\alpha_{\rm rr}}{4\pi R}
    \approx 8.1 \times 10^{44} \,{\rm s}^{-1}\,
    R_{15}^{-1} 
    \nonumber \\
    &\times 
    \alpha_{{\rm rr},-14}
    \left(\frac{\dot{M}}{10^{-5}\,M_\odot\,{\rm yr}^{-1}}\right)^2 
    \left(\frac{v_{\rm w}}{100\,{\rm km \,s}^{-1}}\right)^{-2}
    .
\end{align}
Eq.~\ref{eq:Appendix_Rs_w_1} implies a sharp transition: for low ionizing luminosities (low $Q_0$) only a small region $\sim R_{\rm s}-R \ll R$ ahead of the shell would be photoionized, while for $Q_0 \geq Q_{\rm crit}$ the entire wind would be completely ionized ($R_{\rm s} \to \infty$ is implied by negative values in eq.~\ref{eq:Appendix_Rs_w_1}).
This occurs because, for a wind density profile, the volume-integrated recombination rate is dominated by material at small radii. Once photoionization manages to overcome this interior region ($R_{\rm s} \gtrsim 2R$) then the rest of the wind would be easily photoionized as well.
This can also be seen by observing that the condition $Q_0 \gtrsim Q_{\rm crit}$ for ionization breakout (eq.~\ref{eq:Appendix_Rs_w_1}) is nearly identical to the requirement $d_{\rm s} \gtrsim R$ for ionization breakout from a shell (eq.~\ref{eq:Appendix_ds}), if we consider a shell of density $n_e = \rho_{\rm w}(R)/\mu_e m_p$.

The expression above is however only valid in steady state. The recombination timescale is $t_{\rm rec} = (\alpha_{\rm rr} n_e)^{-1} \propto r^2$ in a wind density profile, so at large radii recombination becomes very long and the steady-state assumption must break down.
Specifically, for a transient source of ionizing X-ray radiation that is active for duration $t_X$, steady-state requires that $t_{\rm rec} < t_X$.
There is a critical radius $R_{\rm rec}$ at which $t_{\rm rec}(R_{\rm rec}) = t_X$,
\begin{align}
\label{eq:Appendix_Rrec}
    R_{\rm rec} 
    &= \left(\frac{\dot{M} \alpha_{\rm rr} t_X}{4\pi \mu_e m_p v_{\rm w}}\right)^{1/2}
    \approx 4.7 \times 10^{14}\,{\rm cm}\,
    \alpha_{{\rm rr},-14}^{1/2}
    \\ \nonumber
    &\times 
    \left(\frac{\dot{M}}{10^{-5}\,M_\odot\,{\rm yr}^{-1}}\right)^{1/2} 
    \left(\frac{v_{\rm w}}{100\,{\rm km \,s}^{-1}}\right)^{-1/2}
    \left(\frac{t_X}{100\,{\rm d}}\right)^{1/2}
    .
\end{align}

At radii $r > R_{\rm rec}$ the recombination time is longer than the transient duration and steady-state cannot be achieved.
The photoionization breakout condition in this regime is altered from eq.~(\ref{eq:Appendix_Rs_w_1}). When the recombination time is longer than $t_X$ one must consider the total {\it number} of ionizing photons produced by the source, rather than the rate. The radius out to which wind material can be ionized is therefore set by equating $Q_0 t_X$ to the total number of available free electrons enclosed within this radius, $\int^{R_{\rm s}} 4\pi n_e r^2\,dr$.
This gives
\begin{align}
    R_{\rm s} = Q_0 t_X \frac{\mu_e m_p v_{\rm w}}{\dot{M}}
    = R \frac{Q_0}{Q_{\rm crit}} \left(\frac{R}{R_{\rm rec}}\right)^{-2}
\end{align}
which has no dependence on the recombination rate. This is expected since recombination is long and does not play a role in this regime.

Combining the two regimes above, we find that the overall photoionization (`Stromgren') radius for a wind density profile can be expressed as
\begin{equation}
\label{eq:Appendix_Rs_final}
    R_{\rm s} = R
    \begin{cases}
    \left( 1 - \frac{Q_0}{Q_{\rm crit}} \right)^{-1}
    &, \,0<R_{\rm s}<R_{\rm rec}
    \\
    \frac{Q_0}{Q_{\rm crit}} \left(\frac{R}{R_{\rm rec}}\right)^{-2}
    &, \,{\rm else}
    \end{cases}
\end{equation}
where $Q_{\rm crit}$ and $R_{\rm rec}$ are given by eqs.~(\ref{eq:Appendix_Qcrit},\ref{eq:Appendix_Rrec}).
X-rays will be photoelectrically absorbed by the wind only if both $R<R_{\rm bf}$ (bound-free absorption optical depth is large) and $R_{\rm s} < R_{\rm bf}$ (radiation cannot photoionize its way out to the optically-thin region).
In any other case (other orderings of $R$, $R_{\rm bf}$, $R_{\rm s}$ and $R_{\rm rec}$) X-rays will be free to escape to an external observer.
In practice, the condition $R_{\rm s}<R_{\rm bf}$ is usually set by the bottom case in eq.~(\ref{eq:Appendix_Rs_final}; the non steady-state case), so that
\begin{align}
\label{eq:Appendix_wind_breakout_condition}
    \left(\frac{L_X}{10^{41}\,{\rm erg\,s}^{-1}}\right) &\left(\frac{t_X}{100\,{\rm d}}\right)
    < 3.2 \times 10^{-5} \,
    \nu_{\rm keV}^{-5/3}
    \\ \nonumber
    &\times
    \left(\frac{\dot{M}}{10^{-5}\,M_\odot\,{\rm yr}^{-1}}\right)^{2} 
    \left(\frac{v_{\rm w}}{100\,{\rm km \,s}^{-1}}\right)^{-2}
\end{align}
as well as $R<R_{\rm bf}$ (eq.~\ref{eq:Appendix_Rbf}) are required for X-rays to be bound-free absorbed. Above we have taken $Q_0 \sim L_X/h\nu$, where $L_X$ is the luminosity ($\nu L_\nu$) at frequency $\nu_{\rm keV}$.
Eq.~(\ref{eq:Appendix_wind_breakout_condition}) is only satisfied for extremely low luminosities, which are not relevant to the parameter space of interest in our present work (that is---where detectability prospects are more promising), unless $\dot{M}/v_{\rm w} \gg$ than values assumed above.
We therefore conclude that a `normal' (time-steady, low-density) stellar-wind cannot photoelectrically absorb X-rays produced by the CSM shell, and that these X-rays---if manage to escape the dense shell itself---would typically reach an external observer uninhibited.

\section{Pair Production and Relativistic Plasma Corrections}
\label{sec:Appendix_Pairs}

Throughout the main text we have neglected relativistic effects. At shock velocities of interest in standard SNe ($v_9 \sim 1$) relativistic corrections are expected to be minor. However care must be taken---the post-shock electron temperature is mildly relativistic ($k_B T_e \sim m_e c^2$) even for non-relativistic shock velocities (eq.~\ref{eq:Te}). 
Relativistic effects will introduce corrections to estimates of the post-shock temperature, the bremsstrahlung emissivity, and Comptonization. These amount to modest quantitative corrections that affect high-velocity shocks, but do not change the qualitative picture presented. A more serious concern has to do with the possibility of runaway pair-creation in a relativistic post-shock plasma. If the plasma temperature is high enough, then pair-creation processes may be able to produce copious pairs that dominate photon emission and scattering. This would have significant qualitative implications on the shock properties (for example, an initially collisionless shock might be able to generate enough pairs to drive the scattering optical-depth above $c/v$ and ``bootstrap'' its way to becoming a radiation-mediated shock).
In this Appendix we investigate this possibility and find that this scenario cannot be realized for initially pair-poor shocks.

We follow the formalism of \cite{Svensson82,Svensson84} who investigated steady-state pair-equilibrium solutions in mildly-relativistic thermal plasmas. We define a normalized temperature $\Theta \equiv k_B T_e / m_e c^2$ and denote the pair multiplicity as $z \equiv n_+ / n_p$, where $n_+$ is the positron pair density and $n_p$ the proton density.
\cite{Svensson82} showed that the solution phase-space can be characterized in terms of the plasma temperature and `proton' optical-depth (the Thompson optical depth, neglecting pairs), with only weak dependence on density. At low temperatures, there exist two steady-state solutions: a low-$z$ solution and a high-$z$ pair-dominated solution. In our context, there is no initial source of pairs in the system (the pair density in the upstream medium is assumed to be negligible). An initially pair-poor shock of this kind may be capable of producing enough pairs to settle onto the low-$z$ branch. Since the pair-multiplicity in this case remains low, pairs are effectively negligible and may safely be ignored.
However, \cite{Svensson82,Svensson84} show that there exists a critical temperature $\Theta_{\rm crit}(\tau_T)$ above which no equilibrium solution exists (see also \citealt{Guilbert&Stepney85}). At $\Theta > \Theta_{\rm crit}$ annihilation cannot balance pair-creation processes, runaway pair-creation takes place, and the plasma becomes pair-dominated with $z \gg 1$. This would dramatically change the shock properties and alter our previous estimates.

\begin{figure}
    \centering
    \includegraphics[width=0.5\textwidth]{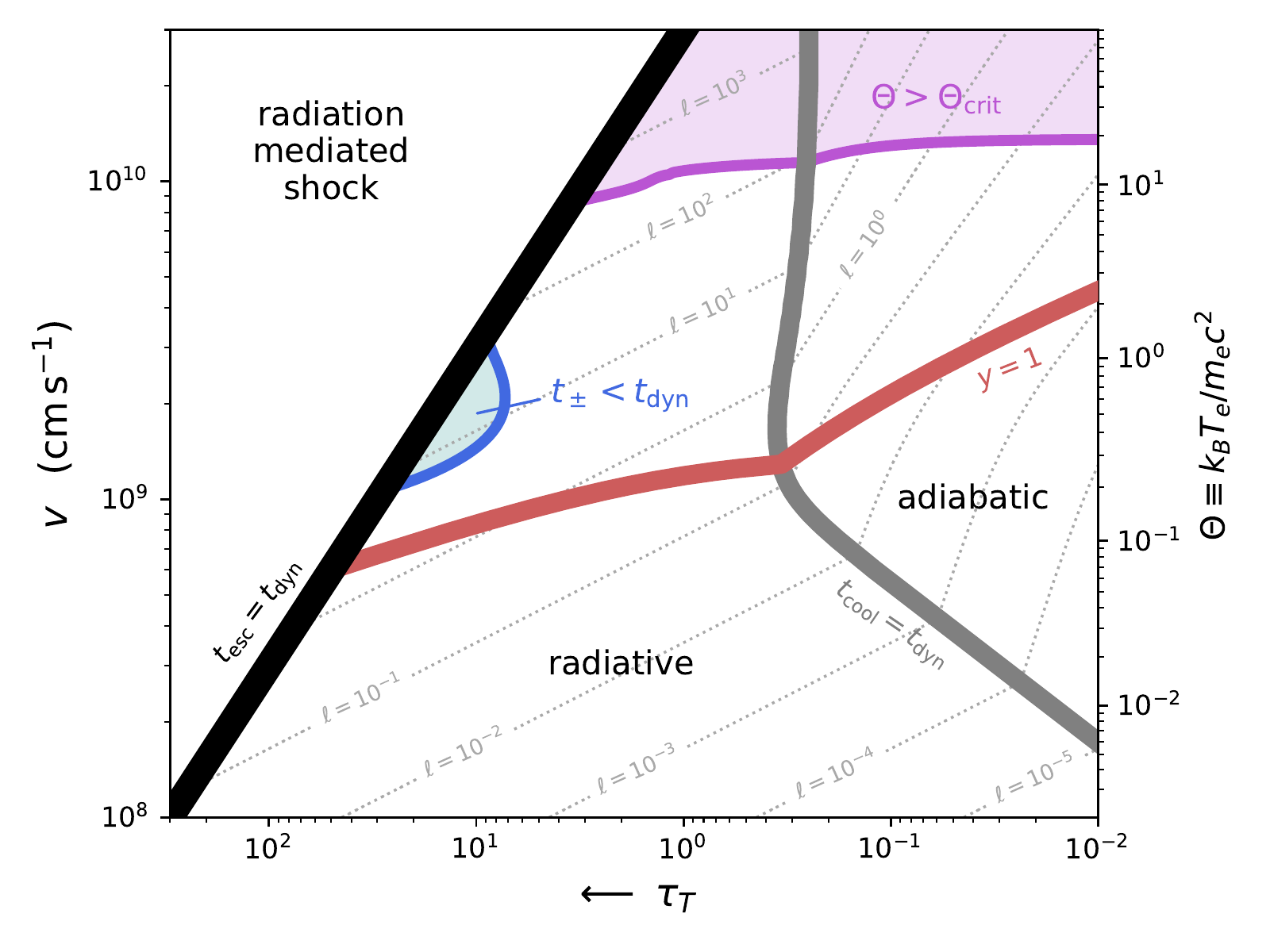}
    \caption{
    Same as Fig.~\ref{fig:PhaseSpace} but accounting for relativistic corrections to the post-shock temperature, bremsstrahlung emissivity, and Comptonization. Thick black, grey, and red curves delineate the shock phase-space into radiation-mediated / collisionless, adiabatic / radiative, and Comptonized / un-Comptonized shocks. These curves are in line with the simpler approximations shown in Fig.~\ref{fig:PhaseSpace}. The parameter space where post-shock electron temperatures exceed the critical pair-equilibrium solution is shaded in purple. Given sufficient time, runaway pair-creation would occur within this region, changing the nature of the shock and emerging radiation. In the scenario under consideration, the pair creation timescale is limited by the shock-crossing (dynamical) time. Regions where the pair-creation timescale is shorter than the dynamical time are shaded in blue. A pair-dominated shock (with $z > 1$) can only occur in regions where the blue and purple shaded areas overlap. The fact that there is no such overlapping region below $\tau_T < c/v$ implies that collisionless shocks cannot reach a pair-dominated state (absent an external source of pairs or radiation).
    Thin dotted grey curves show contours of constant compactness.
    }
    \label{fig:Appendix_pairs}
\end{figure}

We calculate the critical pair-equilibrium solution using the pair creation and annihilation rates $\dot{n}_\pm$, $\dot{n}_A$ from \cite{Svensson84}. Specifically, we include $\gamma\gamma$, $\gamma e$, $\gamma p$, $e e$, and $e p$ pair-creation processes. The photon density (which enters the $\gamma$ processes) is calculated assuming a one-zone steady-state escape model such that $n_\gamma = \dot{n}_\gamma t_{\rm esc}$, and we take electron-proton bremsstrahlung to be the dominant photon-generating process in $\dot{n}_\gamma$. The photon energy distribution is a combined bremsstrahlung + Wien spectrum, with a fraction $f_B$ of photons populating the Wien component, and $f_B$ calculated following Appendix C of \cite{Svensson84} (we adopt $\xi=1$ and $x_{\rm m}=10^{-10}$). At low temperatures ($\Theta < 1$) this Comptonization prescription is consistent with the results we derive in Appendix~\ref{sec:Appendix_Comptonization}.

To cast the results in the shock phase-space of Fig.~\ref{fig:PhaseSpace}, we must translate between shock velocity and CSM column density to the downstream temperature and effective proton optical depth. As discussed in \S\ref{sec:GeneralConsiderations}, the shock velocity governs the immediate downstream temperature. We adopt the temperature prescription $\Theta(v)$ from \cite{Margalit&Quataert21} with $\epsilon_T=1$ (their eqs. 2,3), that extends eq.~(\ref{eq:Te}) to the case of mildly-relativistic shock velocities.
The effective proton optical depth is simply the CSM optical depth $\tau_T$ so long as the shock is adiabatic. The effective optical depth is lower than $\tau_T$ when the shock is radiative because the hot post-shock region subtends only a small fraction $f$ of the total CSM width (see \S\ref{sec:GeneralConsiderations}). This limits the radiated power to the maximum available (the kinetic shock luminosity) such that $f = \min\left( 1, L_{\rm sh} / L_{\rm rad} \right)$.
Here $L_{\rm rad} = \mathcal{C} L_{\rm ff}$, the free-free luminosity $L_{\rm ff}$ is calculated using eqs.~(17-18) of \cite{Svensson82}, and the Comptonization enhancement factor $\mathcal{C}$ is calculated from eq.~C7 of \cite{Svensson84}. These extend $L_{\rm ff}$ and $\mathcal{C}$ used throughout the rest of this paper to mildly-relativistic temperatures.
A root-solving procedure is then used to calculate the effective optical depth at each point, $(1+\tau_{\rm eff}) \tau_{\rm eff} = f(\tau_{\rm eff},\tau_T,\Theta) \times (1+\tau_T) \tau_T$, where $f$ is a function of $\tau_{\rm eff}$ through the Comptonization enhancement factor $\mathcal{C}$.
This approach of replacing $\tau_T \to \tau_{\rm eff}$ ensures a correct calculation of Comptonization and of the photon density $n_\gamma$, taking into account the diffusion time of photons through the full CSM along with the fact that photon production and IC scattering takes place only within a (potentially small) fractional width $f$ of this region.

The results are presented in Figure~\ref{fig:Appendix_pairs}, which shows the shock phase-space accounting for relativistic corrections. The grey curve separates collisionless shocks into adiabatic (right) and radiative (left) cases. The cooling timescale here is calculated using expressions for the bremsstrahlung emissivity and Comptonization from \cite{Svensson82,Svensson84}, as described above. The red curve delineates the parameter-space into Comptonized (above) and un-Comptonized (below) regimes, where the relativistically appropriate Compton-y parameter is taken to be $y = \tau_{\rm eff} (1+\tau_{\rm eff}) ( 4\Theta + 16\Theta^2 )$.
The purple curve shows the critical pair-equilibrium solution, along which $\Theta = \Theta_{\rm crit}(\tau_{\rm eff})$. Above this curve (purple shaded region) no steady-state pair-equilibrium solution is possible. Given sufficient time, runaway pair creation would occur within this region of parameter-space. However in practice, this process turns out to be limited by the available time $\sim t_{\rm dyn}$.

The doubling-time of pairs $\sim z/\dot{z}$ is arbitrarily short if $z \to 0$ or $z \to \infty$, so that the timescale for runaway pair-creation is limited by the doubling rate at $z \sim 1$. We therefore define a critical pair-creation timescale as 
$t_\pm \equiv \left. z n_p / \dot{n}_\pm(z) \right\vert_{z=0.5}$.
The blue curve in Fig.~\ref{fig:Appendix_pairs} shows the contour along which $t_\pm = t_{\rm dyn}$. To the left of this curve (blue shaded region) the pair-creation timescale is shorter than $t_{\rm dyn}$ and the plasma could in-principle reach a pair-dominated state within the shock crossing time.
The system will only manifest this state if both the runaway pair-creation condition ($\Theta > \Theta_{\rm crit}$) and the timescale criterion ($t_\pm < t_{\rm dyn}$) are satisfied. Fig.~\ref{fig:Appendix_pairs} shows that there exists no parameter space where both conditions apply (no intersection of the blue and purple shaded regions) in the collisionless-shock regime. This implies that a pair-dominated state is never realized for such shocks.

We note that the primary reason that runaway pair creation does not take place in the shock scenario is the limit on the radiative luminosity $L_{\rm rad} \leq L_{\rm sh}$. In the AGN context, where this limit is irrelevant, it is well recognized that pair-production can be critical even at low temperatures $\Theta \sim 0.1$ (e.g. \citealt{Fabian+15}). This case corresponds to the high optical-depth branch of the \cite{Svensson84} $\Theta_{\rm crit}$ solution which cannot be realized due to the reduction of the effective optical depth in the radiative shock regime. To further illustrate this, light dotted-grey curves in Fig.~\ref{fig:Appendix_pairs} show contours of constant compactness, $\ell \equiv L \sigma_T / m_p c^3 R$. The cap on the radiated luminosity $L \leq L_{\rm sh}$ can be seen by the kink in these contours at the $t_{\rm esc}=t_{\rm dyn}$ curve. One can map the $\Theta_{\rm crit}$ curve from $\{ \tau_T$,$\Theta \}$ space into an $\{ \ell$,$\Theta \}$ parameter space using these contours, facilitating comparison with e.g. Fig. 1 of \cite{Fabian+15}.


\bibliography{bib}{}
\bibliographystyle{aasjournal}



\end{document}